\documentclass[approved]{PPGCosmoThesis}
\usepackage{epigraph}
\usepackage{bm}
\usepackage{glossaries}
\newtheorem{theorem}{Theorem}

\begin{document}

\title{Accelerated expansion as manifestation of gravity: when Dark Energy belongs to the left}
\author{Leonardo Giani}
\thesisyear{2020}
\supervisor{Prof.~Oliver Piattella}
\supervisoraddress{Universidade Federal do Espírito Santo, Vitória, Brasil}
\cosupervisor{Prof.~Luca Amendola}
\cosupervisoraddress{Institut für Theoretische Physik der Universität Heidelberg, Germany}

\dedicatory{0} 
% The value of 0 does not print the dedicatory page. 
% Here is an example: \dedicatory{To whoever made this wonderful thesis template.}
% If needed, use \\ to break lines.

\banca{
	Prof. Dr. Oliver Fabio Piattella (UFES), orientador,\\ 
	Prof. Dr. Luca Amendola (ITP Heidelberg, Alemanha), co-orientador, \\  
	Prof. Dr. Alexandr Kamenchtchik (UNIBO, Italy), examinador externo,\\
	Prof. Dr. Jorge Zanelli (CECs, Chile), examinador externo, \\
	Prof. Dr. Saulo Carneiro de Souza Silva (UFBA), examinador externo,\\
	Prof. Dr. Hermano Velten (UFOP), examinador externo, \\
	Prof. Dr. Júlio César Fabris (UFES), examinador interno.
	}

 %The \banca content only appears if the option "approved" is used.

\maketitle

%%%%%%%%%%%%%%%%%%%%%%%%% THE CHAPTERS %%%%%%%%%%%%%%%%%%%%%%%%%%

\chapter{Introduction}
\epigraph{\textit{The layman always means, when he says "reality" that he is speaking of something self-evidently known; whereas to me it seems that the most important and exceedingly difficult task of our time is to work on the construction of a new idea of reality.}}{Wolfgang Pauli}

This thesis is presented in candidacy for the degree of doctor of philosophy, and its main goal is to report and collect the scientific results we were able to achieve in the last four years towards the understanding of the fundamental nature of Dark Energy. The latter, whatever it is, became a fundamental ingredient in our description of the Universe after the discovery, in the late 90's, of its accelerating expansion~\cite{Riess:1998cb,Perlmutter:1998np}. In particular, our research focuses on those proposals in which Dark Energy is the manifestation of a different theory of gravitation, i.e. on those geometrical theories in which no new degrees of freedom are introduced to drive the present accelerated expansion.

However, it is a legitimate question to ask why we should consider such things when the current standard model of cosmology, the $\Lambda$CDM, has proven to be in excellent agreement with a number of different observations. From our point of view, there is indeed no completely satisfactory answer to this question since the cosmological constant is the simplest and yet effective candidate of Dark Energy we can think of. It does not introduce any new physics nor changes significantly the behavior of gravity at small scales (where General Relativity has been tested with astonishing precision). On the other hand, there are a number of less satisfactory answers which motivate the quest for a different description of Dark Energy. 

First, General Relativity is a century old and some physicists start to get bored, or at least frustrated, of it. There is no commonly accepted framework in which its quantization can be achieved, and it is extremely difficult to explore its properties in strong gravity regimes. For these reasons Cosmology, in particular at early and late times, is a fertile ground both for testing and speculating on the nature of gravitational interaction.

A slightly more satisfactory reason is that in the last decade we witnessed the appearance of a growing tension between the result of measurements from the local Universe and at early times. Part of the scientific community believe that these tensions are due to systematic, but a lot of people think that they are actually indications of new physics.
At the end of the day, whether there are good reasons for studying Dark Energy or not, it seems to us that the following quote by Weinberg remarkably describes the situation: \textit{"It seems that scientists are often attracted to beautiful theories in the way that insects are attracted to flowers — not by logical deduction, but by something like a sense of smell"}.

Motivated by the above considerations, we decided to dedicate chapters \ref{chapter:LCDM} and \ref{chapter:DE} of this thesis to a review of the $\Lambda$CDM model and of the landscape of Dark Energy candidates. The topics presented there are fairly standard and already covered in many textbooks, and the expert reader may freely decide to skip them.
On the other hand, in our treatment (by no means complete) we tried to privilege our personal perspectives on the topics, which ultimately provide the motivation for the work done in the subsequent chapters. 
The third chapter is also introductory in nature, and aims to review the main motivations which led to consider nonlocal modifications of gravity as Dark Energy candidate, as well as some technicalities typical of this approach.
The remaining chapters of this thesis contain instead a summary of our results. In chapter \ref{chapter:PCNL} we show some interesting cosmological features of the nonlocal model of gravity proposed in Ref.~\cite{Vardanyan:2017kal}, which we studied in Ref.~\cite{Giani:2019vjf}. Here we also try to explain, based on~\cite{Giani:2019xjf}, the apparently coincidental common behavior shared by different nonlocal models in the late stages of the evolution of the Universe. In chapter \ref{chapter:IR} we introduce a novel class of modified gravity models we recently proposed in  Ref.~\cite{Amendola:2020qho}. In chapter \ref{chapter:Lensing} we will discuss how to extract cosmological information from a novel type of observables proposed by us in Ref.~\cite{Piattella:2017uat} in the context of Strong Gravitational Lensing, and how they can be used to test Dark Energy and the Equivalence Principle~\cite{Giani:2020fpz}.
% !TEX root = ../thesis-example.tex
%
\chapter{Overview Of the $\Lambda$CDM model}
\label{chapter:LCDM}

\epigraph{\textit{Indeed, it has been said that democracy is the worst form of government except all those other forms that have been tried from time to time.}}{Winston Churchill}

The discovery of the accelerated expansion of the Universe~\cite{Riess:1998cb, Perlmutter:1998np} is undoubtedly one of the cornerstone of modern cosmology. After roughly two decades, it is commonly accepted that the best description of our universe on cosmological scales relies on the $\Lambda$CDM model. In this chapter we will give a brief introduction of the model,  highlighting its agreement with the main observational evidences and describing the theoretical framework at his foundation.
Finally, we will conclude the chapter mentioning some open problems of the standard model. Most of the material presented here is covered (surely better) in many standard textbooks, and was largely influenced by Refs.~\cite{Piattella:2018hvi,Amendola:2015ksp}, which I recommend for a detailed treatment.
%%%%%%%%%%%%%%%%%%%%%%%%%%%%%%%%%%%%%%%%%%%%%%%%%%%%%%%%%%%%%%%%%%%%%
\section{Theoretical grounds}
\label{sec:Theoretical_framework_LCDM}
\subsection{The Equivalence principle}
\label{subsec:EP}
One of the cornerstone that led Einstein to the formulation of General relativity is the Equivalence Principle. Historically, it is formulated with the statement that \textit{the gravitational and inertial mass are equivalent}, and it is also a pillar of Newtonian theory of gravity. Roughly 300 hundreds years after its verification by Galileo, Einstein realized that one of the consequences of the principle is that no static homogeneous external gravitational field could be detected from physics experiments performed by free-falling observers located in a sufficiently small spacetime region.

In the context of General Relativity, a useful statement of the Equivalence Principle is the following \cite{Weinberg:1972kfs} : \textit{at every spacetime point in an arbitrary gravitational field it is possible to choose a locally inertial coordinate system such that, within a sufficiently small region of the point in question, the laws of nature take the same form as in unaccelerated Cartesian coordinate system in the absence of gravitation}. Usually one refers to the above statement as the \textit{strong Equivalence Principle}, to distinguish it from the aforementioned equivalence between inertial and gravitational mass, which is instead labelled as \textit{weak Equivalence Principle}. 

Experimental tests of the Equivalence Principle, either in its weak of strong version, are of crucial importance for our fundamental understanding of gravity. Indeed, many alternative theories of gravitation result in some kind of violation of the Equivalence Principle, so that the precision within which we can trust its validity can be used to rule out a certain class of models.

%%%%%%%%%%%%%%%%%%%%%%%%%%%%%%%%%%%%%%%%%%%%%%%%%%%%%%%%%%%%%%%%%%%%%
\subsection{Einstein Field Equations}
\label{subsec:EFE_LCDM}
The main goal of a cosmological model is to describe the dynamical evolution of the Universe in agreement with data. In order to relate the dynamics of the Universe to its components, a theory of gravitation is required. The $\Lambda$CDM model assumes that the appropriate description of the gravitational interaction on cosmological scales is given by the Einstein Field Equations (EFE). We prefer to speak of Einstein Field Equations instead of General Relativity because, as showed in Ref.~\cite{BeltranJimenez:2019tjy}, it is possible to obtain the same equations from other geometrical theories that differs from GR at fundamental level. 

The Einstein Field Equations are:
\begin{equation}\label{EFE}
    R_{\mu\nu} - \frac{1}{2}g_{\mu\nu}R + \Lambda g_{\mu\nu} = 8\pi G_N T_{\mu\nu} \; ,
\end{equation}
where $R_{\mu\nu}$ is the Ricci tensor, $R$ its trace, $g_{\mu\nu}$ the spacetime metric, $G_N$ the Newton's constant, $\Lambda$ the Cosmological Constant (CC) and $T_{\mu\nu}$ the energy momentum tensor.
The Ricci tensor and scalars are construed from the Riemann tensor $R_{\mu\nu\lambda\sigma}$, also called curvature tensor, which satisfies the Bianchi identities:\begin{equation}
 R_{\mu \nu \lambda \eta} + R_{\mu \eta \nu \lambda} + R_{\mu \lambda \eta \nu} = 0 \; ,
\end{equation}
\begin{equation}
R_{\mu \nu \lambda \eta ; \sigma} + R_{\mu \nu \sigma \lambda;\eta} + R_{\mu \nu \eta \sigma; \lambda} = 0 \; ,
\end{equation} 
where we use the notation $ A_{\mu;\nu}$ to represent the covariant derivative of $A_\mu$ with respect to $x^{\nu}$.
It is possible to show \cite{Landau:1982dva, Weinberg:1972kfs} using the Bianchi identities that the left hand side of Eqs.~\eqref{EFE} is divergenceless, enforcing the validity of the continuity equation $T^{\mu}_{\nu;\mu}= 0$ in curved spacetime.

\subsection{The Cosmological Principle}
In 1922 the Russian mathematician Alexander Friedmann obtained  an analytical solution of Eqs.~\eqref{EFE} under the assumption that the spacetime is homogeneous and isotropic~\cite{Friedmann1922}. A similar result was obtained 
independently by the Belgian astronomer George Lema\^{i}tre  in 1927~\cite{Lematre1927}, and later on by the American mathematician Howard Robertson~\cite{Robertson} and the British mathematician Arthur Walker~\cite{Walker}. The resulting spacetime is described in terms of a metric usually denoted \textit{FLRW}, named after them. The FLRW line element can be written:
\begin{equation}\label{FLRW}
    ds^2 = -dt^2 + a(t)^2\left(\frac{dr^2}{1-Kr^2} +r^2d\Omega^2 \right) \; ,
\end{equation}
where $\Omega$ is the solid angle, the function $a(t)$ is the scale factor and the constant $K$ is related to the curvature of the spatial slices. A negative, vanishing or positive value of $K$ corresponds respectively to Hyperbolic, Euclidean or Spherical spatial geometry.

The assumptions that at background level the Universe is homogeneous and isotropic are usually referred to as the \textit{Cosmological Principle}. They are also at the core of Newtonian gravity and Galilean relativity, where they are stated as the existence of a universal time and the lack of any preferred direction in space. These definitions on the other hand are not completely satisfactory in the contest of General Relativity, where differential geometry concepts are required to unambiguously define them. A very rigorous definition by Wald is the following, see Ref.~\cite{Wald:1984rg}:
\begin{itemize}
\item[] \textbf{Homogeneity:} \textit{A space-time is said to be homogeneous if $\exists$ a family of 1-parameter  of spacelike hypersurfaces $\Sigma_t$ foliation such that $\forall t$ and  $p,q \in \Sigma_t $ $\exists$ an isometry $I : I(p) \rightarrow q $.    }
    \item[] \textbf{Isotropy:} \textit{a space is spatially isotropic if  $\exists$ at each point a congruence of timelike curves with tangents $u^\alpha$ such that 
    $\forall p \in$  the congruence, given two $s_1^\alpha, s_2^\alpha$ spacelike vectors orthogonal to $u^\alpha \; \exists \; I$ of $g_{\mu\nu} \; : \; s_1^\alpha \rightarrow s_2^\alpha $ leaving $p$ and $u^\alpha$ fixed .  }
\end{itemize}
A slightly less technical definition of the Cosmological Principle  could be instead found in Weinberg's book \cite{Weinberg:1972kfs}: \textit{A globally hyperbolic spacetime is homogeneous and isotropic if}:
\begin{itemize}
    \item[]\textit{i)} Hypersurfaces with cosmic standard time are maximally symmetric subspaces of the whole space-time.
        \item[]\textit{ii)} $g_{\mu \nu}, T_{\mu \nu}$ and all the other cosmic tensor are form invariant with respect to the isometries of these subspaces.
\end{itemize}
We recall that a manifold is \textit{globally hyperbolic} if it possesses a Cauchy surface, i.e. there exists a surface which every causal curve on the manifold crosses exactly once. Roughly speaking this means that a surface exists from which, once specified the initial conditions, it is possible to track past and future evolution of the causal curves through the field equations. A space of dimension $D$ is \textit{maximally symmetric} if it admits $D(D+1)/2$ independent Killing vector fields. A Killing vector field is a vector field on a Riemannian or Pseudo-Riemannian manifold  that preserves the metric. Finally, an \textit{isometry} is a bijective function in a metric space that preserve distances.

We hope the reader could forgive the latter brief technical digression on the cosmological principle, but once a proper definition was given  we are now able to highlight some of its consequences.
First we notice that no concept from General Relativity was used to define the cosmological principle. Indeed, it is an assumption (which is well motivated from the observational point of view, as we will discuss later) independent of the specific metric theory of gravitation we are considering. We also note that the existence of a preferred foliation of the spacetime in terms of a time parameter implies the existence of a privileged class of observers, i.e. free falling observers, whose clocks measure the cosmic time. This could be misleading from the perspective of General Relativity, because of general covariance and of the Equivalence Principle. The main point is that the goal of cosmology is to describe our Universe, which is just a particular realization, or solution, of the Einstein Field Equations (or any alternative metric theory of gravitation), and although the EFE are generally covariant, a particular solution of them does not have to be. 
%%%%%%%%%%%%%%%%%%%%%%%%%%%%%%%%%%%%%%%%%%%%%%%%%%%%%%%%%%%%%%
\subsection{The Friedmann equations}
Computing Eqs.~\eqref{EFE} for the FLRW metric \eqref{FLRW} we obtain the Friedmann equations:
\begin{align}\label{Feq1}
    H^2 + \frac{K}{a^2} - \frac{\Lambda}{3}= \frac{8\pi G_N}{3}T_{00} \; , \\
    g_{ij}\left(H^2 + 2\frac{\ddot{a}}{a} + \frac{K}{a^2} - \Lambda\right)= -8\pi G_N T_{ij} \; , \label{Req1}
\end{align}
where we have defined the Hubble function $H(t) = \dot{a}/a$. Note that the high symmetry of the cosmological principle restricts the allowable choices of $T_{\mu\nu}$. Since the FLRW metric depends only on time, the same must hold for the components of the energy momentum tensor. Furthermore, due to spatial isotropy, the $0i$ components must vanish. Finally, since the left hand side of Eq.~\eqref{Req1} is proportional to $g_{ij}$ the same must be true for $T_{ij}$.

Usually in cosmological applications we consider perfect fluids, which satisfy the above listed properties and can be written in general as:
\begin{equation}\label{Perf.Fluid}
T_{\mu\nu} = \left(\rho(t)  + P(t) \right)u_{\mu}u_{\nu} + P(t)g_{\mu\nu}\; ,
\end{equation}
where we have defined the rest energy density of the fluid $\rho$ and the pressure $P$. We have also introduced the 4-velocity $u_{\mu}$, which is normalized by definition so that $g_{\mu\nu}u^{\mu}u^{\nu} = -1$. Thus, in the comoving frame, where the fluid is at rest, we have $u_{i}=0$ and $u_0 = \frac{1}{\sqrt{|g^{00}|}}$.
Note that Eqs.~\eqref{Feq1},\eqref{Req1} are not completely independent; indeed, since in General Relativity the zero component of the EFE is a constraint equation that contains only first derivatives, Eq.~\eqref{Feq1} contains only the first time derivative of the scale factor. It is possible to obtain Eq.~\eqref{Req1} combining the derivative of Eq.~\eqref{Feq1} with the continuity equation of the fluid:
\begin{equation}\label{ContinuityEq}
\dot{\rho}(t) + 3 H\left(\rho + P\right) =0 \; .
\end{equation}
If we consider barotropic fluids it is possible to relate the pressure and the density through the Equation of State (EoS):
\begin{equation}
    P(t)= w \rho (t) \;
\end{equation}
where $w$ is the EoS parameter.
%%%%%%%%%%%%%%%%%%%%%%%%%%%%%%%%%%%%%%%%%%%%%%%%%%%%%%%%%%%%%%%
\subsection{Analytical solutions of the Friedmann equations}
It is possible to obtain analytical solutions of Eqs.~\eqref{Feq1},\eqref{Req1} for barotropic fluids by solving 
the continuity equation Eq.~\eqref{ContinuityEq}. Indeed, we have:
\begin{equation}
    \frac{\dot{\rho}}{\rho} = -3\frac{\dot{a}}{a}\left(1+w\right) \; ,
\end{equation}
from which:
\begin{equation}\label{eossol}
    \rho = a^{-3(1+w)}\;.
\end{equation}
For most cosmological applications the parameter space for the equation of state parameter $w$ is very simple; one usually consider pressureless non-relativistic matter with $w=0$, also dubbed \textit{dust}, and relativistic matter with $w=1/3$, denoted \textit{radiation}, which includes for example photons and neutrinos.
Let us consider a model of flat Universe $K=0$ filled with a perfect fluid defined by Eq.~\eqref{eossol}. In this case we can rewrite Eq.~\eqref{Feq1} in terms of the scale factor only:
\begin{equation}
    3\frac{\dot{a}^2}{a^2} = 8\pi G_N a^{-3(1+w)} \; ,
\end{equation}which solved with respect to the scale factor gives:
\begin{equation}\label{singlefluidscalefactor}
    a(t) = \left[\sqrt{6\pi G_N}(1+w) t \right]^{\frac{2}{3(1+w)}} \; .
\end{equation}

\subsubsection{Single fluid models}
Let us now focus on the evolution of the Universe for particular solutions of Eq.~\eqref{singlefluidscalefactor} relevant for cosmological purposes. If only a species is present, it is straightforward to integrate the latter equation and obtain analytical solutions for the Hubble function.  In Table  \ref{Table I} is reported the behavior of the scale factor, the density and the equation of state parameter for a flat Universe dominated by matter, radiation and Cosmological Constant. 
\begin{table}
    \centering
    \begin{tabular}{c||c|c|c}
          &  Radiation & Dust & CC \cr
         \hline
         $a(t)$ & $t^{1/2}$ &$t^{2/3}$ &$e^{Ht}$  \\
         
         $\rho(a)$ &$a^{-4}$ &$a^{-3}$ &$\Lambda/3$\\
         
         $ w $ & $1/3$ &$0$ &$-1$
\end{tabular} 
    \caption{Analytical solutions for the scale factor $a(t)$, the density $\rho(a)$ and the EoS parameter $w$ for a flat FLRW Universe during radiation, matter and Cosmological Constant (CC) domination.}
    \label{Table I}
\end{table}
It is important to realize that in an expanding Universe, since $a(t)$ is a growing function, the density function of matter and radiation is decreasing.  Thus, if the Universe contains only  these species, they will eventually dilute and the Hubble function approaches $H(t) \rightarrow 0$. In these models,  the Universe approaches a stable Minkowski attractor in the future. On the other hand, if a Cosmological Constant is present, in the future the Universe reaches a stable de Sitter attractor and the scale factor starts to grow exponentially.

\subsubsection{Einstein Static Universe}
When Einstein was considering cosmological applications of its theory he had in mind a static Universe with $\dot{a}=\ddot{a}=0$. From Eq.~\eqref{Feq1}, since both $\Lambda$ and $\rho$ are positive, we must impose $K = 1$, i.e. a closed Universe. Moreover, the acceleration equation implies:
\begin{equation}
    \rho + 3P = 0 \; ,
\end{equation}
and since $\rho$ is positive definite, the only possibility is that there must be something with negative pressure that compensates. This was the main motivation that brought Einstein to introduce a Cosmological Constant $\Lambda$ into its equations. Indeed, since $\rho_\Lambda = -P_\Lambda$ we have:
\begin{equation}
    2\rho_\Lambda =\rho_m\; ,
\end{equation}
which is a critical point of the dynamical equations. On the other hand such a point is unstable, and depending on the sign of a small perturbation the Universe evolves into  a de Sitter or a Minkowski critical point.

\subsubsection{The standard model}
\begin{figure}[h]
     \includegraphics[width=13cm]{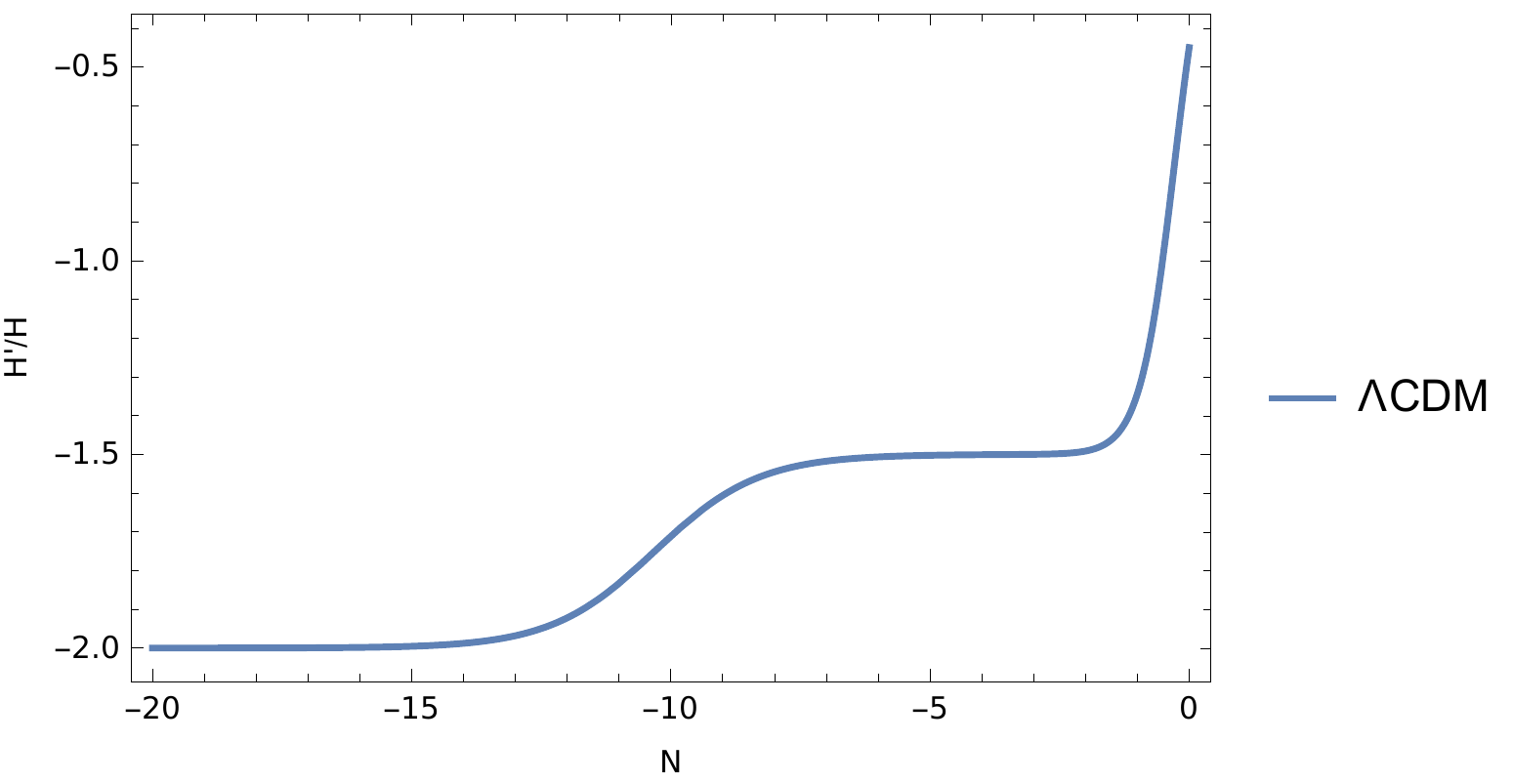}
     \caption{The evolution of $\xi = H'/H$ in the $\Lambda$CDM model}
     \label{xsilcdm}
 \end{figure}

The $\Lambda$CDM model describe a FLRW flat Universe filled with a mixture of dust, in the form of Cold Dark Matter and baryons, radiation and a Cosmological Constant. As we will show later, there are observational evidences that justify the hypothesis of spatial flatness. The Hubble function is given by the first Friedmann equation:
\begin{equation}\label{FriedLCDM}
    H(z) =\sqrt{ \Omega_K^0\left(1+z\right)^2 + \Omega_{m}^0\left(1+z\right)^3 + \Omega^0_{rad}\left(1+z\right)^4+\Omega_{\Lambda}^0} \; ,
\end{equation}
\begin{figure}[h]
	\includegraphics[width=13cm]{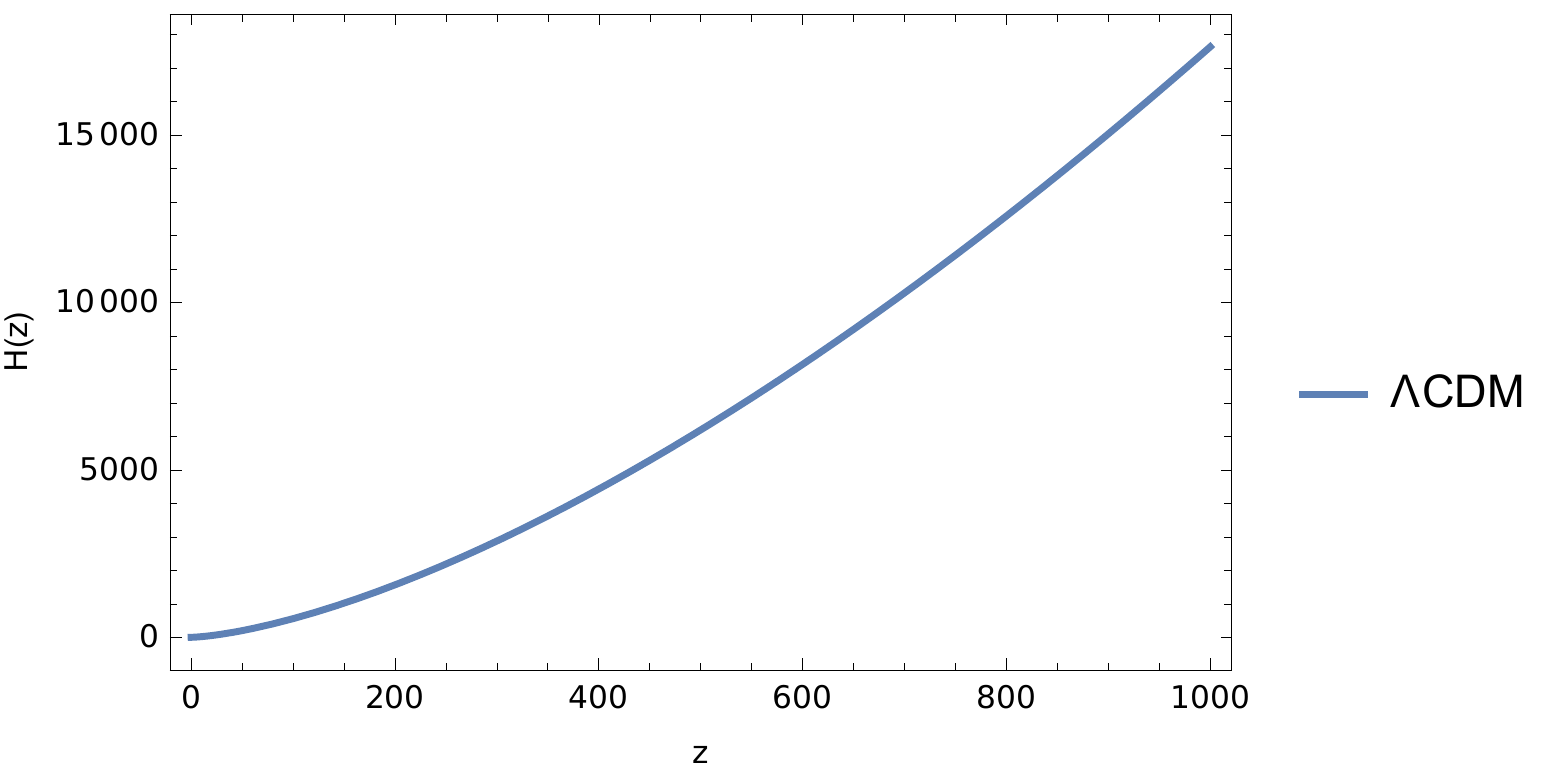} \\[2mm]\caption{The Hubble diagram as function of redshift for the $\Lambda$CDM model }\label{HubblediagramLCDMz}
	\end{figure}
where the $\Omega^0_i \;$ are the present day densities of the species $i$, and where $\Omega_m = \Omega_{CDM} + \Omega_{baryons}$ is the total matter. In Fig.~\ref{HubblediagramLCDMz} the Hubble function for the $\Lambda$CDM model is plotted as a function of the redshift $z = 1/a -1 $ for $\Omega_m^0 \approx 0.3$, $\Omega_r^0 \approx 10^{-5}$ and $\Omega_{\Lambda}\approx 0.7$. If we define the normalized energy density of a species $x$ as $\Omega_x = 8\pi G_N\rho_x \Omega^0_x/3H^2$, it is possible to rewrite the Friedmann equation in the form:
\begin{equation}
    1= \Omega_m + \Omega_r + \Omega_\Lambda \; .
\end{equation}
\begin{figure}[h]
     \includegraphics[width=15cm]{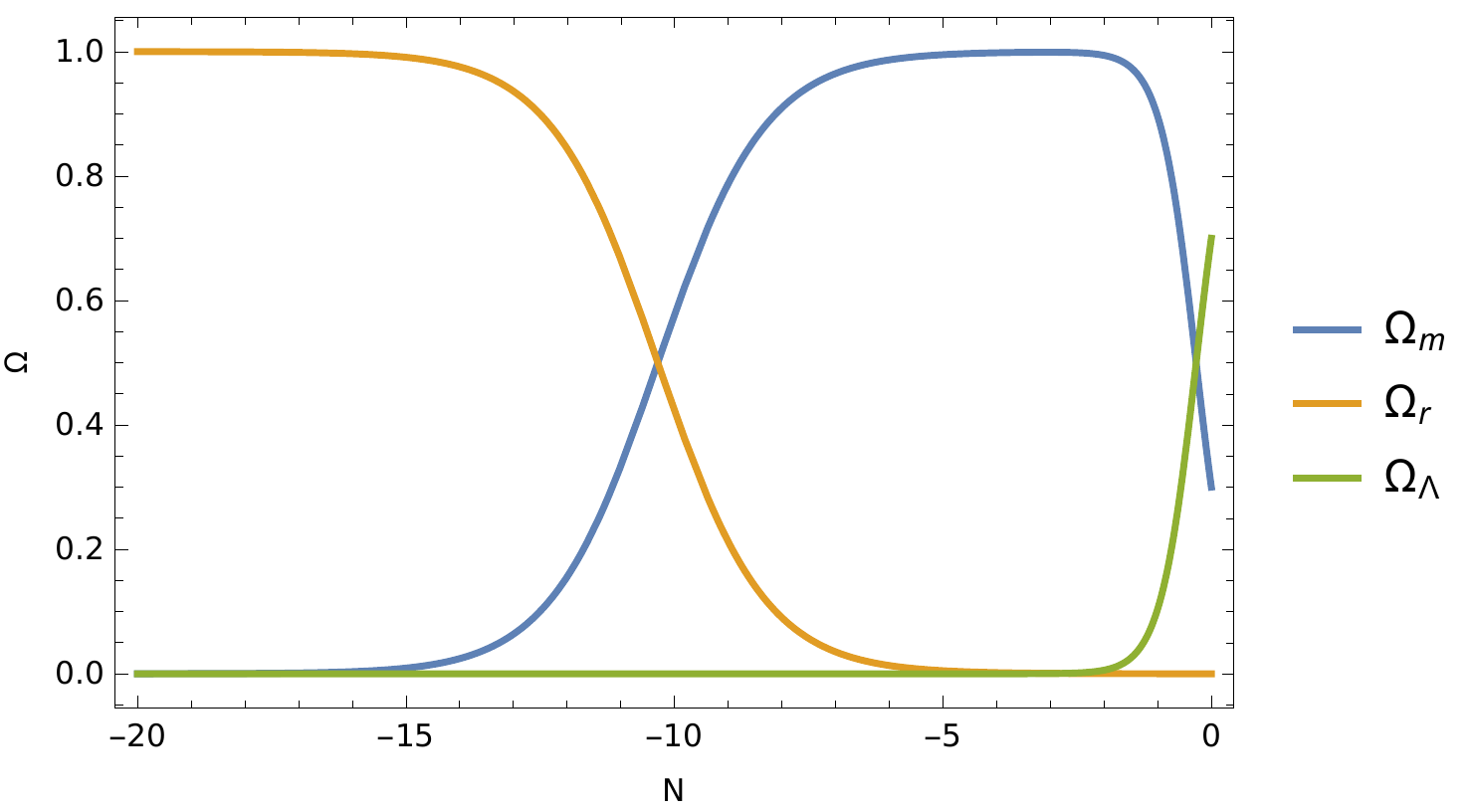}
     \caption{The evolution of the energy densities of the species in the $\Lambda$CDM model}
     \label{Sumrule}
 \end{figure}
 In Fig.~\ref{Sumrule} are plotted the normalized energy densities in terms of the e-fold time parameter $N=\log a$. It is straightforward to realize that the Universe evolution could be divided into different epochs, during which one species is dominant with respect to the others and determinate the rate of expansion. Since radiation dilutes the fastest it will be dominant at earlier times, followed then by matter and finally by the Cosmological Constant. In Fig.~\ref{xsilcdm} the logarithmic time derivative of the Hubble function is given in terms of $N$. For $N \leq -15$, in the radiation dominated epoch,  $H \sim \sqrt{\rho_{rad}}$, so that $H'/H \approx -2$. Then radiation dilutes, and around $N\approx -10$ its density equals the matter one. When matter starts to dominate the Hubble factor behave as $\sqrt{\rho_m}$, so that $H'/H \approx -3/2$. Finally, around today, matter dilutes and its density equals the one of the Cosmological Constant, so that $H'/H$ tends asymptotically to $\rightarrow 0$ in the future.
 
 %%%%%%%%%%%%%%%%%%%%%%%%%%%%%%%%%%%%%%%%%

\section{Observational facts in support of the $\Lambda$CDM model}
\label{sec:related:sec3}
\subsection*{Observational dataset}
The most important cosmological probes that support the $\Lambda$CDM model are the Cosmic Microwawe Background (CMB), type Ia Supernovae observations, and Baryonic Acustic Oscillations (BAO). Combined, they favor a model of flat Universe $|\Omega_k^0| \leq 0.003 $ where the total matter density today $ \Omega_m^0 =\Omega_{DM}^0 + \Omega_{b}^0$ is of order $\Omega_m^0 \approx 0.3$ and the Cosmological Constant value contributes to roughly $\Omega_{\Lambda}^0\approx 0.7$, with the radiation energy density of order $\Omega_r^0 \approx 10^{-5}$ \cite{Aghanim:2018eyx, Abbott:2018wog,Scolnic:2017caz}.

\subsection{Age of the Universe}
In a FLRW background it is possible to compute the age of the Universe $t_0$ by integrating the Hubble function:
\begin{equation}\label{ageofuniverse}
    t_0 = \int_0^\infty \frac{dz }{\left(1+z\right)H(z)} \; .
\end{equation}
It is straightforward to realize that the main contribution to the above integral comes from recent times, i.e. small redshift $z$. In this regime we can neglect the contribution of radiation, whose density is of order $10^{-5}$. Using the first Friedmann equation to eliminate $\Omega_m^0= 1- \Omega_\Lambda^0$ we compute the value of the integral in Eq.~\eqref{ageofuniverse} as function of the Cosmological Constant value $\Omega_\Lambda^0$.
 \begin{figure}[h]
     \includegraphics[width=13cm]{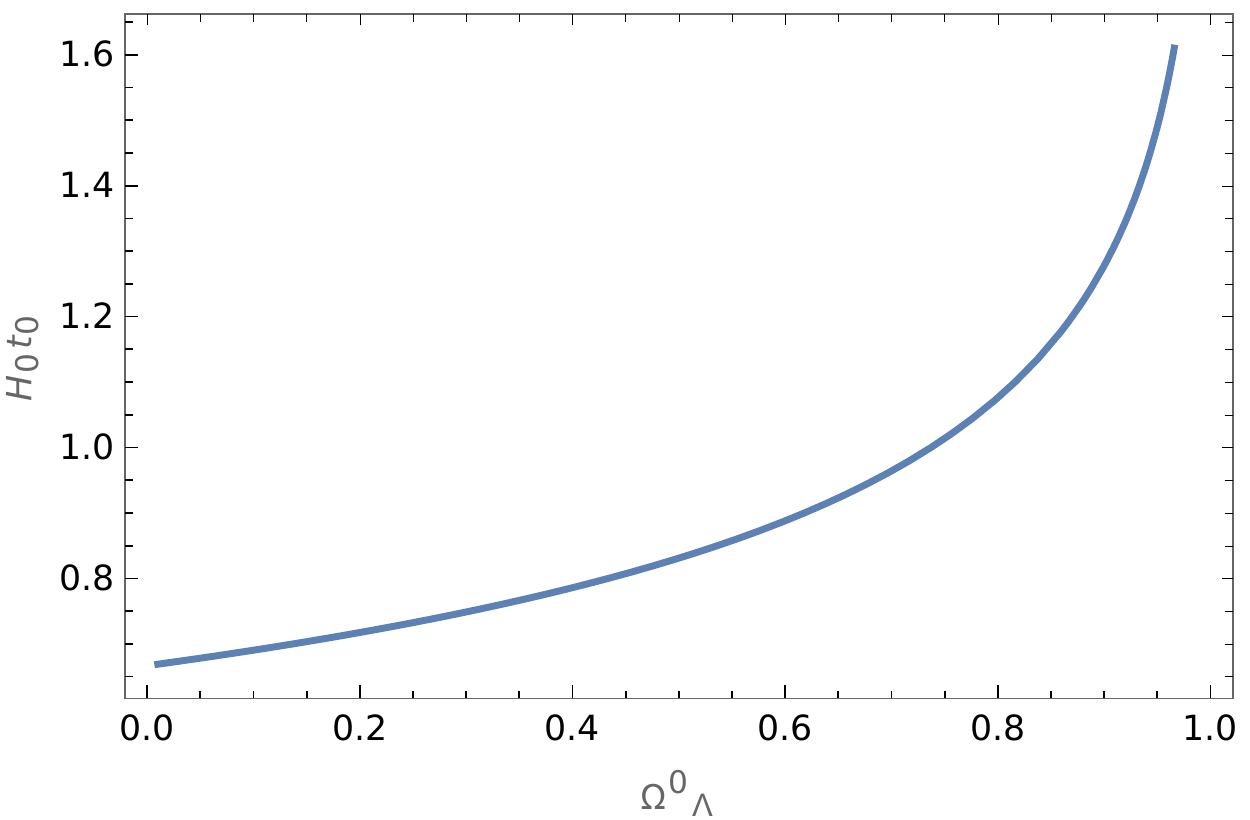}
     \caption{The age of the Universe in unit of $H_0^{-1}$ for a flat FLRW Universe filled with matter and Cosmological Constant depending on the value of $\Omega_\Lambda^0$ }
     \label{ageuniverse}
 \end{figure}In Fig.~\ref{ageuniverse} the result of the integral assuming flatness is reported in units of $H_0^{-1}$, and one can appreciate that for small values of the Cosmological Constant, i.e. without Dark Energy, the Universe is younger, while in the opposite limit $\Omega_m^0\rightarrow 0$ the integral in Eq.~\eqref{ageofuniverse} diverges and the Universe is eternal, as expected for a de Sitter Universe which is free of initial singularity. Even if we do not know exactly the age of the Universe, we can constrain it from below with the age of the oldest object we observe in the sky. Of the utmost importance in this respect are the globular clusters, i.e. clusters of $10^5-10^7$ stars with a high density of around $10^3$ stars for $ly^{-3}$ that share the same age and the same chemical composition, usually found in the galactic halo. These stars are remnants of galaxy formation, and are among the oldest objects observed in the sky, see Ref.~\cite{Trenti:2015zja}.
It is possible to infer their age by using spectroscopic mesurements, i.e. by studying their abundance of heavy elements. It is believed that in the early Universe there were mostly Hydrogen and Helium produced during the Big Bang Nucleosynthesis, as a result the stars that were produced at the time should lack heavy elements. The oldest globular clusters found are dated around 11 Gyr, which correspond roughly to $t_0 H_0 \sim 0.8$ in Fig.~\ref{ageuniverse}. These considerations already rule out a model of flat Universe which contains only matter, signaling the necessity for some form of DE.

\subsection{Structure formation and DM}
Incorporating $\Lambda$ in the total density $T_{00}$ in Eq.~\eqref{Feq1} we can write:
\begin{equation}
    H^2 = \frac{8\pi G \rho_{tot}}{3} - \frac{K^2}{a^2} \; ,
\end{equation}
and we can define the critical density:
\begin{equation}
    \rho_{crit}= \frac{3H^2}{8\pi G} \;,
\end{equation}
which is the value of $\rho$ such that $K=0$. Observations indicate that the today total density $\rho_{tot}^0$ is of order:
\begin{equation}
    \rho_{tot}^0= \frac{3H_0^2}{8\pi G}  \sim \rho_{crit} \sim 10^{-29} g/cm^3 \;.
\end{equation}
This means that the Universe is spatially flat and that its average density is of around 10 protons per cubic meter. On the other hand the existence of baryonic compact objects, like us, indicates that the Universe contains highly nonlinear regions which are uniformly distributed according to the cosmological principle. Baryon's perturbations at recombination, around $z = 1100$, were proportional to the CMB fluctuations which are of order $10^{-4}$. By solving the perturbations equations for $\rho_b$ at linear order during the matter dominated epoch we know that matter overdensities grow linearly with the scale factor. This in turn imply that today, $z \sim 0$, these fluctuations should be of order $10^{-1}$ and thus still linear. We can conclude that if only baryonic matter is present, its perturbations from the recombination would not have been in time for growing non-linearly and form compact objects. On the other hand, if another matter species decoupled from photons prior to recombination soon enough, it would be able to catalyze baryonic structure formation. Thus Dark Matter is a crucial ingredient for structure formation.

\subsection{Disc galaxies rotation curves}
Spiral galaxies, like the one in which we locate ourselves, are common objects in the Universe. The distribution of luminous matter is peaked in the center and, using Newtonian arguments and assuming spherical symmetry, one expects that the centrifugal force is compensated by the gravitational attraction:
\begin{equation}
    \frac{v^2}{r} = \frac{G M(r)}{r^2} \; .
\end{equation}
Since the mass contained within a radius $r$ is proportional to the volume $r^3$,
the velocity of the stars in the galaxy drops down as we move to higher $r$ as $v \propto r$.
On the other hand observations are not compatible with the above simple profile, see for example \cite{Navarro:1996gj}, and show instead that the velocity of stars in the outer arms of spiral galaxies approaches a constant value. This problem is known as the \textit{flatness of velocity curve of stars}, and can be explained by assuming a different distribution of matter from the visible one, thus invoking the presence of a “dark” matter species. 
 
 \subsection{Type Ia supernovae observations}
 
 During the 1998 Riess \textit{et al.} \cite{Riess:1998cb} and Perlmutter \textit{et al.} \cite{Perlmutter:1998np} realized through type Ia Supernovae observations that the rate of expansion of the Universe is accelerating. Supernovae are extremely bright stellar explosions which occur in the last stages of a massive star evolution or during the nuclear fusion of a white dwarf. The brightness of these astronomical transient events is comparable with the one of an entire galaxy, and last for several weeks or months. The classification of Supernovae is made via spectroscopic measurements and depends on which absorption lines are present. If there is no Hydrogen line in the spectrum they are classified as type I supernovae, and type II otherwise. If they contain an absorption line of singly ionized silicon they are classified as Ia, whereas they are classified Ib if they contain Helium. Finally if they lack both Helium and Silicon they are classified Ic. Type Ia supernovae are of the utmost importance in cosmology because their absolute luminosity is roughly constant at the peak of brightness. They are formed in stellar binary systems containing a white dwarf that increases its mass by absorbing gases from the companion, eventually causing it to exceed the Chandrasekhar limit and triggering the explosion. For their properties the type Ia Supernovae are called standard candles, and observing them at various redshifts it is possible to reconstruct the cosmological evolution. Indeed, it is well-known that the apparent magnitudes of two sources $m_i$ are related with their apparent fluxes $\mathcal{F}_i$:
 \begin{equation}
     m_1 -m_2 = -\frac{5}{2}\log_{10}\left(\frac{\mathcal{F}_1}{\mathcal{F}_2}\right) \; .
 \end{equation}
 From the apparent flux of a source and its absolute luminosity $L_s$ it is possible to define the luminosity distance $d_L$:
 \begin{equation}
     d_L^2= \frac{L_s}{4\pi \mathcal{F}} \; ,
 \end{equation}
 finally, the apparent and the absolute magnitude of a source $m$ and $M$ are related as:
 \begin{equation}
     m - M = 5\log_{10}\left(\frac{d_L}{10 pc}\right) \; ,
 \end{equation}
 i.e. the absolute magnitude of a source is defined as the magnitude that it would have if observed at a distance of $10 \;$ pc. For type Ia supernovae $M$ is roughly constant and equal to $\sim -19$,  so that the apparent magnitudes $m_1$ and $m_2$ of two of them can be related to their distances: 
 \begin{equation}
     m_1 -m_2 = 5\log_{10}\left(\frac{d_{L_1}}{d_{L_2}}\right) \; .
 \end{equation}
 The theoretical prediction for the luminosity distance in a FLRW Universe is:
 \begin{equation}
     d_L(z) = \frac{c\left(1+z\right)}{H_0 \sqrt{\Omega_K^0}}\sinh{\left(\sqrt{\Omega_K^0}\int_{0}^{\infty}d\bar{z}\frac{ H(0)}{H(\bar{z})}\right)} \; ,
 \end{equation}
 which for a flat Universe $K=0$ and small values of $z$ can be expanded at second order as:
 \begin{equation}
     d_L(z) \sim \frac{c}{H_0}\left[z + \left(1-\frac{H'(0)}{2H_0}\right)z^2 + ....  \right] \; ,
 \end{equation}
 and if radiation is negligible $\Omega_r \sim 0$ takes the simple form:
 \begin{equation} \label{lumdistlowz}
     d_L (z) = \frac{c}{H_0}\left[z + \frac{1}{4}\left(1-3\omega_{DE}\Omega_{DE}^0 \right)z^2\right] \; .
 \end{equation}
 From Eq.~\eqref{lumdistlowz} it is straightforward to realize that the presence of DE, remembering that $w_{DE} < 0$, pushes $d_L$ to higher values with respect to the case without it. 
 
\subsection{Cosmic microwave background observations}
In the Big Bang paradigm the early Universe was filled with a dense plasma of baryons electrons and photons. Baryons and electrons, which have opposite charge, interact via Coulomb forces. Photons instead bounce between electrons via Thompson scattering.  As the Universe expands its temperature drops, and electrons and baryons merge to produce hydrogen atoms in the so-called epoch of \textit{recombination}, around $z \sim 1100$. As a consequence, the density of free electrons falls and the Thompson scattering of the photons becomes inefficient. When this happens, the photons are able to escape the baryon plasma and to propagate freely until today, cooling down to a temperature of around $\sim 2.7$ Kelvin. This radiation is called Cosmic Microwave Background (CMB), and is one of the most precious sources of cosmological information.  
The CMB was predicted by Alpher and Gamow, see Ref.~\cite{1948PhRv...74..505G,Alpher:1948ve}, already in the late '40 and it has been detected for the first time by Penzias and Wilson in 1965 \cite{Penzias:1965wn}. While at the time they were able to detect only the background temperature, we are nowadays able to observe with great precision anisotropies in the CMB of order $\sim 10^{-5} K$ through satellite experiments like Planck \cite{Aghanim:2018eyx}. The temperature-temperature (TT) power spectrum of the CMB is usually studied in spherical harmonics and its modes are labeled with the harmonic number $\ell$. 
 \begin{figure}[h]
     \includegraphics[width=13cm]{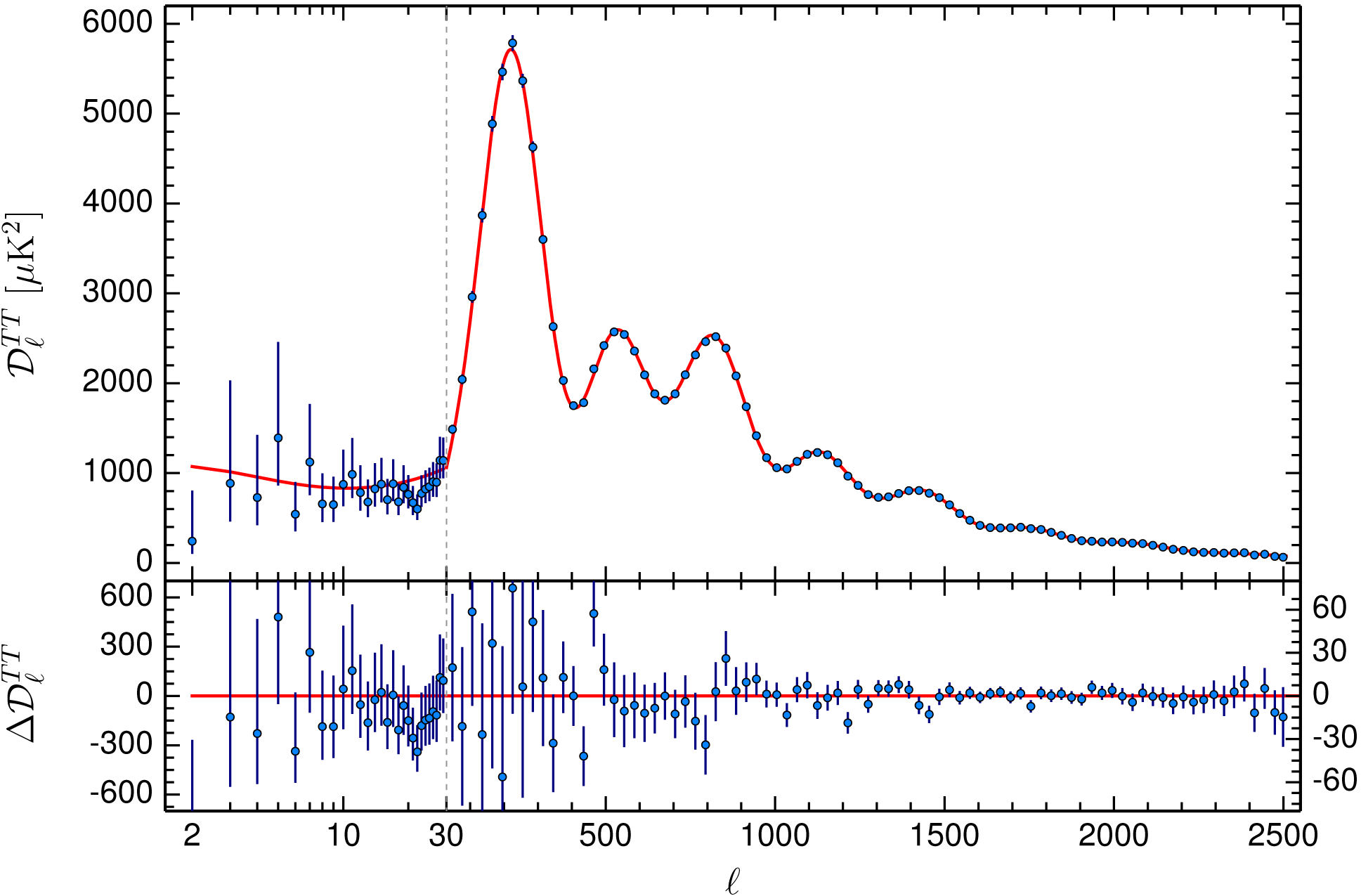}
     \caption{The TT angular Power spectrum of the CMB as a function of the angular scale $\ell$ measured by Planck 2015 \cite{Ade:2015xua} with the residual errors. The solid line represent the theoretical prediction for the $\Lambda$CDM model and the dots the observed data.  Picture taken from \href{https://www.cosmos.esa.int/web/planck/picture-gallery}{https://www.cosmos.esa.int/web/planck/picture-gallery}.   }
     \label{CMBTT2015}
 \end{figure}
 In Fig.~\ref{CMBTT2015} is reported the CMB angular power spectrum for TT anisotropies, together with the theoretical prediction for the $\Lambda$CDM model. The position and the amplitude of the peaks of Fig.~\ref{CMBTT2015} strongly constrain the energy densities of the species in the $\Lambda$CDM model today. For example, the amount of total matter and the ratio of the densities can be extrapolated  measuring position and amplitudes of the first three peaks. For a detailed description of the impact on the TT CMB power spectrum of the cosmological parameters see for example Refs.~\cite{Lesgourgues:2018ncw, Lesgourgues:2013qba}.

\section{Open problems of the $\Lambda$CDM model }
Even being the most accepted paradigm to describe the cosmological evolution of the Universe, the $\Lambda$CDM suffers because of some theoretical and observational issues which lack a satisfactory explanation. Moreover, even if not properly an issue, the fact that most of the energy density content of the Universe today is composed by dark species, which are undetected directly with laboratory  experiments, is a strong motivation for research and studies beyond the $\Lambda$CDM model.
\subsection{Troubles with the Cosmological Constant}
The Cosmological Constant $\Lambda$ is the simplest natural candidate for Dark Energy. On the other hand, the phenomenological value required by the observations to produce the accelerated expansion is quite challenging to predict from the theoretical point of view. From a quantum field theory (QFT) perspective the behavior of the Cosmological Constant is at a phenomenological level equivalent to the expected behavior of vacuum quantum fluctuations. Unfortunately, the  vacuum fluctuations of the fields described in the standard model of particle physics would result in a value for the Cosmological Constant which span from 123 to 55 orders of magnitude higher depending on the scenario considered. The above incompatibility is usually referred to as the \textit{Cosmological Constant Problem}, see for example Refs.~\cite{Weinberg:1988cp, Martin:2012bt, Padilla:2015aaa} for a detailed account of the problem. Another problem associated with the Cosmological Constant is the so called \textit{Coincidence Problem}, see for example Refs.~\cite{Piattella:2018hvi,Velten:2014nra}. The coincidence relies on the fact that the present day energy density of DE and DM are roughly of the same order. Such an occurrence, if not explained dynamically, would require an extremely severe fine-tuning in the initial condition of the Universe. Indeed, since the Cosmological Constant density is, of course, constant and the DM density dilutes as $\rho_m \sim a^{-3}$, in the early stage of the evolution of the Universe, say the Planck scale for which $a \sim 10^{-32}$, the ratio $\rho_\Lambda/\rho_{m}$ would be of order $\sim 10^{-96}$. This means that the initial condition for the Universe should be set with the astonishing precision of 96 digits; a one digit difference would result today in a factor 10 difference on the respective energy densities, well outside the parameter space allowed by observations.
\subsection{Troubles with CDM}
It turns out from numerical simulations of structure formation that on small scales, around $\sim 1 kpc$, and for mass scales smaller than $M \leq 10^{11}M_\odot$, CDM is not completely satisfactory. For a review on the topic, see for example Ref.~\cite{Bullock:2017xww}. One of the problems that arise in this framework is known as the \textit{cusp/core} problem. Numerical simulations of the $\Lambda$CDM shows that DM halos should present a steep growth of the density profile at small radius, of order $\rho_{Halo}(r) \sim r^{\gamma}$, with $0.8\leq \gamma \leq 1.4$. On the other hand several observations of small galaxies with well measured rotation curves prefer  $0 \leq \gamma \leq 0.5$, showing that pure $\Lambda$CDM simulations are too \textit{cuspy} compared to the observations.

Another issue with CDM which appeared in the late 90' is the \textit{Missing satellites problem}. Simulations show that DM clumps should exist in a broad range of masses and should results in thousands of satellite objects with mass $M \leq 300 M_\odot$ trapped in those clumps. On the other hand, at the time, only a bunch of these satellites were observed. The problem persisted for roughly two decades, but it seems that nowadays, with the improvement of the observations and of the numerical simulations, the missing satellites problem had been turned inside out. Indeed, in the last years astronomers found thousands of low mass objects, which could be too many compared to those predicted by the simulations, see for example Ref.~\cite{Kim:2017iwr}.

Finally, a third known issue of the CDM paradigm is the \textit{Too big to fail} problem. From observations, it seems that galaxies fail to form in the most massive subhalos, while at the same time satellites
of lower mass form in less dense subhalos. This appears to be a contradiction, since most massive satellites should be too big to fail to form in most dense halos while smaller satellites do so in the lighter ones.

\subsection{The flat Universe conspiracy}
\begin{figure}[h]
\begin{center}
     \includegraphics[width=10cm]{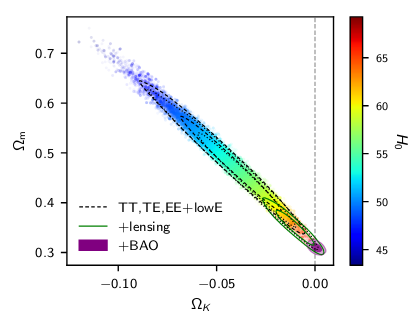}
     \caption{Constraints on a non-flat Universe from CMB angular power spectrum (dashed line), CMB + lensing (solid green), and CMB + lensing + BAO (purple region).  Picture taken from \cite{Aghanim:2018eyx} }
     \label{Okplank}
 \end{center}
 \end{figure}

By looking at the angular spectrum of CMB alone  it seems that a model of Universe with a slightly negative curvature is preferred with respect to a flat one. The situation changes if we also include priors from CMB lensing and BAO, which carry information about $H_0$ and $\Omega_m$. The constraints on the curvature density from Planck 2018 \cite{Aghanim:2018eyx} are reported in Fig.~\ref{Okplank} 

If we assume that BAO, CMB lensing and CMB  
polarization data should not be combined, there is indeed space left for curvature being non-vanishing from Planck 2018 data, see also Ref.~\cite{Handley:2019tkm}. The impact of such a point of view on the cosmological standard model was considered by the authors of Ref.~\cite{DiValentino:2019qzk}, which claims that our current understanding of the Universe could be biased and that would imply a possible crisis for cosmology. As discussed by the authors, the tendency towards a closed Universe could just be a signal of systematic, but is stronger in Planck 2018 than in Planck 2015 \cite{Ade:2015xua}, and could indicate a strong disagreement between CMB power spectrum and BAO measurements. However, it must be noted that there is a strong degeneracy in the CMB power spectrum between the curvature $\Omega_K$ and the lensing amplitude $A_{lens}$. If the Universe is closed, data favor a higher amount of Dark Matter, which in turn enhance the lensing effect allowing for a better fit to the data at lower multipole. Whether there is a conspiracy for a flat Universe or not, the results of \cite{DiValentino:2019qzk} show the kind of dangers hidden behind the corner in the era of high precision cosmology.  

\subsection{Cosmological tensions on $H_0$ and $\sigma_8$}

The history of cosmology is strongly entangled with the history of one of its parameters, the value of the Hubble factor today. Indeed, while the first measurement of $H_0$ by Hubble buried the philosophical preconceptions about a static Universe, its measure today possibly uncovers and targets the Achilles heel of the $\Lambda$CDM model.

\begin{figure}[h]
\begin{center}
     \includegraphics[width=10cm]{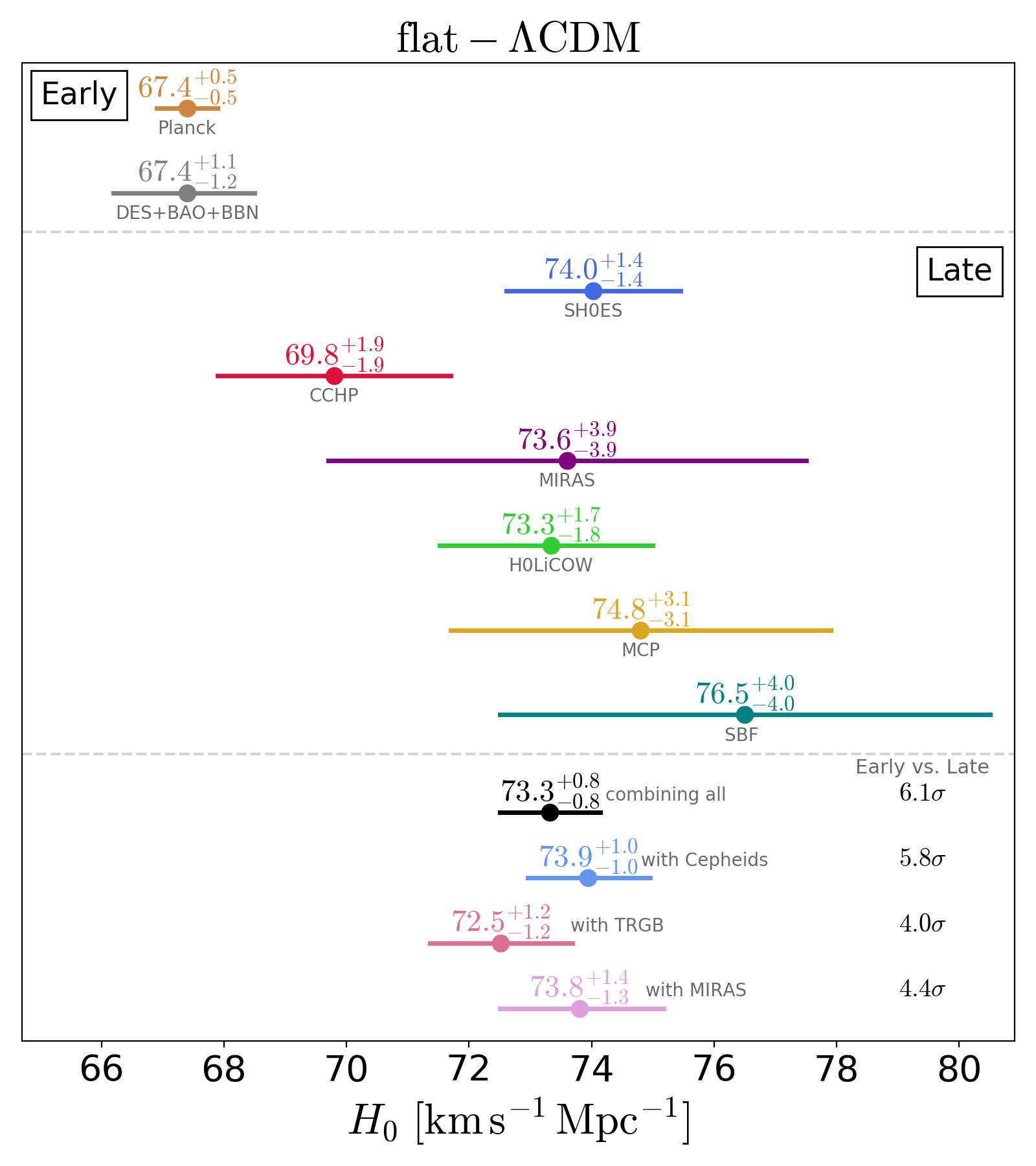}
     \caption{The tension on the value of $H_0$ arising from late and early time measurements. Figure taken from \cite{Verde:2019ivm}.   }
     \label{H0tension}

 \end{center}
 \end{figure}

We can classify brutally most of the sources of cosmological information in two groups, i.e. measurements of early and Late-times Universe. With Late-times Universe sources we refer to those measurements performed at low redshift, like for example type Ia supernovae or strong lensing time delays. With early time Universe measurements we refer mostly to CMB and BAO observations. State of the art experiments seem to indicate that measurements of $H_0$ from late and early Universe are in disagreement and this tension is estimated to be significantly above $4 \sigma$ \cite{Bernal:2016gxb,Verde:2019ivm,Martinelli:2019krf}.  In Fig.~\ref{H0tension}, taken from \cite{Verde:2019ivm}, measurements of $H_0$ from different experiments and their combined results are reported, and the tensions quantified.

Another cosmological parameter suffers from the same kind of issue, the $\sigma_8$ parameter. It refers to fluctuations of the matter density on scales of $8 h^{-1} Mpc$:
\begin{equation}
    \sigma_8^2 = \int_0^\infty \frac{dk}{k} \left[\frac{3j_1(8k)}{8k}  \right]^2 \Delta^2(k) \; ,
\end{equation}
where $j_1=sin(x)/x -cos(x)$ is proportional the first order spherical Bessel function and $\Delta^2(k)$ is the dimensionless matter power spectrum $\Delta^2 =k^3 \mathcal{P}_m (k)/2\pi^2$. 
 It seems to be difficult to solve both the $\sigma_8$ and the $H_0$ tensions within the same framework. One of the reasons is that a higher value of $H_0$ could be obtained with new physics that reduces the size of the sound horizon $r_s$ with early time modifications, whilst in order to tackle the $\sigma_8$ tension one needs to suppress the linear power spectrum of matter at Late-times or decrease $\Omega_m$. These modifications usually point towards opposite directions, making it difficult to relieve both tensions within the same framework. 
 Several modifications of gravity were proposed in order to alleviate the tensions, see for example Refs.~\cite{DiValentino:2017oaw} and \cite{Sola:2018sjf} where interactions between DM and neutrinos and dynamical DE were considered, or Ref.~\cite{Barros:2018efl} where a scalar quintessence field couples to DM, reducing its amount and clustering at Late-times thus alleviating the $\sigma_8$ tension. 
 A promising new approach to the problem emerged in recent years, which tackles the tensions by employing the machinery of DE models proposed to explain the accelerated expansion of the Universe together with a Cosmological Constant $\Lambda$. This was done for example in \cite{Sola:2019jek} and \cite{Sola:2020lba} for the Brans-Dicke model, fitting the data better than $\Lambda$CDM.   
 
 \subsection{Conclusion}
 An Occam's razor logic makes the $\Lambda$CDM model the most successful description of the Universe as we know it.
 At the cost of six parameters we are able to explain a plethorae of observations coming from a very broad landscape of physic ranging from astrophysical to cosmological scales. However, out of these 6 parameters, two are so obscure that we need to label them as \textit{dark}, and the situation is even worse when we realize that the darkness fills roughly the 95\% of the Universe. Moreover, beyond the challenging nature of DM and DE, in the past few years state of the art observations disclose a Pandora's box of  inconsistencies of the $\Lambda$CDM model which cosmologists are now forced to deal with, above all the cosmological tensions on $H_0$ and $\sigma_8$. With our current understanding of the Universe under siege, it is of the utmost importance to look with fresh eyes and open mind to alternatives of the $\Lambda$CDM model. An army of scientists grouped in surveys is currently working on new experiments and exploring the consequences of different models. Maybe in 20 years from now they will still rely on a Cosmological Constant and Cold Dark Matter to describe the evolution of the Universe, or maybe they will have the luck of witnessing the appearance of new physics. Whether this is the case or not, these are exciting times to live for cosmologists. 
% !TEX root = ../thesis-example.tex
%
\chapter{Dark Energy Bestiarium}
\label{chapter:DE}

\epigraph{\textit{The miracle of physics that I'm talking about here is something that was actually known since the time of Einstein's general relativity; that gravity is not always attractive}}{Alan Guth}

The goal of this chapter is to give an overview of the possible Dark Energy models beyond the standard cosmological model, i.e. we would like to present a DE \textit{Bestiarium}.\footnote{This choice of terminology is inspired by chapter 9 of Profumo's book \cite{Profumo:2017hqp}: \textit{Bestiarium: A Short, Biased Compendium of Notable Dark Matter Particle Candidates and Models}} Keeping the analogy with biology, in the first part of the chapter we will attempt to classify the models based on their \textit{taxonomy},\footnote{In biology, taxonomy, from Ancient Greek taxis, meaning 'arrangement', and -nomia, meaning 'method', is the science of naming, defining (circumscribing) and classifying groups of biological organisms on the basis of shared characteristics} i.e. on how they look from the mathematical point of view.
In the second part we will try instead to address their \textit{ethology},\footnote{The term ethology derives from the Greek words ethos, meaning "character" and -logia, meaning "the study of". In Biology refers to the scientific and objective study of animal behaviour}  i.e. the way in which different DE models affect observable quantities.

\section{Taxonomy of Dark Energy}
\subsection{The Lovelock Theorem}

General relativity has been proven to be, at least at some scales, our best description of the gravitational interaction. The Einstein Field Equations have the nice properties of being local and covariant differential equations of the metric and its first and second derivatives, and linear in the latter. One could ask whether there are alternatives to  the EFE in order to describe the interplay between matter and geometry of the space-time. Of the utmost importance in this direction is Lovelock's theorem, see Refs.~\cite{Lovelock:1971yv,Lovelock:1972vz}, which can be enunciated as follows~\cite{Clifton:2011jh}:
\begin{theorem}
The only possible second-order Euler-Lagrange expression obtainable in a four dimensional space from a scalar density of the form $\mathcal{L}= \mathcal{L}(g_{\mu\nu})$ is:
\begin{equation}
    E^{\mu\nu}= \alpha \sqrt{-g}\left[R^{\mu\nu}-\frac{1}{2}Rg^{\mu\nu}\right] + \sqrt{-g} g^{\mu\nu}\Lambda\; ,
\end{equation}
where $\alpha$ and $\Lambda$ are constant.
\end{theorem}
Note that the above theorem is a statement about the form of the field equations resulting from a scalar Lagrangian, not a statement about the form of the Lagrangian itself, which then could be different from the standard Einstein Hilbert action.

To rephrase it in other words, Lovelock's theorem states that if we want a geometrical theory of gravity in 4 dimensions arising from a scalar Lagrangian of the metric, the only possibility are the Einstein field equations plus a cosmological constant. The importance of the above result is that it clearly indicates which kind assumptions we have to relax in order to obtain a gravitational theory different from general relativity plus a cosmological constant. Indeed the only options left are:
\begin{itemize}
    \item Increasing the number of degrees of freedom, i.e. considering other fields together with the metric tensor $g_{\mu\nu}$.
    \item Include higher order derivatives 
    in the field equations
    \item Consider spaces with dimension $N \neq 4$.
    \item  Giving up Lagrangian formulations
    \item Abandoning locality and Lorentz invariance of the field equations.
\end{itemize}

\subsection{Increasing the number of degrees of freedom}
\subsubsection{Scalar-tensor theories}

The most general Lagrangian containing a tensor field and a scalar field which gives second order equation of motion was discovered by Horndenski in 1974 in Ref.~\cite{Horndeski:1974wa}. Later on, it was rediscovered in the context of the so called generalized Galileon theories, see for example Ref.~\cite{Deffayet:2009mn}, and the equivalence between the two theories was shown in Ref.~\cite{Kobayashi:2011nu}. The general form of the Horndeski Lagrangian can be written :
\begin{equation}
\label{HL}
\begin{split}
    \mathcal{L}\left(g_{\mu\nu} , \phi\right) =& G_2\left(\phi, X \right) + G_3 \left(\phi, X \right)\Box \phi + G_4\left(\phi, X \right) R + G_{4,X}\left(\phi, X \right)\left(\Box \phi - \phi^{\mu\nu}\phi_{\mu\nu}\right)+\\
    &G_5\left(\phi, X \right)\phi^{\mu\nu}R_{\mu\nu} -\frac{G_{5,X}}{6}\left[\left(\Box \phi\right)^{3} - 3\Box\phi \phi^{\mu\nu}\phi_{\mu\nu} + 2\phi_{\mu\nu}\phi^{\nu \lambda}\phi^{\mu}_{\lambda} \right] \; ,
\end{split}
    \end{equation}
    where $X= \nabla_\mu \phi \nabla^{\mu} \phi/2$ , $G_{i,X} = \partial_X G_i$ and $\phi_{\mu\nu} = \nabla_\mu \nabla_\nu  \phi$. 
    By taking appropriately the free functions $G_i$ of the above Lagrangian one is able to reproduce any second order scalar tensor theory as a specific case. Choosing $G_4 = \frac{M_{Pl}^2}{2}$ and $G_i = 0$ for $i \neq 4$ reproduces the Einstein Hilbert action. The function $G_2$ can account for any free Lagrangian of the scalar field, for example quintessence.  Note that in Eq.~\eqref{HL} the functions $G_3$ and $G_5$ must have an $X$ dependence, otherwise they can be absorbed into $G_2$ and $G_4$ up to a total derivative.  Note also that general Lagrangians of the Ricci scalar, i.e. $f(R)$ theories, belong to the Horndeski family since they can be cast in a scalar tensor form by defining $\phi = df/dR$ and performing a Legendre transformation of the action functional. The same apply for other geometrical theories which result in second order equation of motion; for example also a non minimally coupled Gauss-Bonnet is contained in the Hordenski Lagrangian~\cite{Kobayashi:2011nu}.

    \subsubsection{Generalized Proca Theories}
    
    Another option is to increase the number of degrees of freedom by means of a vector field. The main advantage of this approach over a multi-scalar field theory is that it generally results in a richer dynamics. This family of theories is called generalized Proca theories, and were recently proposed in Ref.~\cite{Heisenberg:2014rta}. Previous attempts of introducing vector fields in a gravitational context were made already in the 2000's, with the goal of modelling anisotropic Dark Energy, see Refs.~\cite{Koivisto:2007bp,Himmetoglu:2008zp}. The Lagrangian of the generalized Proca theories is:
    \begin{equation}
    \label{GPL}
    \begin{split}
        \mathcal{L}_{GP}&= -\frac{1}{4}F^{\mu\nu}F_{\mu\nu} + G_2(X) + G_3 (X) \nabla_\mu A^{\mu}+ \\
        &G_4(X)R + G_{4,X}\left[\left(\nabla_{\mu}A^{\mu}\right)^2 + c_2 \nabla_{\rho}A_{\sigma}\nabla^{\rho}A^{\sigma}   +\left(1-c_2\right)\nabla_{\rho}A_{\sigma}\nabla^{\sigma}A^{\rho} \right] +\\
        &G_{5}(X) \mathcal{G}_{\mu\nu}\nabla^{\mu}A^{\nu} -\frac{1}{6}G_{5,X}\left[\left(\nabla_{\mu}A^{\mu}\right)^3 -3d_2 \nabla_\mu A^{\mu}\nabla_{\rho}A_{\sigma}\nabla^{\rho}A^{\sigma} -3(1-d_2)\nabla_\mu A^{\mu}\nabla_{\rho}A_{\sigma}\nabla^{\sigma}A^{\rho}\right.+\\
        &\left.
        \left(2 - 3d_2\right)\nabla_{\rho}A_{\sigma}\nabla^\gamma A^{\rho}\nabla^{\sigma}A_{\gamma} + 3d_2\nabla_{\rho}A_{\sigma}\nabla^\gamma A^{\rho}\nabla_{\gamma}A^{\sigma}\right] \; ,
    \end{split}
    \end{equation}
    where $F_{\mu\nu} = \nabla_\mu A_\nu - \nabla_\nu A_\mu$ is the Maxwell tensor and $X= A_\mu A^{\mu}/2$. The $G_i$'s are free functions of the Proca field and $\mathcal{G}_{\mu\nu}$ is the Einstein tensor. Note that in the above Lagrangian the usual $U(1)$ symmetry of the vector field is broken, i.e. $A^{\mu}$ is not Abelian. The above property becomes useful and interesting for cosmological implications, indeed it allows for an isotropic background evolution, see for example Ref.~\cite{DeFelice:2016yws}. 
    It is interesting to note that in this class of theories, on FLRW background, the vector field equation allow for constant solutions which are of de Sitter type, thus being potentially capable of describe the Late-times accelerated expansion of the Universe as well as an inflationary epoch.
    
    \subsubsection{Scalar Vector Tensor (SVT) theories}
    It is possible to construct a consistent covariant theory which combines together scalar, vector and tensor interations, known as SVT theories, see Ref.~\cite{Heisenberg:2018acv}. The resulting theory is richer than a theory built just from the Horndeski and generalized Proca theories. Indeed, it allows for the vector and the scalar fields to interact in non-trivial way. Thus, beyond the standard scalars of Horndeski and Proca theories, the free functions appearing in the SVT Lagrangian depend also on the scalars:
    \begin{eqnarray}
        X_2 = -\frac{1}{2}A^{\mu}\nabla_{\mu}\phi \; , \qquad
        Y_1 = \nabla_\mu \phi \nabla_\nu \phi F^{\mu\alpha}F^{\nu}_\alpha \; , \qquad Y_2 =\nabla_\mu \phi A_{\nu}F^{\mu\alpha}F^{\nu}_\alpha \; .
    \end{eqnarray}
    In the full SVT Lagrangian we also have terms containing the double dual Riemann tensor:
    \begin{equation}
        L^{\mu\nu\alpha\beta} = \epsilon^{\mu\nu\rho\sigma}\epsilon^{\alpha\beta\gamma\delta}R_{\rho\sigma\gamma\delta} \; ,
    \end{equation}
where the $\epsilon$'s are the Levi Civita symbols in 4 dimensions.

    The general form of the Lagrangian is given in Ref.~\cite{Heisenberg:2018acv} and the background and perturbed equations are developed in Ref.~\cite{Heisenberg:2018mxx}. They depend in general on whether the $U(1)$ invariance of the Proca field is broken or not. It turns out that the resulting theories are useful in Dark Energy applications and bouncing scenarios, allowing for an accelerated epoch of expansion while being capable of producing transient contracting phases, which can avoid the appearance of cosmological singularities.
    
    \subsubsection{Bimetric gravity}
    
    It is also possible to increase the number of degrees of freedom by introducing a new metric tensor field $f_{\mu\nu}$. This class of theory is usually known as \textit{bimetric gravity} and it was formulated as an attempt to generalize massive gravity models,  see Refs.~\cite{deRham:2014zqa,Hassan:2011zd,Volkov:2011an}. The action functional in the original formulation takes the form:
    \begin{equation}
        S = M_g^2\int\sqrt{-g}d^4x R_g + M_f^2\int \sqrt{-f}d^4x R_f + M_{eff}m^2\int \sqrt{-g}d^4x\mathcal{L}_{int} \;,
    \end{equation}
    where $M_i$ and $R_i$ are the Planck masses and Ricci scalars relative to the  metric $f$ and $g$. $M_{eff} = \left( M_g^{-1} + M_f^{-1}\right)^{-1}$ is the effective Planck mass and $m^2$ a mass term associated to the massive graviton of the metric $f$. The interaction Lagrangian is given by:
    \begin{equation}
        \mathcal{L}_{int}= \frac{1}{2}\left(X^2 - X^{\mu\nu}X_{\mu\nu}\right) + \frac{c_3}{3!}\epsilon_{\mu\nu\rho\sigma}\epsilon^{\alpha\beta\gamma\sigma}X^{\mu}_{\alpha}X^{\nu}_{\beta}X^{\rho}_{\gamma} + \frac{c_4}{4!}\epsilon_{\mu\nu\rho\sigma}\epsilon^{\alpha\beta\gamma\delta}X^{\mu}_{\alpha}X^{\nu}_{\beta}X^{\rho}_{\gamma}X^{\sigma}_{\delta} \; ,
    \end{equation}
    where $X^{\mu}_{\nu} = \delta^{\mu}_{\nu} + \gamma^{\mu}_{\nu}$, with $\gamma^{\mu}_{\nu}$ defined by the relation $\gamma^{\mu}_{\sigma}\gamma^{\sigma}_{\nu} = g^{\mu\sigma}f_{\sigma\nu}$. The above form of the interaction Lagrangian has been proposed by the authors to avoid the appearance of Boulware-Deser ghost in both the $g$ and $f$ metrics.  Bimetric models offer a rich and interesting phenomenology for cosmological implications and have been studied extensively in the literature, see for example Refs.~\cite{Akrami:2015qga,Akrami:2013pna}
    %%%%%%%%%%%%%%%%%%%%%%%%%%%%%%%%%%%%%%%%%%%%%%%%%%%%%%%%%%%%%%%%%%%%%%
    \subsection{Including higher order derivatives}
    
    It is well known that General Relativity is not renormalizable from the point of view of QFT due to the fact that the Newton constant has dimensions of an inverse squared mass $G_N \propto m^{-2}$. One of the main historical reasons to consider higher order derivative theories is that they modify the propagator improving its UV behavior. Let us consider for example the introduction of a $R_{\mu\nu}R^{\mu\nu}$ term; in this case the propagator can be symbolically written as:
    \begin{equation}
    \label{HDpropagator}
    \frac{1}{k^2}+\frac{1}{k^2}G_Nk^4\frac{1}{k^2}+ \frac{1}{k^2}G_Nk^4\frac{1}{k^2}G_Nk^4\frac{1}{k^2} + ... = \frac{1}{k^2 - G_N k^4} \; .
\end{equation}
The propagator of Eq.~\eqref{HDpropagator} is dominated at high Energy by the $k^{-4}$ term and its UV behavior is improved. On the other hand we can rewrite it as:
\begin{equation}
     \frac{1}{k^2 - G_N k^4} = \frac{1}{k^2} -\frac{1}{k^2 - 1/G_N} \;,
\end{equation}
 from which we can see that it decomposes in the standard graviton mode $k^{-2}$ together with the $-1/\left(k^2 - G_N^{-1}\right)$ mode, which has negative sign and thus corresponds to a ghost. 
 
 The appearance of ghost modes is a recurring theme in higher order derivative theories and is strongly related to Ostrogradsky instability, see Refs.~\cite{Motohashi:2014opa,Woodard:2015zca}.
 To briefly illustrate how Ostrogradsky instability works let us consider a non degenerate Lagrangian containing second order derivatives  $\mathcal{L}(x,\dot{x}, \ddot{x})$. The associated Euler Lagrange equations are:
\begin{equation}
    \frac{\partial \mathcal{L}}{\partial x} - \frac{d}{d t} \frac{\partial \mathcal{L}}{\partial \dot{x}} + \frac{d^2}{d t^2} \frac{\partial \mathcal{L}}{\partial \ddot{x}} =0\; , 
\end{equation}
and the non degeneracy conditions implies $\frac{\partial^2 \mathcal{L}}{\partial \ddot{x}^2} \neq 0$. Ostrogradsky showed that if we choose  the following 4 canonical coordinates:
\begin{equation}
\label{Ostrocoord}
    Q_1 = x \qquad \; , Q_2 = \dot{x} \; , \qquad P_1 = \frac{\partial\mathcal{L}}{\partial \dot{x}} - \frac{d}{d t }\frac{\partial\mathcal{L}}{\partial \ddot{x}} \; , \qquad P_2 = \frac{\partial \mathcal{L}}{\partial \ddot{x}} \; ,
\end{equation}
it is possible to perform a Legendre transformation and obtain the Hamiltonian:
\begin{equation}
\label{OstroHam}
    H = P_1 Q_1 + P_2 A\left(Q_1, Q_2, P_2  \right) - \mathcal{L}\left(Q_1, Q_2, P_2 \right) \; ,
\end{equation}
where the function $A\left(Q_1, Q_2, P_2  \right)$ is obtained inverting Eqs.~\eqref{Ostrocoord} for $\ddot{x}$. 
The Hamiltonian~\eqref{OstroHam} is linear in the canonical momentum $P_1$ and thus is  unbounded from below, i.e. the system is unstable. Note that the only assumption made here is invertibility of the Lagrangian with respect to $\ddot{x}$, so in order to have well-behaved higher order derivative theories we should consider only degenerate Lagrangians.
An interesting example of a degenerate theory is given by the beyond Horndeski Lagrangian, also known as Degenerate Higher Order Scalar Tensor (DHOST) theories, see Refs.~\cite{Gleyzes:2014dya,Langlois:2017mdk} for a detailed discussion. A similar construction can be done also for vector tensor theories, and we end up with the beyond generalized Proca theories, see Ref.~\cite{Heisenberg:2016eld}.

\subsection{Increasing the number of dimensions}

A way out from Lovelock's theorem that was explored by Lovelock himself is to consider a geometrical description of the gravitational interaction in more than four dimensions. On the other hand, we have no observational evidence of the presence of such extra dimensions, thus higher dimensional theories need a mechanism that hides or compactifies these dimensions at scales which do not emerge in standard experiments.
Higher dimensional theories have attracted a lot of interest in the past decades, with the most popular being probably string theory. To give a flavour of the potential of higher dimensional theories, we will discuss briefly here the Kaluza-Klein model, which inspired and motivated subsequent works on compactifications of higher dimensions and unifications of the fundamental interactions.
We will also introduce the Lanczos-Lovelock gravity, which is essentially a generalization of General Relativity to an arbitrary number of dimensions.

\subsubsection{Kaluza-Klein model}
The first higher dimensional extension of General Relativity was suggested by Kaluza in Ref.~\cite{Kaluza:1921tu}, in a similar framework as in a previous attempt by Nordstrom~\cite{Nordstrom:1988fi}.  In this model it is possible to obtain both Maxwell and Einstein equations in four dimensions from a geometrical vacuum theory with a fifth dimension. The five dimensional metric, $\tilde{g}_{ab}$, becomes here a function of the standard four dimensional metric $g_{\mu\nu}$ plus a vector field $A_\mu$ and a scalar field $\phi$, and could be written as:
\begin{equation} \tilde{g}_{ab} = \begin{pmatrix}
g_{\mu\nu} + \phi^2 A_{\mu}A_{\nu} & \phi^2A_{\mu} \\
\phi^2A_{\nu} & \phi^2
\end{pmatrix} \; .
\end{equation}
Kaluza imposed on the metric $\tilde{g}_{ab}$ the so-called cylinder condition, i.e. that it does not  depend on the fifth coordinate $\partial \tilde{g}_{ab}/\partial x^5=0$.
The five dimensional vacuum Einstein field equations reduce to:
\begin{equation}
R_{\mu\nu} - \frac{1}{2} g_{\mu\nu}R = \frac{k^2\phi^2}{2}T^{EM}_{\mu\nu} -\frac{1}{\phi}\left[ \nabla_\mu(\partial_\nu \phi) - g_{\mu\nu} \Box \phi \right] \; ,
\end{equation}
while the Maxwell field equations are:
\begin{equation}
    \nabla^\mu F_{\mu\nu} = -3\frac{\partial^\mu \phi}{\phi}F_{\mu\nu} \; . 
\end{equation}
The so called \textit{Kaluza's miracle} is that the standard four dimensional Einstein and Maxwell field equations, with the electromagnetic term appearing in the former as a source term, are recovered in the limit $\phi = 1$. However, the latter condition is not consistent with the Klein Gordon equation for the scalar field:
\begin{equation}
    \Box \phi = \frac{k^2\phi^3}{4}F_{\mu\nu}F^{\mu\nu} \; .
\end{equation}
A lot of criticism was made to Kaluza's proposal due to the cylindrical condition, i.e. the introduction of a fifth dimension that plays no role in the dynamics. To overcome this issue, Klein suggested in Ref.~\cite{Klein:1926tv} a mechanism of compactification of the fifth dimension, demanding that it has the topology of a circle $S^1$ of very small radius $r$. Thus, the whole spacetime has topology $R^4 \times S^1$ and physical fields must depend on the fifth dimension only periodically.  

\subsubsection{Lovelock gravity}
The topological space's shape of an object is identified by a constant number $\xi$, called Euler number, or Euler characteristic, regardless of the way in which the space is bent. The Euler characteristic in $2n$ dimensions can be written as the integral of the Euler density $\mathcal{R}^n$ which reads:
\begin{equation}
    \mathcal{R}^n = \frac{(2n!)}{2^n}\delta^{\mu_1}_{[\alpha_1}\delta^{\nu_1}_{\beta_1}\delta^{\mu_2}_{\alpha_2}\delta^{\nu_2}_{\beta_2}...\delta^{\mu_n}_{\alpha_n}\delta^{\nu_n}_{\beta_n]}\prod_{r=1}^n R^{\alpha_r\beta_r}_{\mu_r\nu_r}\; ,
\end{equation}
where the square bracket indicate antisymmetrization.
The Lovelock Lagrangian is the sum of the Euler densities:
\begin{equation}
    \mathcal{L} = \sqrt{-g}\sum_{t=0}^{n} \alpha_t \mathcal{R}^t \; ,
\end{equation}
and yields to conserved second order Euler Lagrange equations of motion, see for example Refs.~\cite{Lovelock:1971yv,Padmanabhan:2013xyr} for a detailed derivation. Expanding the above Lagrangian up to second order we obtain:
\begin{equation}
    \mathcal{L} = \sqrt{-g}\left[\alpha_0 + \alpha_1 R + \alpha_2\left(R^2 + R_{\mu\nu\rho\sigma}R^{\mu\nu\rho\sigma} - 4R_{\mu\nu}R^{\mu\nu}\right) + \mathcal{O}(R^3)   \right]\; ,
\end{equation}
which shows that at zero and first order the Lovelock Lagrangian reproduces the standard Einstein Hilbert action plus a cosmological constant, while from the second order term inside the bracket we appreciate that it contains the Gauss-Bonnet gravity term. Note that in four dimensions the second and higher order terms become trivial and we are left with standard GR. 

\subsection{Abandoning Lagrangian formulations}

Several proposals of modified gravity are based on ad hoc modifications of the EFE which are not derivable from an action functional, often with interesting cosmological applications. To illustrate the potential of this kind of modifications we will briefly present two theories belonging to this class, the Rastall gravity and the $RT$ nonlocal model.

\subsubsection{Rastall gravity}

Following the idea that the stress energy tensor $T_{\mu\nu}$ could be not conserved in curved spacetime, Rastall proposed in Ref.~\cite{Rastall:1973nw} the following modification of the Einstein field equations:
\begin{equation}
    R_{\mu\nu} -\frac{1}{2}g_{\mu\nu}R = 8\pi G\left( T_{\mu\nu} - \frac{\gamma - 1 }{2}g_{\mu\nu}T \right) \; ,
\end{equation}
with a non conserved continuity equation for $T_{\mu\nu}$:
\begin{equation}
    T^{\mu\nu}_{;\mu} = \frac{\gamma-1}{2}T^{\; \nu} \; . 
\end{equation}
It has been shown that Rastall gravity is very interesting from the cosmological point of view, being able to reproduce $\Lambda$CDM at the background level, see for example Ref.~\cite{Batista:2011nu}, while being different at perturbative level and resulting in a type of Dark Energy capable of clustering.

It is a matter of debate if it is possible or not to derive the Rastall equations from a Lagrangian density.
In the 80', in Ref.~\cite{1984NCimB..80...42S}, the Rastall field equation where obtained by a variational principle of a Lagrangian density, but the latter was not a scalar Lagrangian and thus the derivation is not completely satisfactory.
Some more recent attempts were made in Refs.~\cite{Santos:2017nxm,DeMoraes:2019mef}, where the field equations were obtained as a particular case of an $f(R,T)$ theory of the type $f(R,T) = f_1(R) + f_2(T)$, or from a matter Lagrangian non minimally coupled to gravity. However, some criticism emerged since for $f(R,T)$ theories of this type has been claimed that the $f_2(T)$ type term should be included in the matter Lagrangian and not in the gravitational part, see for example Ref.~\cite{Fisher:2019ekh}. 
It was also suggested in Ref.~\cite{Visser:2017gpz} that Rastall gravity is  equivalent to general relativity and Rastall's stress–energy tensor corresponds to an artificially isolated part of the physical conserved one. This point of view, however, was criticized in Ref.~\cite{Darabi:2017coc} and the debate is still open.

\subsubsection{The $RT$ model}
\label{sec:NLF}

The $RT$ model was proposed in Ref.~\cite{Maggiore:2013mea} and consist of a nonlocal modification of the EFE involving the inverse d'Alembertian of the Ricci scalar $\Box^{-1} R$:
\begin{equation}
    R_{\mu\nu}-\frac{1}{2}g_{\mu\nu} R -\frac{m^2}{3}\left(g_{\mu\nu}\Box^{-1}R\right)^T = 8 \pi G_N T_{\mu\nu} \; ,
\end{equation}
 where the  superscript $T$ denotes the extraction of the transverse part, which is in itself already a nonlocal operation.
 
 We will discuss in detail nonlocal modifications of gravity in chapter \ref{chapter:NL}, for the moment we will just mention that the within the $RT$ model one is able to reproduce a viable cosmological history both at background and perturbative level. Contrary to other nonlocal models with similar features, the $RT$ model is also compatible with experiments at Solar System scales, in particular Lunar Laser Ranging constraints \cite{Belgacem:2018wtb}, making it very appealing despite the lack of a Lagrangian formulation.
 %%%%%%%%%%%%%%%%%%%%%%%%%%%%%%%%%%%%
 \subsection{Giving up locality and Lorentz invariance}  
 Another class of theories that escapes Lovelock's theorem is based on modifications of gravity which  include nonlocal terms or which broke explicitly Lorentz invariance. We will discuss in detail the former in Sec.~\ref{chapter:NL}, while we present here as prototypical examples of the latter class of theories the Unimodular and the Ho\v{r}ava-Lifshitz gravities.
 %%%%%%%%%%%%%%%%%%%%%%%%%%%%%%%%%%%%%%%%%%%%%%%%%%%%%%%%%%%%%%%%%%
 \subsubsection{Unimodular gravity}
 The ideas behind Unimodular Gravity (UG) are almost as old as GR itself, and were considered by Einstein already in Refs.~\cite{Einstein:1916vd,1952prel.book..189E}. From the mathematical point of view, UG is equivalent to standard GR with the following gauge choice, called Unimodular condition:
 \begin{equation}\label{UGeq}
     \sqrt{-g}=\epsilon_0 \; ,
 \end{equation}
 where $\epsilon_0$ is a fixed scalar density which provides a fixed volume elements. Thus, UG is essentially GR with less symmetry, being invariant only with respect to the restricted group of diffeomorphisms respecting the Unimodular condition. The interesting property of UG is that, at classical level, its field equations coincide with the traceless EFE. Then, taking into account them together with the Bianchi identities, one obtain the standard EFE with a cosmological constant appearing as an integration constant.
 Quantum corrections to the energy-momentum tensor of matter which are of the form $Cg_{\mu\nu}$, where $C$ is a constant over spacetime, do not contribute to the traceless EFE. In particular, vacuum fluctuations in the trace of the energy-momentum tensor of matter do not affect the metric.  With the latter interpretation the cosmological constant does not couple to gravity, and consequently UG solves the cosmological constant problem, see Ref.~\cite{Weinberg:1988cp}.  
 Several generalizations of UG have been proposed, see for example Ref.~\cite{Henneaux:1989zc}, where the right hand side of Eq. \eqref{UGeq} is equal to the divergence of a vector density field, or Ref.~\cite{Barvinsky:2017pmm} where an ADM decomposition of the spacetime is assumed with the requirement that the lapse is a function of the determinant of the spatial metric $N=N(\gamma)$ only.
 Of course UG and its generalizations differ from GR at a quantum level, and their quantization is an active field of research,  see for example Refs.~\cite{Alvarez:2015pla,Smolin:2009ti,Padilla:2014yea,Fiol:2008vk,Shaposhnikov:2008xb}.
 %%%%%%%%%%%%%%%%%%%%%%%%%%%%%%%%%%%%%%%%%%%%%%%%%%%%%%%%%5     
 \subsubsection{Ho\v{r}ava-Lifshitz gravity}

 The model was suggested in Ref.~\cite{Horava:2009uw} as a viable candidate of quantum gravity, and is inspired  by physics of condensed matter systems. Its characteristic feature is that space and time are treated at fundamental level on a different ground, in such a way that they scale anisotropically in the UV limit. The degree of anisotropy between space and time is measured by the anisotropic parameter $z$, and the resulting theory is power counting renormalizable for certain values of $z$. The starting point of the construction is that the line element has the following ADM shape: \begin{equation}
ds^2 = -N^{2}dt^2 + g_{ij}\left(dx^i + N^{i}dt\right) \left(dx^j + N^{j}dt\right),
\end{equation}
in which $N$ is the shift, $N^i$ the lapse and $g_{ij}$ is the spatial metric.
In GR we have the gauge freedom of representing the line element in this way foliating the space-time in terms of spacelike surfaces $\Sigma_{t}$, whilst in Ho\v{r}ava gravity the above decomposition is not just a choice of coordinates but  rather the fundamental structure of the spacetime.
The kinetic term of the action is given by:
\begin{equation}
S_{kin} = \frac{1}{g_{K}}\int d^{3}x d^t N \sqrt{g}\left[ K_{ij}K^{ij} -\lambda K^2 \right] \; ,
\end{equation}
where the main difference with respect to standard GR is in the constant parameter $\lambda$, which must be unity if we demand Lorentz invariance. The potential part of the action, due to the anisotropic scaling, allow for the presence of higher order derivative terms of the \textit{spatial} Ricci tensor $P_{ab}$, defined in terms of the spatial metric $g_{ab}$. To achieve power counting renormalizability in 3+1 dimension we need $z=3$, which implies that we can have term up to cubic order in the 3D Ricci tensor and its spatial derivatives. The specific form of the potential depends on the formulation of Ho\v{r}ava gravity we are considering. In the original formulation of Ref.~\cite{Horava:2009uw} it is given by:
\begin{equation}
\begin{split}
V_{HL} =& \frac{-g^2_{K}}{2\omega^4}C^{ij}C_{ij} + \frac{g^2_{k} \mu }{2\omega^2}\epsilon^{ijk}P_{il}\nabla_{j}P^{l}_{k} -\frac{g^2_{K}\mu^2}{8}P^{ij}P_{ij}\\
&+ \frac{g^2_{K}\mu^2}{8(1-3\lambda)}\left( \frac{1 - 4\lambda}{4}P^2 + \Lambda P -2\Lambda^2 \right) \; ,
\end{split}
\end{equation}
where $C_{ij}$ is the Cotton tensor and $ g_k$ and $\omega $ are coupling parameters.
At long distances this potential is dominated by the last two terms, the cosmological constant and the spatial curvature, and the theory flows in the infrared to $z=1$ so that Lorentz invariance is accidentally restored.
There is a very interesting phenomenology arising from Ho\v{r}ava gravity in cosmological applications. It has been shown that certain choices of the potential are able to mimic DM, see Ref.~\cite{Mukohyama:2009mz}. It is also possible to seed cosmological perturbation without inflation, see Ref.  \cite{Mukohyama:2010xz}, realizing bouncing scenarios, see Ref.~\cite{Brandenberger:2009yt}, and model Dark Energy, see Refs.~\cite{Saridakis:2009bv,Carloni:2010nx,Calcagni:2009ar}.

\section{Ethology of Dark Energy}
As we saw in the previous section, there is a theoretically broad landscape of Dark Energy candidates. Thus, it is of the utmost importance to have a framework in which to study the impact of each particular theory on cosmological or astronomical observables. The standard approach consist of studying the specific form that a bunch of observed parameters takes in a modified gravity model and compare it with experimental data.

\subsection{The $\eta$ and $Y$ parameters }
A given theory of DE which allows for a background compatible with the accelerated expansion of the Universe will have, at perturbative level, some impact on smaller scales. 
For cosmological implications one is usually interested  only in scalar perturbations, and it is convenient to work in the Newtonian gauge:
\begin{equation}
    ds^2 = -\left(1+2\Psi\right)dt^2 + a^2(t)\left(1+2\Phi\right)\delta_{ij}dx^idx^j \; ,
\end{equation}
where $\Psi$ and $\Phi$ are two scalar functions, i.e. the two gravitational potentials. One should study the perturbed modified EFE for this metric and compare with the ones of standard GR.
For many purposes it is useful to work in the quasi static approximation (QSA), i.e. within the assumptions that spatial derivatives dominate over time ones. This approximation is valid only on scales well inside the Hubble Horizon, $k/a H \gg 1$, see Ref.~\cite{Amendola:2019laa} for a detailed discussion about the scope of validity of the QSA.  From the modified EFE we obtain the two generalized Poisson equations in Fourier space for the potentials $\Psi$ and $\Phi$ in the case of pressureless matter:
\begin{eqnarray}
&k^2\Phi = \frac{1}{2} Y(k,t) \eta (k,t)\rho_m(t)\delta_m (k,t) \; , \\
&k^2\Psi = \frac{1}{2} Y(k,t) \rho_m(t)\delta_m (k,t) \; , 
\end{eqnarray}
where we have defined the anisotropic stress parameter $\eta = -\Phi/\Psi$ and the $Y$ parameter, which describes an effective gravitational coupling $G_{eff}$  and measures deviations from the Newton constant $G_N$ for matter. Both these parameters can be constrained by observations; for example $\eta$ has been constrained to be $\eta \leq 10^{-5}$ on solar system scales by the Cassini spacecraft, see Refs.~\cite{Will:2014kxa,Uzan:2010pm,Bertotti:2003rm}. It is also possible to constrain $Y$ and its time derivative at various scales, see for example Refs.~\cite{Hofmann:2018myc,DeglInnocenti:1995hbi,Verbiest:2008gy,Giani:2020fpz}.

\subsection{Linear theory of structure formation}
As we saw before, several Dark Energy models can be expressed in the form of a scalar-tensor theory, i.e. they belong to the general class of Horndeski theories. Through an effective field theory approach (EFT) for the Horndeski theories, it was shown in \cite{Bellini:2014fua} that the cosmological information about linear perturbation theory can be encoded in four parameters:
\begin{itemize}
    \item $\alpha_K$ is a parameter related to the kinetic energy of the scalar field and to it contribute all the $G_i$ functions of the Hordenski Lagrangian \eqref{HL}. It is also called \textit{Kinecity} 
    \item $\alpha_B$ is a parameter related to the clustering properties of DE. It is also called \textit{Braiding} and comes from the mixing of the kinetic terms of both the metric and the scalar field. To it contribute the functions $G_3, G_4$ and $G_5$.
    \item $\alpha_M$ encodes the effects of a varying effective Planck mass $M$ and generates anisotropic stress. To it contribute $G_4$ and $G_5$
    \item $\alpha_T$ is related to the velocity of propagation of tensor modes. It leads to the emergence of anisotropic stress by modifying the Newtonian potential $\Psi$ even in absence of scalar perturbations. To it contribute both $G_4$ and $G_5$.
\end{itemize}

A similar EFT approach could be applied also to Generalized Proca Theories, see Eq.~\eqref{GPL}, and SVT theories. 

As an example of the capability of the method, let us consider the measurements of $\alpha_T$ made possible from the event GW170817 and its electromagnetic counterpart. Since no significant deviation on the velocity of the gravitational waves was detected with respect to the value predicted by GR, the observation suggests $\alpha_T = 0$. This in turns implies $G_4 = const$ and $G_5 = 0$, thus ruling out roughly half of the Hordenski theories, see Refs.~\cite{Creminelli:2017sry,Kase:2018aps,Bordin:2020fww,Ezquiaga:2017ekz,Dalang:2019rke}. The same applies for the $G_4$ and $G_5$ function of generalized Proca  and SVT theories.
There is however still a caveat in the above argument, which relies on the fact that the event GW170817 comes from a fairly close distance, and thus we only got information about the value of $\alpha_T$ from Late-times observations. It was showed in Ref.~\cite{Amendola:2018ltt} that it is possible to have a class of theories, which exhibits scaling behavior, capable of reaching dynamically an attractor solution compatible with $\alpha_T = 0$. 

\subsection{Equation of state of Dark Energy}
We do know that the equation of state parameter $w_{DE}$ of Dark Energy in the case of a cosmological constant behave as the one of vacuum energy, and has the value $w_{\Lambda} = -1$. Current observations are compatible with this value, but do not exclude a wider parameter space with enough accuracy. It must be stressed however that a measure of $w_{DE}$ alone cannot tell us too much about the fundamental nature of Dark Energy, see for example Refs.~\cite{Amendola:2016saw, Amendola:2012ky}. On the other hand, a precise measurement of $w_{DE}$ could be used to rule out particular Dark Energy candidates. For example, a measurement of $w \neq -1$ with enough statistical accuracy  would rule out a pure $\Lambda$CDM scenario. Most of $DE$ models predict a value for $w_{DE} > -1$, thus a measurement in this direction would not be particularly enlightening about the nature of DE. On the other hand, a statistically significant measure of $w_{DE} < -1$ at any time of the cosmological evolution would carry a lot of information about gravitational physics. Such a regime, called \textit{phantom}, it is indeed associated to the fact that either gravity is not minimally coupled to matter, or that DE is not a perfect fluid which can interact with other species. Indeed, it is a well known fact that a perfect fluid or a minimally coupled scalar field in a phantom regime would carry ghost or gradient instabilities, see Ref.~\cite{Dubovsky:2005xd}.  

\subsection{Variation of the electromagnetic coupling $\alpha_{EM}$}

As we mentioned before for the case of the Newton constant $G_N$, some models of Dark Energy result in a violation of the equivalence principle.
The specific case of a possible variation of the fine structure constant was discussed by Bekenstein in the pioneering work~\cite{Bekenstein:1982eu}. Bekenstein, at the time, concluded that tests of the equivalence principle rule out spacetime variability of $\alpha_{EM}$ at any level.
In the last decades, however, huge improvements have been made from the experimental point of view, leading to tight constraint on the variation of $\alpha_{EM}$, see for example Refs.~\cite{Uzan:2010pm, Leal:2014yqa,Holanda:2015oda,Pinho:2016mkm,Hees:2020gda}, and  claims of statistical evidences of $\alpha_{EM}$ variations, see for example Refs.~\cite{Murphy:2003hw,Webb:2010hc}. For the above reasons, the subject is nowadays very popular and a violation of the equivalence principle could potentially confirm or rule out several alternative theories of gravity \cite{Tino:2020nla}. One could in general distinguish between variations on the value of $\alpha_{EM}$ on large or local scales. On local scales the variation of $\alpha_{EM}$ is related to the local gravitational field, see for example Refs.~\cite{Bekenstein:2009fq, Shaw:2006zs, Barrow:2014vva}.  On cosmological scales these could be motivated by a modification of the gravitational theory due to the dynamical behavior of the DE field and were extensively studied in the literature, see for example Refs.~\cite{Olive:2001vz,Marra:2005yt,Barrow:2009nt, Barrow:2011kr,Barrow:2013uza,Sloan:2013wya,Graham:2014hva,vandeBruck:2015rma,Fritzsch:2016ewd}. 
 It is important to note that the connection between DE and the variation of $\alpha_{EM}$ is of great importance from the observational point of view, since it is indeed possible to relate constraints on $ \Delta\alpha$ to constraints on DE parameters, see for example Refs.~\cite{Calabrese:2013lga,  Martins:2019ebg, Martins:2015jta}.

% !TEX root = ../thesis-example.tex
%
\chapter{Nonlocal gravity}
\label{chapter:NL}

\epigraph{\textit{I am a Quantum Engineer, but on Sundays I Have Principles.}}{John Stewart Bell}

Most of our research in the last years focused on nonlocal modifications of gravity as DE candidates. We mention this class of theories in the previous chapter, since giving up from locality is a possible escape route from Lovelock's Theorem. Amongst the possible choices of nonlocal modifications, particularly interesting for DE applications are those which introduce the inverse d'Alembert operator acting on the Ricci scalar $\Box^{-1} R$. In this chapter we will briefly review this particular branch of modified gravity models, with emphasis on their fundamental motivations and their general features. In particular, we will introduce the Deser Woodard (DW), the $RR$\footnote{The $RR$ model takes its name from the structure of its Lagrangian term $R\frac{1}{\Box}R$.} and the VAAS\footnote{This model takes its name from the initials of authors Vardanyan, Akrami, Amendola and Silvestri.} nonlocal models. The first two are amongst the most popular and analyzed models belonging to this class, while the latter has been proposed recently in Ref.~\cite{Vardanyan:2017kal}.

\section{ Motivation}

 Nonlocalities emerged from quantum mechanics already in the early stage of its formulation, see for example Ref.~\cite{Einstein:1935rr}, and Ref.~\cite{Wiseman_2006} for a nice historical review. Phenomena observed experimentally like the quantum entanglement and the Ahronov-Bohm effect show indeed that an effective description of quantum mechanics, or rather of reality, must be nonlocal.
 As it is widely known, it is difficult to construct a consistent theory of quantum gravity starting from GR, which in order to be renormalizable requires the introduction of infinite counterterms. Thus, just like the Fermi description of the weak interaction, one could think that GR is just an effective geometrical description of a most deep underlying theory, and it makes sense to look for phenomenological modifications of the EFE that arise from quantum effects. The idea that the latter could be used to explain the nature of DE or other open problems of the $\Lambda$CDM is particularly intriguing, and the \textit{Leitmotiv} of many nonlocal theories of gravity. The standard recipe is to postulate an ansatz for the functional form of the nonlocal modification motivated by fundamental physics. Then one studies the phenomenology of the modification at background and perturbative level and test it against observations.
 
 \subsection{The quantum effective action}
Nonlocal effects could naturally arise when we move from the classical action functional of a given field theory to its quantum effective action. We will briefly review here the construction of the quantum effective action and its generalization to curved background following Ref.~\cite{Belgacem:2017cqo}.

To begin with, consider a scalar field $\varphi$ with classical action $S\left[\varphi\right]$ in a flat space of dimension $D$. Once we introduce an auxiliary, classical source $J(x)$ we can define the generating functional of the connected Green's function $W[J]$:  
\begin{equation}\label{vevscalar}
    e^{i W\left[J\right]} = \int D\varphi e^{i S\left[\varphi\right] + i\int J \varphi} \; ,
\end{equation}
where the path integral measure $D\varphi$ denotes integration over all the possible configurations of the field $\varphi$ and $\int J\varphi$ is a shortcut for the integral $\int d^Dx J(x) \varphi(x)$, so that the spatial dependence of the source has been integrated out.
Functional variation of $W[J]$ with respect to the source gives the vacuum expectation value of the field $\varphi$ in presence of a source, i.e.:
\begin{equation}\label{4.2}
    \frac{\delta W[J]}{\delta J(x)} = \langle 0|\varphi(x) |0 \rangle_{J}\equiv \phi[J] \; ,
\end{equation}
where we have defined the scalar field $\phi$ to indicate the vacuum expectation value of $\varphi$ as function of the source $J$.
The quantum effective action $\Gamma[\phi]$ is a function of the vacuum expectation value and is defined as the Legendre transform:
\begin{equation}
    \Gamma[\phi] \equiv W[J] - \int \phi J[\phi] \; ,
\end{equation}
where $J[\phi]$ is obtained by inverting Eq.~\eqref{4.2}.
By varying the quantum effective action $\Gamma$ we obtain:
\begin{equation}\label{eomEA}
    \frac{\delta \Gamma[\phi]}{\delta \phi(x)} = -J(x) \; ,
\end{equation}
where the implicit spatial dependence of $\phi[J(x)]$ has been exploited. As we can see, since on the right hand side of Eq.~\eqref{eomEA} we have the source $J(x)$, the variation of the quantum effective action gives directly the equation of motion for the vacuum expectation value of the field.
A useful path integral representation of the quantum effective action is:
\begin{equation}
\begin{split}
    e^{i\Gamma[\phi]} &= e^{i W[J] - i\int\phi J}\\
    &= \int D\varphi e^{i S[\varphi + \phi] - i\int \frac{\delta \Gamma[\phi]}{\delta \phi}\varphi} \; ,
\end{split}
\end{equation}
which explicitly shows that the quantum fluctuations of the field $\varphi$ have been integrated out in the quantum effective action $\Gamma[\phi]$, which is instead a functional of the vacuum expectation value and the source only.
It is straightforward to generalize the above construction on curved background described by a metric $g_{\mu\nu}$ using a semi-classical approach, i.e. treating the metric at a classical level while the other fields as quantum objects. 
The representation of quantum effective action then becomes:
\begin{equation}
    e^{i \Gamma\left[g_{\mu\nu},\phi\right]} = e^{i S_{EH}}\int D\varphi e^{i S_m\left[g_{\mu\nu,\phi \varphi}\right] - i\int \frac{\delta \Gamma\left[g_{\mu\nu, \phi}\right]}{\delta \phi}\varphi} \; ,
\end{equation}
from whose variation we obtain the semi-classical EFE $G_{\mu\nu} = \langle 0|T_{\mu\nu}|0\rangle$.
\subsection{The QED example}
To understand which kind of nonlocal modifications could appear in the quantum effective action let us consider the case of Quantum Electrodynamics (QED).
The quantum effective action takes the form, see for example Ref.~\cite{Dalvit:1994gf}:\footnote{When we integrate out the quantum fluctuations of the electron and restrict ourselves to terms involving the photon field only for simplicity}
    \begin{equation}
        \Gamma_{QED} = -\frac{1}{4}\int d^4 x \left[F_{\mu\nu}\frac{1}{e^2\left(\Box\right)}F^{\mu\nu} + \mathcal{O}\left(F^4\right) \right] \; .
        \end{equation}
    where $e^2 \left(\Box \right)$ is called \textit{form factor}. When the electron mass is small with respect to the relevant energy scale we have:
\begin{equation}
    \frac{1}{e^2 \left(\Box \right)} \simeq \frac{1}{e^2\left(\mu\right)} - \frac{1}{\left(12\pi\right)^2} \log{\left(\frac{-\Box}{\mu^2}\right)} \; ,
\end{equation}    
where the $\mu$ is the renormalization scale and $e^2\left(\mu \right)$ the renormalized charge. 
Nonlocality emerges because of the logarithm of the d'Alembert operator $\Box$. It is defined by its integral representation:
    \begin{equation}
        \log{\left(\frac{-\Box}{\mu^2}\right)} = \int_0^{\infty} dm^2\left[\frac{1}{m^2 + \mu^2} - \frac{1}{m^2 - \Box } \right] \;,
    \end{equation}
where nonlocality emerges due to the appearance of the $\Box^{-1}$ operator.    
In the above example we have explicitly shown how from the classical Lagrangian of QED we could obtain nonlocal contributions due to the running of the coupling constant.

\section{Technical stuff}
\subsection{Localization}
The main character of the nonlocal class of theories we are considering here is the inverse of the d'Alembert operator acting on the Ricci scalar $\Box^{-1}R$. In order to perform calculations involving this quantity an extremely useful trick was developed in Ref.~\cite{Nojiri:2007uq}, which makes it is possible to cast these  models in the form of a scalar tensor theory. 
The starting point is to define an auxiliary scalar field $U = -\Box^{-1}R$ whose Klein Gordon equation immediately follows:
\begin{equation}
\label{KGlocalized}
    \Box U = -R \; .
\end{equation}
Thus a general Lagrangian density containing an arbitrary function  $f(\Box^{-1} R)$ could be rewritten as:
\begin{equation}
    \mathcal{L}_{NL} = f(U) + \lambda\left(\Box U + R\right) \, ,
\end{equation}
where we introduced the Lagrange multiplier $\lambda$.
If negative powers of the d'Alembert operators appear in the original action, the above procedure can be iterated and the  theory is mapped in a multi-scalar tensor theory. For example, if the Lagrangian contains a term  $ \Box^{-2}R$, as in the model proposed in Ref.~\cite{Amendola:2017qge}, it can be localized by introducing 4 coupled auxiliary fields  with their respective Lagrange multipliers:
\begin{eqnarray}
   & \Box U =- R \;, \qquad \Box S = -U \; ,\qquad \Box Q = -1 \;, \qquad \Box L = -Q \; .
\end{eqnarray}

It is important to properly carry on the procedure of localization without introducing modifications of the original theory. The equivalence between the two formulations was debated after the papers~\cite{Dodelson:2013sma,Park:2012cp} and~\cite{Nersisyan:2017mgj}, which were analyzing structure formation in the DW model and obtained initially different results. It turns out that  the analysis of Ref.~\cite{Dodelson:2013sma} was not correct, but the arising discussion about the localization procedure helped to outline its possible stability issues and the appearance of ghosts~\cite{Park:2017zls,Park:2019btx}

\subsection{Degrees of freedom and stability in the localized formulation}
\label{sec:dofstability}
 The introduction of auxiliary fields naturally rises the question of whether there are or not new degrees of freedom generated by the nonlocal operator $\Box^{-1}R$. The question is subtle and some care must be taken in the procedure of localization. Moreover, by looking at Eq.~\eqref{KGlocalized}, it is straightforward to realize that the kinetic energy of the auxiliary field would be of negative sign, i.e a ghost.
 
 To properly understand this point, let us consider the Lagrangian of a massive Proca field:
\begin{equation}\label{Lproca}
    \mathcal{L}_{Proca} = -\frac{1}{4}F^{\mu\nu}F_{\mu\nu} - \frac{1}{2}m^2 A^{\mu}A_{\mu} \; ,
\end{equation}
which describe a massive boson, with the $U(1)$ gauge invariance broken by the mass term.
It has been shown in Ref.~\cite{Dvali:2006su} that the above Lagrangian is equivalent to the following nonlocal but gauge invariant one:\footnote{As long as we impose that the inverse d'Alembertian is defined in terms of the retarded Green's function only to obtain casual solutions.}
\begin{equation}\label{Lprocanl}
    \mathcal{L}= -\frac{1}{4}F_{\mu\nu}\left(1-\frac{m^2}{\Box}\right)F^{\mu\nu}.
\end{equation}
 If we now proceed with the localization procedure and define an auxiliary field $U^{\mu\nu} =\Box^{-1}F^{\mu\nu}$, the latter would introduce new degrees of freedom, which are surely not present in the Lagrangian~\eqref{Lproca}. 
In order to avoid the appearance of such spurious degrees of freedom, in the localization procedure we have to select only a particular family of solutions of Eq.~\eqref{KGlocalized}. 
The general solution for the field $U$ would be the sum of the homogeneous and the particular one: $U = U_{hom} + U_{par}$. In order to avoid any further propagating degrees of freedom, during the localization procedure we should specify that $U$ is not the most general solution defined by $\Box^{-1}R$, but only a 
particular solution selected by fixing its boundary conditions. Following this recipe we avoid the appearance of ghosts associated to $U$ after quantization. Of course the above procedure does not prevent the theory from developing instabilities at classical level, which however do not necessarily imply a pathological behavior of the theory. In particular, if such instabilities emerge on cosmological scales and at Late-times, they could be able to drive the present accelerated expansion of the Universe. 
\section{Nonlocal models}

In this section we will briefly review a bunch of different nonlocal models proposed in the last years which are able to provide a viable cosmological history at background level, and are thus potentially very interesting Dark Energy candidates.

\subsection{The Deser Woodard model}
One of the most popular nonlocal gravitational theory was proposed in Ref.~\cite{Deser:2007jk} and it is known as Deser Woodard (DW) model.
The fundamental idea is to incorporate nonlocal effects without postulating a priori any specific form of the nonlocal modification. This is achieved by introducing a free function of the inverse d'Alembertian of the Ricci scalar, called distortion function, and reconstruct it in such a way that it produce a background cosmological history identical to the one of the $\Lambda$CDM without cosmological constant. Then one can study the perturbative regime and test its compatibility with observations.
For a review on the main features of the model we address the reader to Ref.~\cite{Woodard:2014iga}; the issue of ghosts is studied in Ref.~\cite{Park:2019btx}.
A detailed study of its dynamics is performed in Ref.~\cite{Koivisto:2008xfa}, while its  Newtonian limit is studied in Ref.~\cite{Koivisto:2008dh}. The effects of such kind of modification for structure formation are studied in Refs.~\cite{Park:2012cp,Dodelson:2013sma,Nersisyan:2017mgj}, while constraints from observational datasets are found in Ref.~\cite{Amendola:2019fhc}. Finally, an improved version of the model has been recently proposed in Ref.~\cite{Deser:2019lmm}.

The Lagrangian of the model is:
\begin{equation}
    \mathcal{L} = \mathcal{L}_{EH} + Rf\left(\frac{1}{\Box}R\right) \; ,
\end{equation}
and it could be mapped in a localized scalar tensor theory by introducing the auxiliary fields, see Ref.~\cite{Nojiri:2007uq}:\footnote{We are using a different definition for the field $U$ with respect to the one of Ref.~\cite{Nojiri:2007uq}. The original ones are obtained by making the substitutions $U \rightarrow -U$, $\Bar{f}\rightarrow-\Bar{f}$. }
\begin{eqnarray}
        \Box U = -R \; ,\\
        \Box V = \Bar{f}(U)R \; ,
\end{eqnarray}
where $\Bar{f}$ indicates the derivative of the distortion function $f$ with respect to $U$. The EFE on FLRW flat background  using the e-fold number $N$ as time parameter are:
\begin{eqnarray}
        \left(1 + f - V\right) = -\frac{U'V'}{6} - f' + V' +\frac{\Omega_R + \Omega_M}{h^2} \; , \label{CosmoDW1}\\
        \left(2\xi + 3\right)\left(1 + f - V\right) = V'' - f'' + \left(V' - f'\right)\left(2 + \xi\right) + \frac{U'V'}{2} - \frac{\Omega_R}{h^2} \; ,\label{CosmoDW2}
\end{eqnarray}
and the KG equations for the auxiliary fields are:
\begin{eqnarray}
  U'' + \left(3 + \xi\right)U' = 6\left(2 + \xi\right) \;, \label{KGDWU}\\
  V'' + \left(3 + \xi\right)V' =-6\left(2 + \xi\right)\Bar{f} \;. \label{KGDWV}
\end{eqnarray}
To solve the above system of equations without introducing ghost we need to fix the initial conditions for the auxiliary fields. 
If we impose the latter in such a way that they are compatible with a radiation-dominated epoch, $h_i^2 \sim \Omega_{Ri}$ and $\xi_i \sim -2$, Eqs.~\eqref{CosmoDW1} and \eqref{CosmoDW2} provide the two following constraints:
\begin{eqnarray}
        f_i - V_i = -\frac{1}{6}U_i'V_i' - f_i' + V_i' \; , \label{ICCosmoDW1} \\
        -f_i + V_i = -V_i' -f_i'' + \frac{U_i'V_i'}{2} \label{ICCosmoDW2} \; ,
\end{eqnarray}
where in Eq.~\eqref{ICCosmoDW2} we used Eq.~\eqref{KGDWV} evaluated at $\xi_i = -2$. 

As expected, the value of $U_i$ is unconstrained since it appears on the field equations only through the function $f(U_i)$; to compute the time derivative of the latter we use the chain rule $f' =  \Bar{f}U'$, so that:
\begin{equation} \label{f''}
   f'' = \Bar{\Bar{f}}U'^2 -\Bar{f}U'' \; .
\end{equation}
Evaluating Eq.~\eqref{f''} at $N = N_i$ we get:
\begin{equation} \label{f_i''}
    f_i''= \Bar{\Bar{f_i}}{U_i'}^2 - \Bar{f}U_i' \; ,
\end{equation}
where we have used Eq.\eqref{KGDWU} with $\xi_i = -2$.

In Ref.~\cite{Deffayet:2009ca} it was developed a technique to reconstruct the distortion function starting from any cosmological history. For the $\Lambda$CDM, the best analytical approximation for the distortion function was computed in Ref.~\cite{Park:2017zls} and is given by:
\begin{equation} \label{fDW}
    f(U) = 0.243\left[ \tanh{\left(0.348Z + 0.033Z^2 + 0.005Z^3\right) } - 1\right]\; ,
\end{equation}
where $Z= -U + 16.7$. 
Note that the above distortion function satisfies the condition $f(U_i) \simeq 0$ by choosing $U_i = 0$.

\subsection{The $RR$ model}
\label{subsec:RR}
Another very popular nonlocal theory is the $RR$ model, proposed by Maggiore and Mancarella in Ref.~\cite{Maggiore:2014sia}. The model attempts to ascribe the accelerated expansion of the Universe to nonlocal modifications of the quantum effective action caused by the appearance of a new mass scale $m^2$ dynamically generated in the infrared. For a complete review on the model we address the reader to Ref.~\cite{Belgacem:2017cqo}; the cosmological perturbation theory and the impact on structure formation are studied in Ref.~\cite{Dirian:2014ara}. A dynamical system analysis of the model was performed numerically in Ref.~\cite{Nersisyan:2016hjh}, while in  Ref.~\cite{Dirian:2016puz} the model is tested against observation and compared using a bayesian approach with the $\Lambda$CDM.

In this theory one adds to the usual Einstein-Hilbert Lagrangian a nonlocal modification of the form:
\begin{equation} 
    \mathcal{L}= \mathcal{L}_{EH} -\frac{1}{6}m^2 R\frac{1}{\Box^2}R \;,
    \end{equation}
and it is possible to localize the theory by introducing the auxiliary fields:
\begin{eqnarray}
        \Box U = -R \; , \\
        \Box S = -U \; . 
\end{eqnarray}
Defining the dimensionless quantity $V = H_0^2S$ and varying the action we obtain the following background EFE, written in terms of the e-fold number $N$, see Ref.~\cite{Nersisyan:2016hjh}:
\begin{eqnarray}
        h^2 = \frac{\Omega_{M}^0e^{-3N} + \Omega_{R}^0e^{-4N} +\frac{\gamma}{4}U^2}{1 + \gamma\left(-3V -3V' + \frac{1}{2}U'V'\right)} \; , \label{CosmoRR1} \\
        \xi = \frac{\frac{-3\Omega_M -4\Omega_R}{h^2} + 3\gamma\left(\frac{U}{h^2} + U'V' -4V'\right)}{2\left(1-3\gamma V\right)} \label{CosmoRR2} \; ,
\end{eqnarray}
where we have defined $\gamma \equiv m^2/9H_0^2$. The KG equations of the auxiliary fields are instead:
\begin{eqnarray}
        V'' + V'\left(3 + \xi\right)= \frac{U}{h^2}\; , \label{KGRRV} \\
        U'' + U'\left(3+\xi \right)= 6\left(2 + \xi \right) \label{KGRRU} \; .
\end{eqnarray}
In order to avoid the introduction of new degrees of freedom we have to fix properly the initial conditions. Compatibility with radiation domination  $h_i^2 \sim \Omega_{Ri}$ and $\xi_i \sim -2$ imply that  Eqs.~\eqref{CosmoRR1} and \eqref{CosmoRR2} at some initial time $N = N_i$ become:
\begin{eqnarray}
\frac{U_i^2}{4h_i^2}&=&  -3V_i -3V_i' + \frac{1}{2}U_i'V_i' \; ,\label{ICCosmoRR1}\\
V_i &=& \frac{U_i}{4h_i^2} + \frac{1}{4}U_i'V_i' - V_i' \;  \label{ICCosmoRR2},
\end{eqnarray}
providing two constraints for the four initial conditions required on $V_i, V_i', U_i, U_i'$.
\subsection{The VAAS model}
\label{subsec:VAAS}
This model was proposed in Ref.~\cite{Vardanyan:2017kal}, where the possibility of a nonlocal interaction term in a bimetric theory of gravity was investigated for the first time.
The action is the following:
\begin{eqnarray}
	S = \frac{M_{Pl}^2}{2}\int d^4x\sqrt{-g}R + \frac{M_f^2}{2}\int d^4x\sqrt{-f}R_f\nonumber\\ -\frac{M_{ Pl}^2}{2}\int d^4x\sqrt{-g}\alpha\left(R_f\frac{1}{\Box}R + R\frac{1}{\Box}R_f\right) + S_{m}[g,\Psi]\;, 
\end{eqnarray}
where $\Psi$ is a shortcut notation for all the matter fields, including CDM, and $f_{\mu\nu}$ is the auxiliary metric which does not couple to matter. It turns out, from computing the Bianchi constraints, that $R_f$ must be constant and thus the
action become:
\begin{equation}
    	S = \frac{M_{Pl}^2}{2}\int d^4x\sqrt{-g}\left(1 + m^{2}\frac{1}{\Box}\right)R + S_{m}[g,\Psi]\;,
\end{equation}
and the arising field equations can be cast as follows:
\begin{eqnarray}\label{FieldequationsNL}   
&(1 - 2\alpha V)G_{\mu\nu} + m^2(1 - U/2)g_{\mu\nu} + 2\alpha\nabla_\mu\nabla_\nu V + \alpha \nabla^\rho V\nabla_\rho Ug_{\mu\nu}\nonumber\\ &- 2\alpha\nabla_{\mu}U\nabla_{\nu}V = \frac{1}{M_{Pl}^2}T_{\mu\nu}\;,
\end{eqnarray}
where the auxiliary metric $f$ enters only through $m^2 \equiv -2\alpha R_f$ and all the other geometrical quantities are computed from the  metric $g_{\mu\nu}$ . The two auxiliary fields $U$ and $V$ were introduced in order to localize the theory and satisfy the following KG equations:
\begin{equation}\label{auxiliaryeqsNL}
	\square U = R\;, \qquad \square V = -\frac{1}{2\alpha}m^2\;.
\end{equation}
Later on, in Ref.~\cite{Amendola:2017qge}, an equivalent formulation of the theory was obtained motivated by  nonperturbative lattice quantum gravity. 
The background cosmology was numerically studied in Ref.~\cite{Vardanyan:2017kal}, where the compatibility with the standard cosmological history of the $\Lambda$CDM is showed. The EFE and the Klein gordon equations in a flat FLRW background, written in terms of the e-fold number $N$, are:
\begin{eqnarray}
&\label{FriedeqVAAS} 3\tilde V + \frac{m^2U}{2H^2} + 3\tilde V' + \frac{U'\tilde V'}{2} = \frac{\rho}{M_{Pl}^2H^2}\;,\\
&\label{acceqVAAS} -\tilde V\left(3 + 2\xi\right) + \frac{m^2}{H^2}(1 - U/2) + \tilde V' + \frac{U'\tilde V'}{2} = \frac{1}{M_{ Pl}^2H^2}P\;,\\
&\label{KGVAASU}U'' + \left(3 + \xi\right)U' + 6\left(2 + \xi\right) = 0\;,\\ 
&\label{KGVAASV}\tilde V'' + \left(3 + \xi\right)\tilde V' = -\frac{m^2}{H^2}\; ,
\end{eqnarray}
where we have defined $	\tilde V \equiv 1 - 2\alpha V$.
Imposing initial conditions compatible with a radiation-dominated era, i.e. $h_i^2 \sim \Omega_{Ri}$ and $\xi_i \sim -2$, the EFE for a flat FLRW at some initial time $N=N_i$ read:
\begin{eqnarray}
        \tilde{V_i} + \frac{\gamma U_i}{6h^2} +\tilde V_i' +\frac{1}{6}U_i'\tilde V_i' = 1 \; , \label{ICCosmoVAAS1} \\
        \tilde V_i - \frac{\gamma}{h_i^2}\left(1 - \frac{U_i}{2}\right) + \tilde V_i' +\frac{1}{2}U_i'\tilde V_i' = 1 \; . \label{ICCosmoVAAS2}
\end{eqnarray}
The latter equations provide two constrains among the four initial conditions on the auxiliary fields $U_i,U_i',\tilde{V}_i,\tilde{V}_i'$.

\chapter{Personal Contribution: \\
Nonlocal gravity}
\label{chapter:PCNL}

\epigraph{\textit{The worthwhile problems are the ones you can really solve or help solve, the ones you can really contribute something to. No problem is too small or too trivial if we can really do something about it.}}{Richard Feynman}{}

In this chapter I will present the results of my research about nonlocal models of gravity. In particular, the first part of the chapter is devoted to the results of Ref.~\cite{Giani:2019vjf} on VAAS gravity. The second part of the chapter addresses instead the study of the general features of the Late-times asymptotic equation of state for a general class of nonlocal models, following the results of Ref.~\cite{Giani:2019xjf}.

\section{Dynamical system analysis and Newtonian limit of VAAS gravity}

In Ref.~\cite{Vardanyan:2017kal} the cosmological behavior of VAAS gravity was analyzed numerically and found to be compatible with the one of $\Lambda$CDM, but an analytical understanding of its dynamics was not addressed. In Ref.~\cite{Giani:2019vjf} we tried to fill this gap using a dynamical system analysis approach, revealing a number of interesting features.
In particular, we addressed the existence of critical points and their stability, and studied in a qualitative but analytical way the Late-times dynamics of the model.
We also studied the impact on small scales of the nonlocal modification, i.e. we have studied its Newtonian limit on solar system scales and within the quasi static approximation (QSA), showing explicitly the existence of static solutions and the changes in the perturbations equation for the density contrast.

\subsection{Critical points}
Defining $X\equiv\dot{U}$ and $Y\equiv\dot{V}$ it is possible to write down Eqs.~\eqref{FieldequationsNL} and \eqref{auxiliaryeqsNL} for a flat FLRW background in the form of a closed dynamical system:
\begin{subequations}\label{dsfinitedistance2}
\begin{eqnarray}
    \dot{H} &=&\frac{1}{1 - 2 \alpha V} \left[\frac{\rho - P}{2 M_{ Pl}^{2}} + \frac{m^{2}}{2}\left(1 - U \right) + 2\alpha H Y\right] - 3 H^{2}\;,\\
    \dot{\rho} &=& -3 H(\rho + P)\;,\\
    \dot{X} &=& - 3 H X - 6 H^{2} - \frac{6}{1 - 2 \alpha V} \left[\frac{\rho - P}{2 M_{ Pl}^{2}} + \frac{m^{2}}{2}\left(1 - U \right) + 2\alpha H Y\right]\;,\\
    \dot{Y} &=& \frac{m^{2}}{2 \alpha} - 3 H Y \;,\\
    \dot{U} &=& X\;,\\
    \dot{V} &=& Y\;.
\end{eqnarray}
\end{subequations}
We also have to keep into account the following constraint coming from the first Friedmann equation:
\begin{equation}\label{modFriedeq2}
	(1 - 2\alpha V)3H^2 + \frac{m^2U}{2} - 6H\alpha Y - \alpha XY = \frac{\rho}{M_{ Pl}^2}\;.
\end{equation}
We define as critical point of the dynamical system a point in the phase space for which the right hand side of equations \eqref{dsfinitedistance2} vanish. The details of the analysis of the above system are reported in Appendix \ref{AppendixA}, from which we can draw the following conclusions:
\begin{itemize}
	\item For $m^2 = 0$ and $U = const$, the only critical point at finite distance represents Minkowski space.
	\item For $m^2 = 0$, $U = -4Ht$ and $H$ constant, the only critical point at finite distance represents a de Sitter space.
	\item For $m^2 \neq 0$ there are no critical points at finite distance;
	\item At infinite distance we found an unstable hyperplane of critical points of Minkowski type.
\end{itemize}
The above results are particularly interesting because $m^2$ is the free parameter of the theory that set the strenght of the nonlocal modification. The fact that for $ m^2 \neq 0$ we do not have stable critical points reflects and confirms the classical instability typical of these models we were talking about at the end of subsection \ref{sec:dofstability}.
%%%%%%%%%%%%%%%%%%%%%%%%%%%%%%%%%%%%%%%%%%%%%%%%%%%%%%%%%%%%%%
\subsection{Qualitative dynamics}
\begin{figure}[h]
\begin{center}
     \includegraphics[width=12.5cm]{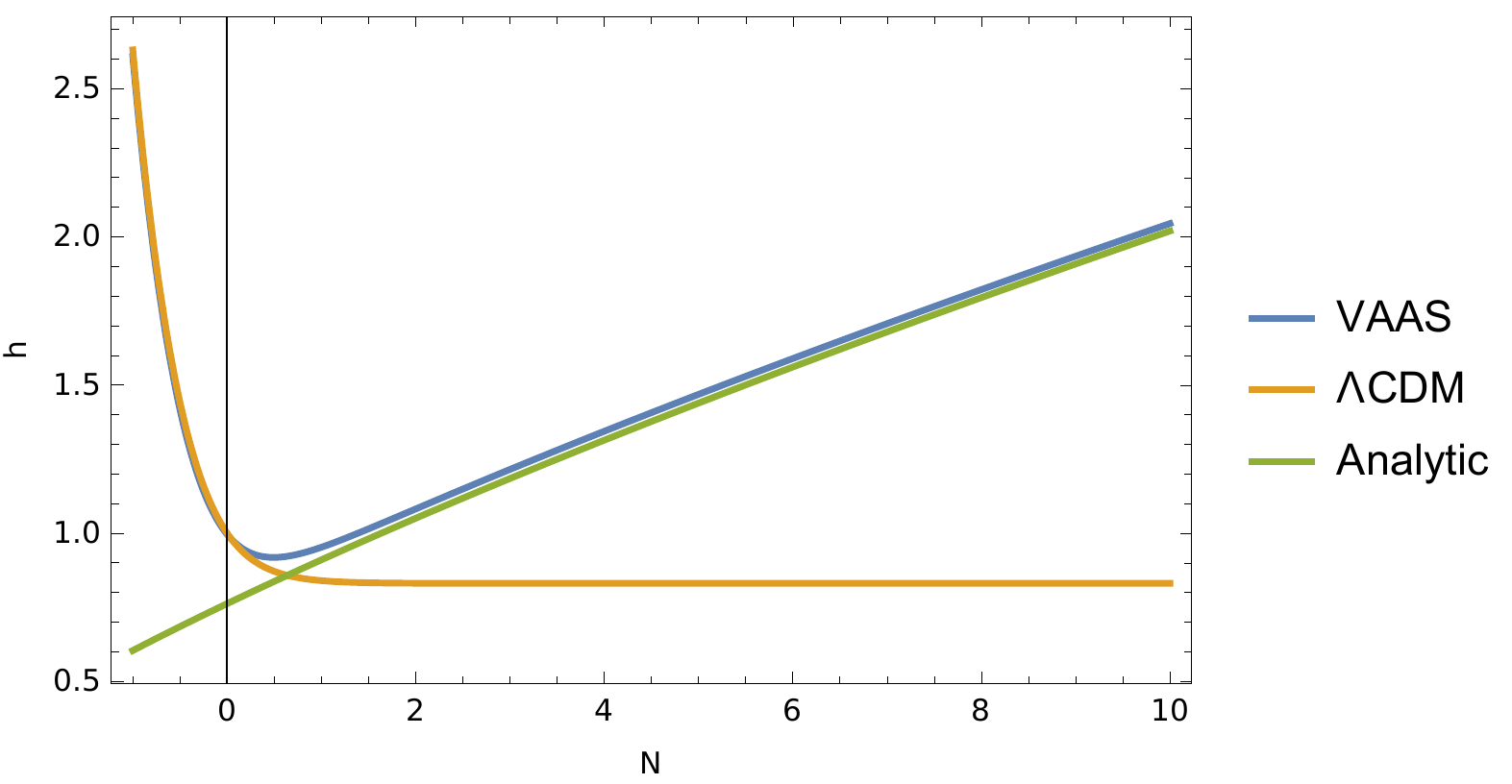}
     \caption{Behavior of the VAAS model and the analytical approximation of Eq.\eqref{HVAAS} compared to $\Lambda$CDM for $m^2 = 0.232 H_0^2$.}
     \label{VAAS dynamic}
 \end{center}
 \end{figure}

By looking at Eqs.~\eqref{FriedeqVAAS} and \eqref{acceqVAAS} we see that there is a complicated interplay between the auxiliary fields $U, V$ and the Hubble function $H$, which makes not trivial the qualitative understanding of the dynamical behavior of the system. 
To get some insight, let us begin with the Klein Gordon Eqs.~\eqref{auxiliaryeqsNL} written in terms of $X \equiv\dot{U}$ and $Y\equiv\dot{V}$:
\begin{eqnarray}
 X' + \left(3 + \xi\right)X + 6\left(2 + \xi\right) = 0\;,\\ 
 Y' + \left(3 + \xi\right)Y = -\frac{m^2}{H^2}\;.
\end{eqnarray} 
The latter have the formal solutions:
\begin{eqnarray}\label{Usol}
	X(N) = C_1e^{-F(N)} - 6e^{-F(N)}\int^N_{N_i} d\bar Ne^{F(\bar N)}[2 + \xi(\bar N)]\;,\\
	Y(N) = C_2e^{-F(N)} - e^{-F(N)}\int^N_{N_i} d\bar Ne^{F(\bar N)}\frac{m^2}{H^2(\bar N)}\;,
\end{eqnarray}
with $C_1$ and $C_2$ integration constants and:
\begin{equation}\label{fndef}
	F(N) \equiv \int^N_{N_i}d\bar{N}[3 + \xi(\bar{N})] \;.
\end{equation}
As we discussed in \ref{sec:dofstability}, it is important to fix the initial conditions in order to avoid the appearance of spurious propagating degrees of freedom. From the formal solutions of $X$ and $Y$ we see that:
\begin{equation}
	C_1 = X(N_i)\;, \qquad C_2 = Y(N_i)\;,
\end{equation} 
and choosing vanishing initial conditions for the fields, which are compatible with the constraints  of Eqs.~\eqref{ICCosmoVAAS1} and \eqref{ICCosmoVAAS2}, we obtain:
\begin{eqnarray}
	X(N) = -6e^{-F(N)}\int^N_{N_i} d\bar Ne^{F(\bar N)}[2 + \xi(\bar N)]\;,\\
	Y(N) = -e^{-F(N)}\int^N_{N_i} d\bar Ne^{F(\bar N)}\frac{m^2}{H^2(\bar N)}\;.
\end{eqnarray}
It is straightforward to realize from the above equations that if $\xi > -2$, which is compatible with the standard cosmological evolution, then both $X$ and $Y$ are always negative. It is also easy to prove the following inequality:\footnote{The details of the calculation are reported in Appendix \ref{appendixB}}
\begin{equation}\label{Xinequality}
    X + 6 > 0 \; .
\end{equation}
%by rewriting the formal solution for $X$ as:
%\begin{equation}
	%X(N) = -6e^{-F(N)}\int^N_{N_i} d\bar Ne^{F(\bar N)}[3 + %\xi(\bar N)] + 6e^{-F(N)}\int^N_{N_i}d\bar Ne^{F(\bar N)}\; ,
%\end{equation}
%and recasting the first integral as:
%\begin{equation}
%	X(N) = -6e^{-F(N)}\int^N_{N_i} d\bar N\frac{d(e^{F(\bar N)})}{d\bar N} + 6e^{-F(N)}\int^N_{N_i}d\bar Ne^{F(\bar N)}\;,
%\end{equation}
%so that we finally obtain:
%\begin{equation}
%	X(N) +  6 =  6e^{-F(N)} + 6e^{-F(N)}\int^N_{N_i}d\bar Ne^{F(\bar N)}\;,
%\end{equation}
%and the inequality \eqref{Xinequality} follows from the positivity of the right hand side of the latter equation.
It is interesting to study the behavior of $U$ and $V$ during the different phases of the cosmological evolution. The explicit calculations are reported in Appendix \ref{appendixB}, the results of which are:
\begin{itemize}
    \item During the radiation domination (RD), $\xi = -2$, we obtain:
    \begin{eqnarray}
        &U_{RD} = 0 \; , \\
       	&\tilde V_{RD} = -\frac{m^2}{20H_0^2\Omega_{r}^0}e^{4N} - \frac{m^2}{5H_0^2\Omega_{r}^0}e^{5N_i - N} + \frac{m^2}{4H_0^2\Omega_{r}^0}e^{4N_i} + 1\;.
    \end{eqnarray}
    \item During the matter dominated epoch (MD), $\xi = -3/2$, we have:
    \begin{eqnarray}
        &U_{MD} = C_1 - 2N \;, \\
        &\tilde V_{MD} \sim 1 - \frac{m^2}{12H_0^2\Omega_{ m}^0}e^{3N}\;.
    \end{eqnarray}
\end{itemize}
Let us now try to understand how these solutions evolve at Late-times when matter and radiation are diluted enough. Following the reasoning of Appendix \ref{appendixB}, we can conclude that at Late-times (LT) the field $V$ evolve as:
\begin{equation}
    \tilde{V}_{LT} \sim -\frac{m^2 U_{LT}}{6H^2} \; .
\end{equation}
From the latter equation we can estimate then the behavior of $\xi$:
\begin{equation}
	\xi \sim -3 - \frac{1}{\tilde V_{LT}}\frac{m^2U_{LT}}{2H^2} \rightarrow 0\;,
\end{equation}
so that at Late-times $\xi$ vanishes approaching the value it would have in the $\Lambda$CDM scenario, independently of the value of $m^2$. The latter occurrence is not a coincidence, and we will discuss it in detail in section \ref{sec:latetimes}. Using the above asymptotic expressions for $\xi$ we found:
\begin{equation}
    U_{LT} = C_1 - 4N \;,
\end{equation}
from which we finally obtain the evolution of the Hubble factor at Late-times:
\begin{eqnarray}
	&\xi = -\frac{3}{C_1 - 4N} = -\frac{3}{U_{LT}}\;, \\
	&H = C_2|C_1 - 4N|^{3/4} = 3|U_{LT}|^{3/4}\;.\label{HVAAS}
\end{eqnarray}
Fig. \ref{VAAS dynamic} shows the agreement between our analytical approximation for the evolution of the Hubble factor at Late-times and the numerical solution, as well as the agreement between the latter and the $\Lambda$CDM model in the past.

Summarizing, our qualitative analysis shows that if choosing vanishing initial condition deep into the radiation dominated epoch results in the following cosmological evolution:
\begin{itemize}
    \item During the RD epoch the field $U$ is constant and vanishing, while $V \sim e^{4N}$.
    \item During the MD epoch the field $U$ becomes linear in $N$ and starts to grow, while $V \sim e^{3N}$.
    \item At Late-times we found that $V\propto U/H^2$, which in turns implies that $\xi \rightarrow 0$. As a result, $U$ goes linearly as $U \sim -4N$ and $H = 3|U|^{3/4}$, so that $\xi = H'/H \propto U^{-1}$ approaches 0 as $ \sim N^{-1}$.  
\end{itemize}
%%%%%%%%%%%%%%%%%%%%%%%%%%%%%%%%%%%%%%%%%%%%%%%%%%%%%%%%%%%
\subsection{Newtonian Limit}
\subsubsection{First order perturbations}
In order to understand the effects of the VAAS nonlocal term at small scales we will consider first order perturbations of the FLRW metric in the Newtonian gauge, i.e.:
\begin{equation}
    ds^2 = -dt^2\left(1 + 2\Psi(\textbf{x,t})\right) + a(t)^{2}\left(1+ 2\Phi(\textbf{x,t})\right)\delta_{ij}dx^idx^j \; ,
\end{equation}
where $\Psi$ and $\Phi$ are the gravitational potentials. We also need to consider perturbations of the auxiliary fields:
\begin{equation}
    U(\textbf{x},t) = U_0(t) + \delta U(\textbf{x},t) \; , \qquad  V(\textbf{x},t) = V_0(t) + \delta V(\textbf{x},t) \; ,
\end{equation} 
where $U_0$ and $V_0$ are the background solutions of the Klein Gordon equations \eqref{KGVAASU}\eqref{KGVAASV}.
The first order perturbed Friedmann equation in Fourier space is then:
\begin{equation} \label{1order00eq}
\begin{split}
    &\left(1 -2\alpha V \right)\left( 6H\phi_{,0} - 6H^2\psi +2\frac{k^2\phi}{a^2}\right) -6\alpha\delta V H^2 -2m^2\left(1 - \frac{U}{2}\right)\psi +m^2\frac{\delta U}{2} \\
    &+2\alpha\left(\delta V_{,00} -\psi_{,0}V_{,0} -i \frac{k^{i}\psi}{a^2} V_{,i}\right)  -\alpha\left(V_{,}^{0}\delta U_{,0} +  V_{,}^{k}\delta U_{,k} + \delta V_{,}^{0}U_{,0} + \delta V_{,}^{k}U_{,k}\right)\\
    &-2\alpha \psi \left(V_{,}^{0} U_{,0} + V_{,}^{k} U_{,k}\right) -2\alpha\left(\delta U_{,0}V_{,0} + U_{,0}\delta V_{,0} \right) = \frac{\delta \rho}{M_{Pl}^2} \; ,
    \end{split}
    \end{equation}
    while the acceleration equation is:
    \begin{equation}\label{1orderijeq}
\begin{split}
  &  \left(1-2\alpha V \right)\frac{2}{3a^2}k^2\left(\phi + \psi \right) + \left(\frac{k^i k^j}{k^2} -\frac{1}{3}\delta^{ij} \right)\left[-2\alpha\delta VG_{ij} + m^2\left(1 - \frac{U}{2}\right)2a\delta_{ij}\phi -\frac{m^2 \delta U}{2}\delta_{ij}a\right.  \\
  & +2\alpha\left\{\delta V_{,ij} -\delta_{ij}a_{,0}a\delta V_{,0} -\delta_{ij}a^2\left[H +2H\left(\phi - \psi \right) + \phi_{,0}   \right]V_{,0} -i\phi\left(\delta_{ki}k_j +\delta_{kj}k_i -\delta_{ij}k_k\right)V_{,k} \right\} \\
  &+ \alpha a\delta_{ij}\left(V_{,}^{0}\delta U_{,0} +  V_{,}^{k}\delta U_{,k} + \delta V_{,}^{0}U_{,0} + \delta V_{,}^{k}U_{,k}\right)
   +2\phi a\delta_{ij} \alpha \left(V_{,}^{0} U_{,0} + V_{,}^{k} U_{,k}\right) 
 -2\alpha\left(\delta U_{,i}V_{,j} + U_{,i}\delta V_{,j} \right)\bigg]\\
 &= \left(\frac{k^i k^j}{k^2} -\frac{1}{3}\delta^{ij} \right)\frac{\delta T_{ij}}{M^2_{Pl}} \, .
    \end{split}
\end{equation}
Finally, the Klein Gordon equations for the auxiliary fields are:
\begin{align}\label{1orderKGeqU}
\begin{split}
&\delta U_{,\mu}^{\mu} + H \delta U_,^{0} + \psi_{,0}U_{,}^{0} + ik_{i}\psi U_{,}^{i} +\phi_{,0}U_{,}^{0} + i\phi k_{j} U_{,}^{j}\\
&= -12\psi\left( H^2 + \frac{a_{,00}}{a}\right) -\frac{2}{a^2}\nabla^2\psi + 6\phi_{,00} -6H\left(\psi_{,0} -4\phi_{,0}\right)
-\frac{4}{a^2}\nabla^2\phi \, ;
\end{split}\\\label{1orderKGeqV}
&\delta V_{,\mu}^{\mu} + H \delta V_,^{0} + \psi_{,0}V_{,}^{0} + ik_{i}\psi V_{,}^{i} +\phi_{,0}V_{,}^{0} + i\phi k_{j} V_{,}^{j}= 0 .
\end{align}
%%%%%%%%%%%%%%%%%%%%%%%%%%%%%%%%%%%%%%%%%%%%%%%%%%%%
\subsubsection{Solar system scales}
For analyzing physics at solar system scales we will adopt an approach similar to the one of Ref.~\cite{Koivisto:2008dh} and make the following approximations:
\begin{itemize}
    \item We will ignore the cosmological expansion, so we set the scale factor $a\approx1$ and the Hubble factor $H\approx0$.
    \item We look for static spherically symmetric solutions of the gravitational potential. 
    \item We set matter perturbations to 0.
\end{itemize}
Note that since in VAAS gravity there is no Minkowski solution at background level for $m^2\neq 0$, we have to consider  $m^2$ as a perturbative quantity in the following calculations.
The perturbations equations under these approximations become:
\begin{eqnarray}
    &\nabla^2 \Phi = 0 \; , \\
    &\nabla^2 \left(\Phi + \Psi \right) + \alpha\nabla^2\delta V = 0\; , \\
    &\left(\partial^i\partial_j - \frac{\delta^i_j}{3} \nabla^2\right)\left(\Phi + \Psi -2\alpha\delta V\right)= 0 \; , \\
    &2\alpha\nabla^2\delta V = -m^2\;, \\
    &\nabla^2\delta U = -2\nabla^2\left(2\Phi + \Psi\right) \;.
\end{eqnarray}
Solving the above equations for the gravitational potentials we can draw the following conclusions:
\begin{itemize}
    \item The $\Phi$ potential is a solution of the standard Poisson equation and thus has the same form as in GR: $\Phi = GM/r$.
    \item The $\Psi$ potential is instead sourced by the $m^2$ term and its solution is $\Psi = -GM/r + m^2r^2/12$. The term $\propto m^2r^2$ is particularly interesting because it closely resembles the one that would appear in GR for the Schwarzschild-deSitter solution, i.e. the nonlocal term behave similarly to a cosmological constant.
    \item We can write down the post Newtonian parameter $\gamma = -\Phi/\Psi$ = $m^2r^3/12GM$, which can be used to constraint the value of $m^2$. It turns out that the current constraint on $\gamma \leq 10^{-5}$ are satisfied for the value $m^2 \sim H_0^2$ required for a well-behaved cosmological evolution.
\end{itemize}
%%%%%%%%%%%%%%%%%%%%%%%%%%%%%%%%%%%%%%%%%%%%%%%%%%%%%%%%%%%%%%
\subsubsection{Structure formation in the Quasi Static Approximation (QSA)}

In order to address the impact of the nonlocal modifications on the process of structure formation at small scales we will work within the QSA. So we consider scales for which $k^2 \gg H^2$, in such a way that for a given perturbation we can neglect in the first order perturbation equations those terms containing time derivatives, or those proportional to $H$, with respect to terms proportional to $k^2$.
Combining the Euler and the continuity equation for dust we obtain the equation for the density contrast $\delta_M$:
\begin{equation}\label{dcqsaeq.}
    \ddot{\delta}_M + 2H\dot{\delta}_M +\frac{k^2}{a^2}\Psi = 0 \; .
\end{equation}
The equations for $\delta U$ and $\delta V$ are instead:
\begin{equation}\label{pertUVsf}
    \delta U = -2\left(\Psi + 2\Phi\right) \; , \qquad \frac{k^2}{a^2}\delta V = \frac{m^2}{\alpha}\Psi \; .
\end{equation}
    Finally, the EFE are:
    \begin{eqnarray}
        &2\left(1-2\alpha V_0\right)\frac{k^2\Phi}{a^2} + m^2\frac{\delta U}{2} -2m^2\Psi = \frac{\delta \rho_M}{M_{Pl}^2} \; , \\
        &-2\left(1-2\alpha V_0\right)\frac{k^2}{a^2}\left(\Psi + \Phi\right)  - 3m^2\frac{\delta U}{2} -2 \alpha \frac{k^2}{a^2} \delta V= 0 \; .
        \end{eqnarray}
Combining the latter equations and using Eq.~\eqref{pertUVsf} to eliminate $\delta U$ and $\delta V$ we obtain:
\begin{equation}
    -\left[2\left(1-2\alpha V_0\right)\frac{k^2}{a^2} - 6m^2\right]\Phi =\left[2\left(1-2\alpha V_0\right)\frac{k^2}{a^2} - m^2\right]\Psi   \; ,
\end{equation}
from which we can read off the slip parameter $\eta$ as:
\begin{equation}
    \eta \equiv -\frac{\Phi}{\Psi} = \frac{2\left(1-2\alpha V_0\right)k^2 - m^2 a^2 }{2\left(1-2\alpha V_0\right)k^2 - 6m^2 a^2 } \;,
\end{equation}
note that, as expected, the GR result $\Phi = - \Psi $ is recovered for $m^2 \rightarrow 0$.
Using the above expression for $\eta$ in the modified Poisson equation we can define the effective gravitational coupling $Y$ as follows:
\begin{equation}
    1 + Y \equiv -\frac{2k^2\Psi}{3H^2a^2\Omega_M\delta_M} = \frac{\left(1-2\alpha V_0\right)k^4 -3m^2k^2a^2}{\left(1-2\alpha V_0\right)^2k^4 - m^4a^4} \; .
\end{equation}
 Using the latter it is possible to rewrite Eq.~\eqref{dcqsaeq.} as:
 \begin{equation}
     \ddot{\delta}_M +2H \dot{\delta}_M -\frac{3}{2}\left(1+Y\right)\Omega_M\delta_M = 0 \;.
 \end{equation}
 Thus, within the QSA, using the value of $m^2$ required for a compatible background history $m^2\approx H^2 $ we have that the slip parameter is essentially the same as in GR, $\eta \approx 1$, while the effective gravitational coupling is modified by the background  value of the field $V_0$ and gets larger as soon as nonlocality starts to drive the accelerated expansion. For example, using the reference value $m^2/H_0^2 \approx 0.2$ we have that $1-2\alpha V_0 \approx 0.98$, and hence $Y \approx 0.02$, so that the effective gravitational coupling strength is enhanced by 2\% .
 \subsubsection{Lunar Laser Ranging constraints}
 Lunar Laser Ranging measurements provide a constraint on the time variation of the Newton constant, see for example Ref.~\cite{Hofmann:2018myc}. Currently, this constraint is of order:
 \begin{equation}
     \frac{\dot{G}}{G} = 7.71 \pm 7.76 \times 10^{-14} \;, \text{yr}^{-1} = 0.99 \pm 1.06 \times 10^{-3} \frac{0.7}{h_0} H_0 \; ,
 \end{equation}
 where $h_0$ is the Hubble constant expressed in units of 100 km s$^{-1}$ Mpc$^{-1}$.
 As argued in Ref.~\cite{Belgacem:2018wtb}, nonlocal modifications of gravity result in field equations which can be written as:
 \begin{equation}
     G_{\mu\nu} + \Delta G_{\mu\nu} = \frac{1}{M_{Pl}^{2}}T_{\mu\nu} \; ,
 \end{equation}
 where $\Delta G_{\mu\nu}$ accounts for deviations from standard GR. The terms which in the latter are proportional to $G_{\mu\nu}$ will generally result in a modification of the gravitational coupling, which for Eq.~\eqref{FriedeqVAAS} reduce to:
 \begin{equation}
     G_{eff}=\frac{G_{N}}{-2\alpha V} =_{QSA} \frac{G_{N}}{-2\alpha V_0}  \; .
 \end{equation}
 where in the last equality we are considering $k^2 \gg H$, thus we applied the result of the previous section whitin the QSA, which is surely a well justified assumption for scales related to the Earth-Moon distance.
 Taking the time derivative of the above expression we can write:
 
 \begin{equation}\label{geffvaas}
 \frac{\dot{G}_{eff}}{G{eff}}= \frac{2\alpha\dot{V}_0}{1-2\alpha V_0} \; ,
 \end{equation}
 or, using the e-fold number parameter:
 \begin{equation}
     \frac{\dot{G}_{eff}}{G{eff}}= H \frac{\tilde{V}'_0}{\tilde{V}_0} \; ,
 \end{equation}
 and we conclude that Lunar Laser Ranging constraints are satisfied in VAAS only for $m^2/H_0^2 \leq 10^{-3}$. The latter upper bound is two order of magnitude smaller than the value required for reproducing a viable cosmological expansion, thus the LLR experiments rule out VAAS gravity. Indeed, LLR have already ruled out several nonlocal proposal, including the $RR$ and the DW model, with the exception of the $RT$ model which we mention in section~\ref{sec:NLF}. There are however a couple of caveats; if a static solution of the above field equations exist it would be clearly compatible with LLR test, thus one should study whether these static solutions are stable against time perturbations. Moreover, the calculation presented here is based on the assumptions that we can extrapolate the solution we found for linear cosmological perturbations all the way down to Earth-Moon scales, which is not guaranteed a priori and very difficult to prove due to the lack of a satisfactory geometrical description that joins cosmological and local scales. Indeed, the results of Ref.~\cite{Belgacem:2018wtb} rely on the use of the McVittie metric to connect the cosmological solution with the solar system one, but such an assumption could be too strong.

 \section{Late-times asymptotic equation of state in nonlocal gravity}
 \label{sec:latetimes}
 As we already mention, in the VAAS model the asymptotic equation of state of the Universe approaches the value $w_{eff} \rightarrow -1$ as in the $\Lambda$CDM model. This was shown numerically by the author of Ref.~\cite{Vardanyan:2017kal}, and explicitly shown analytically by us in Ref.~\cite{Giani:2019vjf}. 
 It is interesting to note that similar numerical investigations of the $RR$ model and of the one proposed in Ref.~\cite{Amendola:2017qge} showed a similar behavior. Motivated by the latter apparent coincidence, we studied in Ref.~\cite{Giani:2019xjf} the behavior of the Late-times asymptotic effective equation of state for different Lagrangians containing functions of $\Box^{-1} R$. Applying the same technique we developed in Ref.~\cite{Giani:2019vjf} for the VAAS model we show analytically that, under a certain choice of initial conditions, all the models in which a term $\Box^{-1} R$ appears explicitly in the Lagrangian result in an asymptotic equation of state $w_{eff} \rightarrow -1$. This happens because such a term will inevitably diverge in the future, reflecting the classical instability we discussed in Sec.~\ref{sec:dofstability}. 
 As a result, $H^2$ will diverge but  $\dot{H}/H^2 \rightarrow 0$. On the other hand, we found that if the function $f(\Box^{-1}R)$ is chosen in such a way that $|f(\infty)| \leq const$, like for the standard DW model, the asymptotic equation of state will not approach asymptotically the $\Lambda$CDM value.

 \subsection{General scheme}
 The general scheme presented here was developed in Ref.~\cite{Giani:2019vjf} to study the late-times behavior of the model~\cite{Vardanyan:2017kal}. A sketch of the general strategy is the following: we use the fact that the sign of the first derivative of the auxiliary fields is determined by the formal solutions of the KG equations. Then,  imposing initial conditions compatible with radiation and matter domination, we are able to understand qualitatively the evolution of the nonlocal fields when matter sources are totally diluted by imposing consistency with the first Friedmann equation. Finally, we insert the asymptotic solution obtained for the fields and their derivatives into the acceleration equation to compute the asymptotic value of $\xi$.  Note that the scheme presented here is only valid  if we make the crucial assumptions $\xi + 2 \geq 0$, which is reasonable since we fix the initial conditions during the radiation-dominated era, when $\xi = -2$, to which follow a matter-dominated epoch  $\xi = -3/2$.

\subsection{Qualitative behavior of the fields}
First, let us notice that we can extract information about the qualitative behavior of $U$ already from the structure of its Klein Gordon equation.
Indeed, it is straightforward to realize that with the above choice of initial conditions, see Appendix \ref{appendixB} for the explicit computation, the following inequality for $U'$ holds:
\begin{equation}
    0 \leq U' \leq 6 \; ,
\end{equation}
so that choosing non-negative initial conditions would always imply $U \geq 0$. 
Since we know that $U$ has a definite sign, and we have constrained its first derivative, we grossly know its qualitative evolution. Moreover, since the other auxiliary fields are generally defined in terms of $U$, we can conclude that it is possible to obtain a similar amount of information from their Klein Gordon equations.

Another crucial information on the behavior of the system comes from the first Friedmann equation, which could be written:
\begin{equation}
    \left(1+g(N)_{NL}\right)h^2 = \Omega_{R} + \Omega_{M} +\Omega_{NL} \;,
\end{equation}
where the functions $g(N)$ and $\Omega_{NL}$ are the modifications due to  nonlocal terms. 

The initial conditions set the value $g(N_i)$, and since we can estimate the signs of the first derivative of the auxiliary fields entering in $g_{NL}$, using their KG equations we are able to estimate the qualitative behavior of $g(N)$ through the cosmological history, and in particular its asymptotic value at Late-times when matter and radiation are diluted enough.

\subsection{The Late-times asymptotic equation of state for the RR model}
To begin with let us consider the $RR$ field equations \eqref{CosmoRR1},\eqref{CosmoRR2} and \eqref{KGRRV} when matter and radiation density are negligible and define $\tilde{V} \equiv 1 -3\gamma V$: 
\begin{eqnarray}
\tilde{V} &=& \frac{\gamma U^2}{4 h^2} + \tilde{V}'\left(\frac{U'}{6} -1 \right) \; , \label{CosmoRR1A}\\
\xi &=& \frac{1}{2\tilde{V}}\left[\frac{3\gamma U}{h^2} - U'\tilde{V}' + 4\tilde{V}'\right] \; , \label{CosmoRR2A}\\
\tilde{V}'' &+& \tilde{V}'\left(3 + \xi\right) = -\frac{3\gamma U}{h^2} \; \label{KGRRVA} .
\end{eqnarray}
The formal solution of Eq.~\eqref{KGRRVA} for $\tilde{V}'$ compatible with the initial condition $\tilde{V}'(N_i)=0$ is:
\begin{equation}\label{FormalVRR}
    \tilde V' = -3\gamma e^{-F(N)}\int_{N_i}^{N}d\Bar{N}e^{F(\Bar{N})}\frac{U}{h^2}  \; ,
\end{equation}
where $F(N)$ was defined in Eq.~\eqref{fndef}.
From Eq.~\eqref{FormalVRR} it is straightforward to realize that $\tilde{V}'$ is always negative since $U$ is always positive, while imposing vanishing initial conditions for the nonlocal fields at early times determines the initial value $\tilde{V} =1 $. On the other hand from the  right hand side of Eq.~\eqref{CosmoRR1A} we see that $\tilde{V}$ must be positive and so we can conclude that $ 0\leq \tilde{V} \leq 1 $.\footnote{Note that the parameter $\gamma$ can be considered as positive definite since changing its sign corresponds to switch the sign of the nonlocal interaction term in the Lagrangian. In this case it is more convenient to change the sign of the source term in the equation of the auxiliary field $U$,  in such a way that the product $\gamma U$ is positive definite. Here we are neglecting the radiation and matter contributions which are anyway also positive definite.} This last argument tells us then that $\tilde{V}'$ must also vanish at late-times, or it would push $\tilde{V}$ to negative values. On the basis of these considerations we can conclude that at late-times we have
\begin{equation} \label{RRasyVV'}
    \tilde{V} \sim \frac{\gamma U^2}{4 h^2} \; , \qquad    \tilde{V}' \sim 0 \; ,
\end{equation}
and using the above results in Eq.~\eqref{CosmoRR2A} we get:
\begin{equation} \label{AxiRR}
    \xi \sim \frac{3\gamma U}{4h^2} \frac{4h^2}{\gamma U^2} \sim \frac{1}{U} \rightarrow 0 \; ,
\end{equation}
where the last limit holds true since $U$ diverges.\footnote{Note that $U$ cannot reach a constant value since $U' = 0$ is possible only for $\xi = -2$, and we easily see from Eq.~\eqref{AxiRR} that at late-times $\xi > 0$. }
%%%%%%%%%%%%%%%%%%%%%%%%%%%%%%%%%%%%%%%%%%%%%%%%%%%%%%%%%%%%%%%%%%%%%%%%%%
\subsection{The Late-times asymptotic equation of state for the DW model}
In order to study qualitatively the dynamic of the DW model at late-times we have first of all to understand qualitatively the behavior of the free function $f(U)$ defined in Eq.\eqref{fDW}, since it enters directly in the Friedmann equations and also rules the dynamics of the localized field $V$. It is straightforward to realize from Eq.~\eqref{fDW} that $(-2)(0.245) < f < 0$ and $\Bar{f} < 0$ , and that \ $\Bar{f} \rightarrow 0$ when $U \rightarrow \infty$.

The formal solution of \eqref{KGDWV} for $V'$ is:
\begin{equation} \label{FormalVDW}
    V' =  -6e^{-F(N)}\int_{N_i}^{N}d\Bar{N}e^{F(\Bar{N})}\left(2+\xi\right)\Bar{f}  \; .
\end{equation}
Since $\Bar{f} < 0$ and $\xi > -2$ from Eq.~\eqref{FormalVDW} it is straightforward to realize that we have at all times $V'> 0$. Since we impose initial conditions in such a way that during the radiation-dominated epoch $V$ is vanishing, we also can conclude that $V > 0$. At late-times, when matter is completely diluted Friedmann equations \eqref{CosmoDW1} and \eqref{CosmoDW2} become:
\begin{eqnarray}
        V &=& -V'\left(1-\frac{U'}{6}\right)  +\Bar{f}U' + f + 1 \; , \label{CosmoDW1A} \\
        \left(2\xi + 3\right)\left(1+f-V\right) &=& V'' - f'' + \left(V' - f'\right)\left(2 + \xi  \right) + \frac{U'V'}{2}\; \label{CosmoDW2A} .
\end{eqnarray}
Note that the first two terms in the right hand side of Eq.  \eqref{CosmoDW1A} are strictly negative since, $V' > 0$ and $\Bar{f}<0$, while $f+1 > 0$. On the other hand $V' >0$ implies that $V$ is a monotonic function, and we are left with two cases; either $U$ diverges, in which case $\Bar{f} \rightarrow 0$ and $f \rightarrow (-2)(0.243)$, or $U \rightarrow const$, in which case $U' \rightarrow 0$, $f'\rightarrow 0$ and $f + 1 \rightarrow const$. In both cases consistency requires that  $V' \rightarrow 0$, or  $V$ will be a decreasing function and so $V' < 0$. Thus we can conclude that asymptotically:
\begin{equation}
    V \sim \Bar{f}U' +  f + 1 \; .
\end{equation}
Using the above in \eqref{CosmoDW2A} we obtain finally:
\begin{equation}
    \xi \sim \frac{U' - 12 -\frac{f''}{\Bar{f}}}{6-U'}  \; ,
\end{equation}
which is in general non-vanishing.
Note also that:
\begin{equation}
    \frac{f''}{\Bar{f}} = \frac{(\Bar{f}U')'}{\Bar{f}}= U'' + \frac{\Bar{\Bar{f}}}{\Bar{f}}U'^2 \; ;
\end{equation}
 and we can conclude that if $U \rightarrow \infty$ the term $\Bar{\Bar{f}}/\Bar{f} \rightarrow - \infty$, while $U''$ cannot diverge since $0<U'<6$, so in this case the asymptotic effective equation of state $w_{eff} \rightarrow \infty$. On the other hand, if $U \rightarrow const$, we have $U' = U'' \rightarrow 0$ and we are left with $\xi \rightarrow -2$, in such a way that the effective equation of state approaches one of radiation type.
 
We have then shown that in the DW model with the distortion function given by \eqref{fDW} at late-times $w_{eff} \neq -1$.
%%%%%%%%%%%%%%%%%%%%%%%%%%%%%%%%%%%%%%%%%%%%%%%%%%%%%%%%%%%%%%%%%

\subsection{Summary and discussion}
By studying the Late-times asymptotic equation of state for a number of nonlocal models we realized that the auxiliary localized field related to $\Box^{-1}R$ will diverge asymptotically if we impose vanishing or positive initial conditions. The divergence of the latter in turn will push $H\rightarrow \infty$ while the ratio $H'/H \rightarrow 0$.  

 To summarize, the structure of the source terms of the KG equations implies that only the auxiliary field related to the nonlocal term $1/\Box R$ is still dynamical asymptotically, and diverges, while the auxiliary fields related to the Lagrange multipliers used to localize the theories freeze  and approach a constant value.   
 We show how the mechanism works using as an example the $RR$ model and the DW model, since in the latter the divergence of $U$ is hidden in the distortion function $f(U)$, which is regular for $U\rightarrow \infty$. The full computation for the VAAS model and the model proposed in Ref.~\cite{Amendola:2017qge} are reported in Appendix \ref{appendixB}.

It is important to remark that our conclusions strongly depend on the choice of initial conditions. Indeed, our method relies on the observation that by using Friedmann equations we can constrain the sign of the auxiliary fields, while their KG equations provide constraint on the sign of the first derivatives for our choice of initial conditions. 
As an example, let us consider the $RR$ model. If we choose a negative initial condition in Eq.~\eqref{KGRRV} for the field $U$  then $V' >0 $ and Eq.~\eqref{RRasyVV'} does not hold anymore. This situation corresponds to the evolution Path B described in Ref.~\cite{Nersisyan:2016hjh}, for which at late-times $w_{ eff}\rightarrow 1/3$. However, our  analysis still holds for any choice of initial conditions with non-vanishing but positive values of $U$. As discussed in Ref.~\cite{Belgacem:2017cqo}, there are fundamental motivations that justify processes during the inflationary epoch that result in a huge non-vanishing positive values for the field $U$ in the RD epoch.
It is interesting to note that a behavior of the type $w_{Eff}\rightarrow -1$ is remarkably for a model that wants to be competitive with the $\Lambda$CDM. Indeed, in such  models, and in the $\Lambda$CDM, the so called \textit{Coincidence Problem} \cite{Velten:2014nra} is less severe (if not a problem at all, depending on the personal perspective), since at some point of its history, independently of the initial conditions, the Universe always passes through a phase in which the matter and DE densities are of the same order and then DE starts to dominate, which in the standard model in terms of cosmic time accounts for at least the last 3.5 billion years. On the other hand, in the nonlocal models considered here we have to deal with a different sort of coincidence. Indeed, the Hubble function reaches a minimum when the nonlocal fields cosmological density starts to dominate, and the occurrence of this is roughly today. This occurrence looks to us  coincidental at least as the one present in $\Lambda$CDM.

\chapter{Personal Contribution: \\
Ricci Inverse Gravity} 
\label{chapter:IR}

\epigraph{\textit{I don't think there is a final theory of anything. It's theories (turtles) all the way down.}}{Jim Peebles}

In Ref.~\cite{Amendola:2020qho} we proposed a novel class of modified gravity theories based on the inverse of the Ricci tensor $R_{\mu\nu}$, which we call the anticurvature tensor $A_{\mu\nu}$. Taking the trace of the latter we obtain the anticurvature scalar $A$, which can then be used to construct a new type of Lagrangian densities. 
It is interesting to note that with the anticurvature scalar is very simple to write down Lagrangian densities terms with the same dimension as $R$, like for example $A^{-1}$ or $R^2 A$, and thus without introducing new dimensional constants. 

\section{Field equations}
The anticurvature tensor $A_{\mu\nu}$ is defined as the inverse of the Ricci tensor:
\begin{equation}\label{Adefinition}
    A^{\mu\rho}R_{\rho\nu}= \delta^{\mu}_{\nu} \; ,
\end{equation}
from which, taking the trace, we obtain the anticurvature scalar $A=A^{\mu\nu}g_{\mu\nu}$. Note that it is possible to write the inverse of any matrix in terms of its adjugate, i.e. in terms of the matrix itself and the Levi Civita symbols. 
In particular, for the anticurvature tensor we have:
\begin{equation}\label{AR}
    A^{\mu\nu}= 4\frac{R_{\kappa\pi}R_{\lambda\rho}R_{\xi\sigma}\varepsilon^{\mu\kappa\lambda\xi}\varepsilon^{\nu\pi\rho\sigma}}{R_{\alpha\zeta}R_{\beta\eta}R_{\gamma\theta}R_{\delta\iota}\varepsilon^{\alpha\beta\gamma\delta}\varepsilon^{\zeta\eta\theta\iota}}\; .
\end{equation}
From the latter equation we can appreciate that a theory based on the anticurvature scalar is actually a strongly nonlinear, higher order theory of gravity. 

The field equations for a general Lagrangian $f(R,A)$ are:
    \begin{align}
f_{R}R^{\mu\nu}-f_{A}A^{\mu\nu}&-\frac{1}{2}fg^{\mu\nu}+g^{\rho\mu}\nabla_{\alpha}\nabla_{\rho}f_{A}A^{\alpha}_{\sigma}A^{\nu\sigma}-\frac{1}{2}\nabla^{2}(f_{A}A_{\sigma}^{\mu}A^{\nu\sigma}) &\nonumber\\&-\frac{1}{2}g^{\mu\nu}\nabla_{\alpha}\nabla_{\beta}(f_{A}A_{\sigma}^{\alpha}A^{\beta\sigma})-\nabla^{\mu}\nabla^{\nu}f_{R}+g^{\mu\nu}\nabla^{2}f_{R}  =T^{\mu\nu} \; ,\label{eq:master}
\end{align}
where $f_{A,R}$ indicate derivation with respect to the Ricci or anticurvature scalars, see Appendix \ref{Appendix C} for a detailed derivation of the above field equations.\footnote{A code that evaluates the equations of motion for any $f(R,A)$ in a given metric is made publicly available \href{https://github.com/itpamendola/inverse-ricci}{here}}

It is well known that one can recast an $f(R)$ theory in the form of a scalar-tensor theory in the Einstein frame introducing a scalar field non minimally coupled to gravity. 
Usually this is done by defining a scalar field $\phi = df/dR$ and performing a Legendre transformation of the function $f$. Such an approach, however, 
fails here, because $A$ is not a one-to-one function of $R$ and therefore
 $df(A)/dR$ is in general not invertible.
 
 \section{Cosmology}
 
 We are interested in understand which kind of behavior could arise from Eq.~\eqref{eq:master} for cosmological implications.
 Since we are mainly interested in Dark Energy phenomenology, let us begin with a de Sitter ansatz for the metric $g_{\mu\nu}$. Under this assumption $R_{\mu\nu}=  R g_{\mu\nu}/4$, and it is straightforward to compute $A_{\mu\nu}$ from the inverse metric. In this case  all the terms with  derivatives in Eq.~\eqref{eq:master} vanish, and taking the trace one has in vacuum:
\begin{align}
f_{R}R-f_{A}A-2f =0 \; .
\end{align}
Since on de Sitter background $R=12H^2$ and $A=4/(3H^2)$, the latter equation can be easily solved for any $f(R,A)$ model to check whether one gets non-trivial (i.e. $H\not = 0$) solutions that could replace a cosmological constant. For instance, if $f=R-\alpha A$ (where $\alpha$ is a constant with dimensions $H_0^4$) then we see that $H=(\alpha/3)^{1/4}=const$. 

Having seen that it is in principle possible to have de Sitter solutions in this model, let us investigate the behavior in a flat FLRW background. 
In this case we have that the Ricci and anticurvature scalars can be written:
\begin{eqnarray}
    R = 6\left(\dot{H} + 2H^2\right) \;, \\ A=\frac{2(6+5\xi)}{3H^2(1+\xi)(3+\xi)} \;\label{AFLRW} ,
\end{eqnarray}
where $\xi = \dot{H}/H^2$. It is straightforward to realize from Eq.~\eqref{AFLRW} that the anticurvature scalar become singular in the following cases: $H\rightarrow0$,  $\xi \rightarrow -1$, or $\xi \rightarrow -3$.
The fact that for $H\rightarrow 0$ the Lagrangian is ill-defined implies, as expected, the lack of a Minkowski solution for Lagrangians containing positive powers of the anticurvature scalar $A^n$. On the other hand, this does not apply for Lagrangian containing negative powers $A^{-n}$, which are instead singular for $\xi \rightarrow -6/5$.
Thus, just by looking at the shape of the anticurvature scalar in FLRW background we are able to formulate the following no-go theorem for cosmology:
\begin{theorem}[FLRW no-go] If the cosmic evolution passes through any one of these values of $\xi$: $\xi = -1$, $\xi = -3$ or $\xi = -6/5$, either $A$ or $A^{-1}$, or any of their powers, develops a singularity. If during the evolution $A$ passes through \textbf{both} 0 and $\pm \infty$, then any term in the Lagrangian that contains $A^n$, for $n$ positive or negative, will blow up. This behavior will reflect into equations of motion that also contain a singularity at the same cosmic epochs.
\end{theorem}

 Observations~\cite{Abbott:2018wog,Scolnic:2017caz,Ade:2015xua,Aghanim:2018eyx} tell us that the Universe evolved from a decelerated phase with $w_{eff}\approx 0$ (so $\xi\approx -1.5$) into an accelerated phase $w_{eff}\approx -0.7$ (so $\xi\approx -0.45$). Therefore, the cosmic expansion had to pass, at redshifts around unity, through both $\xi=-1$ and $\xi=-6/5$. This demonstrates that $A$ and $A^{-1}$ will both be singular at some epoch between deceleration and acceleration.
Consequently, any Lagrangian that contains additive terms proportional to $A^n$ (e.g. the two simplest scale-free models, $f(R,A)=R+\alpha A^{-1}$ and $f(R,A)=R+\alpha R^2 A$, with $\alpha$ a dimensionless constant) are ruled out as Dark Energy models. Notice also that $R=6H^2( \xi+2)$ so no power of $R$ can cure the singularity. 
In order to see some concrete realizations of the no-go theorem in the following we will briefly illustrate the cosmological behavior for the Lagrangians $f=R+\alpha A^{-1}$ and $f=R+\alpha R^2 A$.

\section{Lagrangian $R+\alpha A^{-1}$}
In this case equations \eqref{eq:master} for a FLRW background reduce to:
\begin{equation}\label{FrIA}
 \rho_t= 3 \alpha  H^2\frac{ (\xi +3)^2 (5 \xi +6)-18 \xi '}{4 (5 \xi +6)^3}+3 H^2 \; ,
\end{equation}
which is the modified Friedmann equations, and to:
\begin{align}
w_t\rho_t=&-\frac{\alpha  H^2 \left[(5 \xi +6) \left((\xi +3)^2 (2 \xi +3) (5 \xi +6)-18 \xi ''\right)+270 \left(\xi '\right)^2-54
   (\xi +2) (5 \xi +6) \xi '\right]}{4 (5 \xi +6)^4}\nonumber \\
   &-2 H^2 \xi -3 H^2 \; ,
\end{align}\label{AccIA}
which is the $(i,i)$ equation. Note that, as follows from the No-go theorem, the singularity at $\xi = -6/5$ appears in both the equations. We have used the subscript $t$ in $\rho_t$ to indicate  the total matter, which of course satisfy the continuity equation $\rho_t'=-3(1+w_t)\rho_t$.  From Eq.~\eqref{FrIA} we can easily define the energy density associated to the anticurvature scalar:
\begin{equation}
     \Omega_A\equiv  -\alpha\frac{    (\xi +3)^2 (5 \xi +6)-18 \xi '}{4 (5 \xi +6)^3} \; .
\end{equation}
We will now consider two different cases, considering pressureless matter only and then adding a cosmological constant

\subsection{Evolution with Dust}
Assuming $w_t = 0$ we find for Eqs.~\eqref{FrIA} and \eqref{AccIA} the following critical points:
\begin{align}
     &\Omega_m =1+\frac{\alpha}{4} \; , \qquad \xi = -\frac{3}{2} \; ,\\
    &\Omega_m=0\,,\quad   \xi_{\pm}=\frac{3(-40-\alpha\pm 6 \sqrt{-\alpha})}{100+\alpha} \; .
 \label{eq:walpha}
\end{align}

\begin{figure}
\includegraphics[width=12cm]{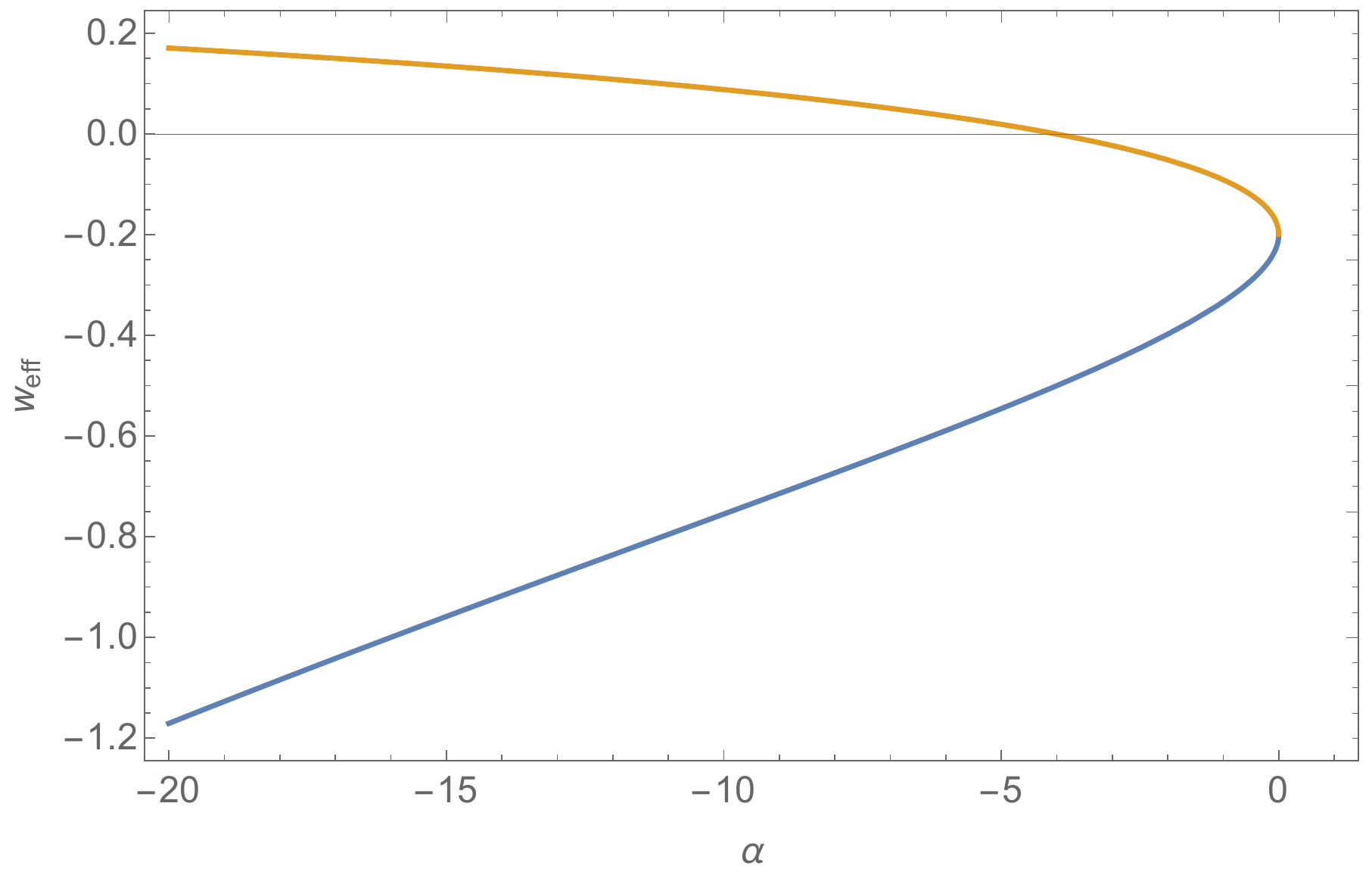}
\caption{The two real branches of the function $w_{ eff}(\alpha)=-2\xi/3-1$ from Eq. (\ref{eq:walpha})\label{fig:walpha}}
\end{figure}

\begin{figure}[tb]
\begin{center}
    \includegraphics[width=10cm]{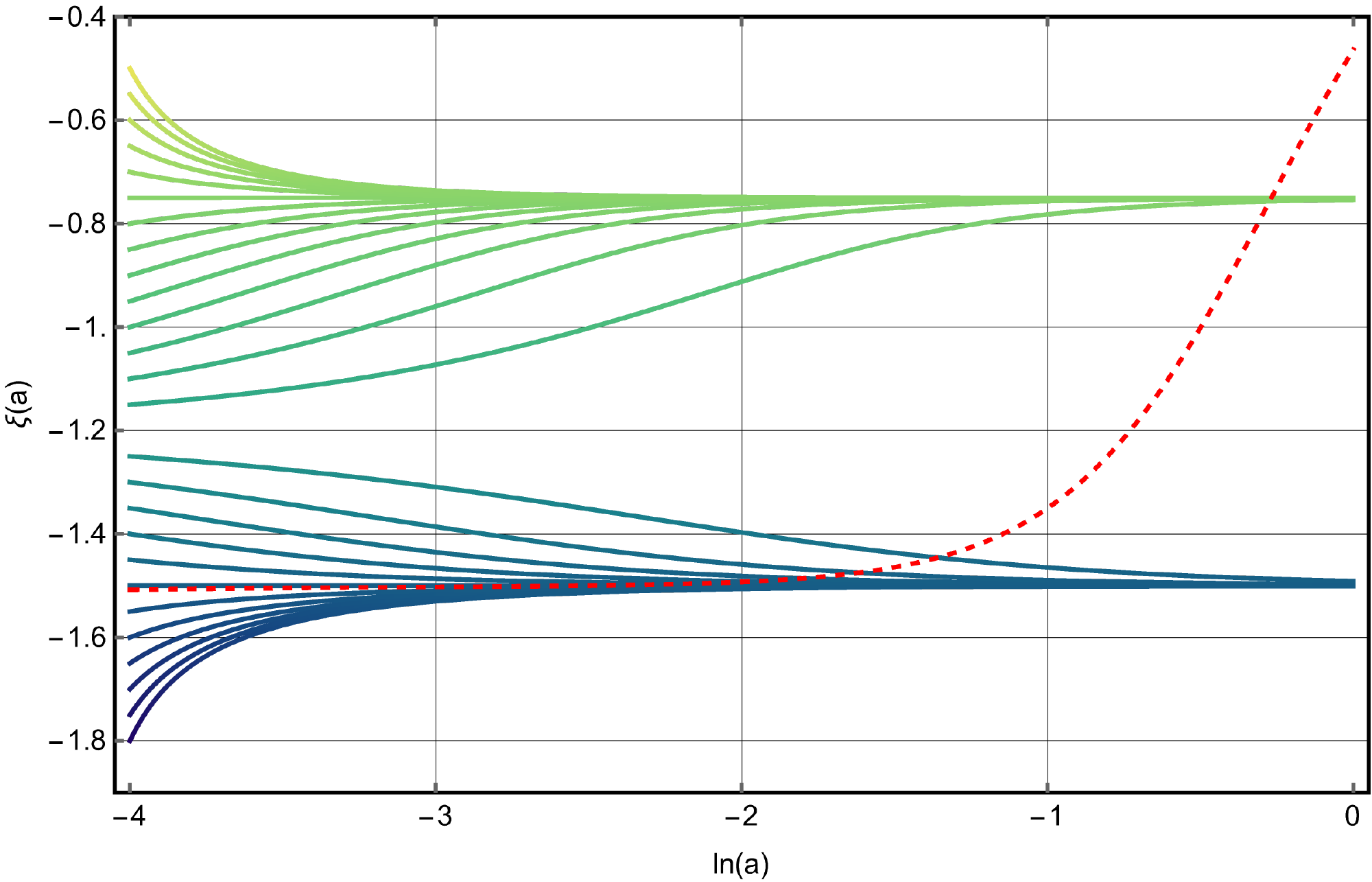}
    \caption{ Numerical solutions $\xi(a)$ of Eq.~\eqref{eq:walpha} in case of $\Omega_m \neq 0$, $\Omega_\Lambda=0$ with $w=0$ and $\alpha=-4$. The  solutions $\xi=-3/2$ and $\xi=-3/4$ are confirmed to be attractors. The divide at $\xi=-6/5$ is also evident. The red dashed line is the $\Lambda$CDM behaviour.}
    \label{fig:ximatter}
\end{center}
\end{figure}

The first critical point corresponds to a matter dominated Universe in which $\Omega_A$ behaves as matter. In particular, if $\alpha = -4$, it corresponds to an empty Universe wich behave as it was filled with dust. The critical points of Eq.~\eqref{eq:walpha} instead admit solutions only for certain values of $\alpha$, see Fig.~\ref{fig:walpha} for a graphical representation of $\xi_+$. We see that for every $\alpha<0$ there are two  real solutions, one above, the other below $\xi=-6/5$, or equivalently $w_{ eff}=-0.2$.
The properties of these solutions seem very interesting for cosmological implications. For instance, for $\alpha\approx -8$, the two solutions correspond to the observed present accelerated value $w_{ eff}\approx -0.67$ 
and to an expansion quite close to a matter dominated era,  $w_{ eff}\approx 0.06$. Analogously,
if $\alpha=-4$, one has $w_{eff}=0$, i.e. an exact matter era evolution without matter, in which the $A$ energy density acts as a form of Dark Matter. The other solution, $\xi_+$, corresponds to $w_{ eff}=-0.5$, i.e. an accelerated solution still marginally compatible with observations. In Fig.~\ref{fig:ximatter} we see the behavior of the Hubble parameter for the particular case $\alpha  -4$. A cosmic evolution that moves from one such solution to the other would be indeed an intriguing possibility, replacing both Dark Matter and Dark Energy with the anticurvature tensor without any new scale nor fine-tuned parameters. However, as a consequence of the no-go theorem, this cannot occur. 

Through a stability analysis of the linearized dynamical system we find that the critical point $\xi_{-}$ is a stable attractor only for $-4\le\alpha\le 0$.
The critical point $\xi_+$ is a stable attractor for $\alpha \leq 0$, while the linear analysis alone cannot assess the stability of the point $\xi = -3/2$. These findings are supported by the numerical investigation shown in Fig.~\ref{fig:ximatter}, so that the cosmic evolution will end up either at $\xi_+$ or $\xi_{-}$, depending on whether the initial $w_{eff}$ is above or below the singularity at  $w_{eff}=-0.2$. The crucial point is that no trajectory can cross  the $w_{eff}=-0.2$ ridge; consequently, as anticipated on general grounds, the cosmic expansion cannot move from a decelerated phase around $w_{ eff}=0$ to an accelerated one around $w_{ eff}\approx -0.7$.

\subsection{Evolution with Dust and a cosmological constant}
\begin{figure}[tb]
    \centering
    \includegraphics[width=10cm]{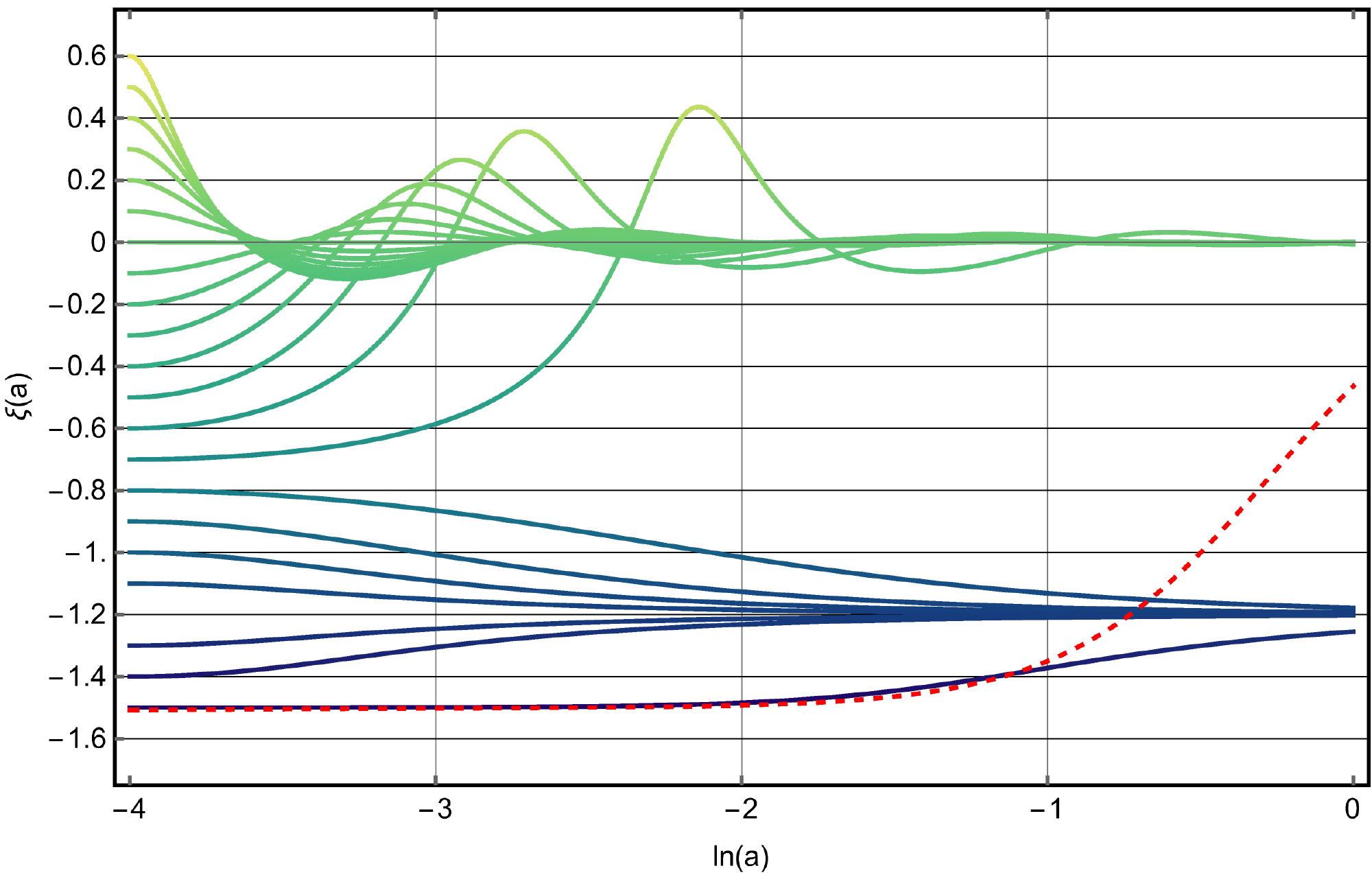}
    \caption{ Numerical solutions $\xi(a)$ of Eq.~\eqref{eq:mattercosmconst} with matter and cosmological constant, for $\alpha=-4$. The upper curves converge toward the de Sitter attractor at $\xi=0$. The lower curves converge towards  the divide line at $\xi=-6/5$, which is now also an attractor. The red dashed line is the $\Lambda$CDM behavior.}
    \label{fig:ximatterlambda}
\end{figure}

If a cosmological constant is present we can combine Eqs.~\eqref{FrIA} and \eqref{AccIA} and obtain:
\begin{equation}
\xi ''= \frac{6 (5 \xi +6)^4 \Omega _m+\xi  (5 \xi +6)^2 \left(9 (\alpha +16)+(\alpha +100) \xi ^2+6 (\alpha +40) \xi \right)+135
   \alpha  \left(\xi '\right)^2-27 \alpha  \left(5 \xi ^2+11 \xi +6\right) \xi '}{9 \alpha  (5 \xi +6)} \; ,
   \label{eq:mattercosmconst}
\end{equation}
which must be solved together with the continuity equation:
\begin{equation}
\Omega_{m}'=-(3+2\xi)\Omega_{m} \; .
\end{equation}
In this case the phase space is more complicated, $\xi_-$ is now always unstable and $\xi_+$ is a stable attractor for $\alpha <-16$. The critical point $\xi = -3/2$ is always unstable, while we found a new critical point $\xi = 0$, which is a stable de Sitter attractor when $-16 \leq\alpha \leq0$. However, the bottom line is the same, as can be immediately gleaned from Fig.~\ref{fig:ximatterlambda},  so the model is ruled out as a candidate for Dark Energy even when a cosmological constant is added, regardless of the value of $\alpha$.

\section{Circumventing the no-go theorem}
Motivated by the interesting phenomenology which could be described within the anticurvature scalar,  we will try now to address some possible escape routes from the no-go theorem. 
\subsection{Modifying the geometry}
To begin with, we should consider possible modifications of the background geometry that allow for a different evolution. If we relax the assumption of a spatially flat Universe the anticurvature scalar become:
\begin{equation}
    A=\frac{2(6+5\xi+3\Omega_k)}{3H^2(1+\xi)(3+\xi+6\Omega_k)} \; ,
\end{equation}
where we can see that the appearance of the energy density associated with the curvature shifts the singularity $\xi = -3$ for positive powers of $A$, while shifts the singularity $\xi = -6/5$ of $A^{-1}$. Thus, in principle the presence of curvature is able to shift the singularity of the Lagrangians outside the range required by the observations. However, since observations suggest that $\Omega_k \sim 0$, this possibility is very unlikely.

Another option is to take into account the contribution from  spatial anisotropies. For illustrative purposes, let us consider as an example a Bianchi I geometry: 
\begin{equation}\label{BImetric}
    ds^2 = -dt^2 +a(t)^2\left(e^{2\beta_x(t)}dx^2 +e^{2\beta_y(t)}dy^2 + e^{2\beta_z(t)}dz^2 \right) \; ,
\end{equation}
where we have defined the averaged scale factor
\begin{equation}
    a(t) = \sqrt[3]{a_x(t) a_y(t) a_z(t)}\; ,
\end{equation}
so that $a_i(t) = a(t)e^{\beta_i}$, and the $\beta_i$ satisfies $\sum_i \beta_i = 0$.
For the sake of simplicity, let us specialize to the case $\beta_x = -\beta_z \equiv \beta$ and $\beta_y = 0$. In this case the anticurvature scalar $A$ reads:
\begin{equation}
    A = \frac{1}{H^2}\left[\frac{4\xi + 6 + \frac{(\beta')^2}{2}}{3\left(3+\xi\right)\left(1+\xi + \frac{(\beta')^2}{6}\right)} + \frac{2(3+\xi)}{\left(3+\xi\right)^2 - \frac{1}{4}\left(\beta'' + \beta'(3+\xi)\right)^2}  \right] \; ,
\end{equation}
which for $\beta'=0$ reduces to the FLRW case.
As we can see, the singularity $\xi = -1$ is shifted by the anisotropic term $(\beta')^2/6$. Note  that also the singularity appearing in $A^{-1}$, $\xi = -6/5$, is in general shifted. For example, if $\beta''$ is negligible, we have that $A^{-1}$ is singular for
\begin{equation}
    \xi = \frac{24-\frac{\beta'^4}{4}}{ 2\beta'^2-20}\approx -\frac{6}{5}(1-\frac{(\beta')^2}{10})\;,\label{betasingular}
\end{equation}
(the last approximate equality being valid for  $\beta'\ll 1$) which recovers the FLRW case for $\beta'=0$, while being regular in $\xi= -6/5$ unless $\beta'^2 = 48/5$, i.e. the two roots of Eq.~\eqref{betasingular} for $\xi = -6/5$. 
This shows that relaxing the assumption of spatial isotropy the singularities occurring in the anticurvature scalar and its inverse can be arbitrarily shifted, but not removed. It is clear however that one needs $\beta'$ of order unity to move the singularity outside the observational range, which on the other hand is not likely to be compatible with experimental data.

To illustrate that, let us naively estimate $\beta$ from the evidence of anisotropic expansion claimed recently in \cite{Migkas:2020fza}, emerged from X-ray observations of galaxy clusters. Here the authors find that the highest and the lowest values observed for the universe expansion rate are $H_{ max} \sim 75 \;$  km/s/Mpc and $H_{ min} \sim 66 \;$  km/s/Mpc. Assuming that the averaged Hubble factor is $H \sim 70 \;$ km/s/Mpc, we obtain
\begin{equation}
    \beta' \sim 0.06 \; ,
\end{equation}
which shows that generally $\beta'$ is constrained from the observations to be too small to shift the singularities of $A$ outside the observational range.

\subsection{Non Polynomial Lagrangians}

Another possibility is to consider Lagrangian densities which are not singular when $A$ or $A^{-1}$ diverge. 
Still considering  scale-free Lagrangians for simplicity, we can choose for example the scalar densities $R+\alpha R\exp[-\beta (RA)^2]$ or $R/(1+\alpha RA)$. In the former case, for example, the Friedmann equation around the critical points, i.e. assuming $\xi'=\xi''=0$, becomes:
 \begin{equation}
     3 H^2 \left(1-\frac{\alpha \mathcal{P}_5(\xi,\beta)}{(\xi +1)^3 (\xi +3)^2}e^{-\frac{16 \beta(\xi +2)^2 (5 \xi +6,)^2}{(\xi +1)^2 (\xi +3)^2}}\right) = \rho_m \; ,
 \end{equation}
 where $\mathcal{P}_5(\xi,\beta)$ is a polynomial of order five in $\xi$ and linear in $\beta$. It is straightforward to realise that the above equation is regular on the poles of the denominator due to the presence of the exponential factor.
 
Another option is to include, as in Gauss Bonnet gravity for the Ricci tensor, scalar combinations of higher order in the anticurvature tensor, like $A^{\mu\nu}A_{\mu\nu}$. In FLRW background the latter looks as follows:
\begin{equation}
    A^{\mu\nu}A_{\mu\nu} = \frac{4}{9H^4}\frac{7\xi^2 + 15\xi + 9}{(\xi+1)^2(\xi + 3)^2} \; .
\end{equation}
We see that it still contains the singularities at $\xi = -3$ and $\xi=-1$, but remarkably it never vanishes, and thus $(A^{\mu\nu}A_{\mu\nu})^{-1}$ can be used to build Lagrangians which are free of this kind of singularities.

\section{Summary and Outlooks}

We have shown that it is difficult to describe Dark Energy using polynomial Lagrangians of the anticurvature scalar because of the no-go theorem. On the other hand, we found that an interesting phenomenology arise already for the simplest choices of $f(R,A)$, and thus Lagrangians that escape the no-go theorem are particularly promising for cosmological model building.

It is important to realize that in this framework we are introducing higher order derivative terms in the Lagrangian, see Eq.~\eqref{AR}, and  then we expect that instabilities will generally occur unless we consider degenerate Lagrangians.  The above stability issues and the no-go theorem should be taken into account when a particular form of $f(R,A)$ is specified, which is the task we address for future works.

\chapter{Personal Contribution: \\
Strong Lensing for testing Gravity and Cosmology}
\label{chapter:Lensing}
\epigraph{\textit{Simplicity is the touchstone in finding new physical laws… If it's elegant, then it's a rough rule of thumb: you're on the right track}}{Kip Thorne}

Lensing effects provide fertile ground for testing gravitational theories from more than a century. In particular, it is well known that weak gravitational lensing by Large Scale Structures can provide useful insights on the nature of Dark Energy. In this chapter we will discuss instead the potential of strong gravitational lensing in achieving a similar task. After a brief review of the main equations governing this phenomenon, we will introduce two novel drift effects proposed by us in Ref.~\cite{Piattella:2017uat}, and discuss the possibility of using them for testing violations of the Equivalence Principle and Dark Energy models~\cite{Giani:2020fpz}.

\section{Overview of Gravitational Lensing}
It is a well known results of geometrical optics that light rays passing through a medium will generally be refracted, as encoded in the Snell's law. Considering the gravitational field in the empty space around its source as a sort of "medium", it is a reasonable expectation that light rays traveling through it will be deflected. It is slightly uncomfortable to give a meaningful explanation of this effect within Newtonian gravity because photons have no rest mass, and thus should be blind to the gravitational interaction. On the other hand, it is a standard approach to study the motion of bodies in a gravitational potential by mean of test particles, i.e. particles whose mass is small enough to ignore their backreaction on the gravitational potential. Thus, considering photons as test particles we expect already in Newtonian gravity the deflection of light rays close to a massive body. This result was indeed obtained by Soldner\footnote{A translation from german of the original article is available \href{https://en.wikisource.org/wiki/Translation:On_the_Deflection_of_a_Light_Ray_from_its_Rectilinear_Motion}{here}} more than one century before Einstein's theory of general relativity.
In GR, instead, the interpretation of the gravitational potential in the empty space as a sort of medium is straightforward, and is logical to conclude that trajectories of massless particles will in general be bent because of the curvature's gradient of the spacetime. 

An effective treatment of the above phenomenon resembles the standard approach of geometrical optics. The source of the gravitational field that deflects light rays is then called \textit{lens}, and the overall effect \textit{gravitational lensing}.
We distinguish between the deflection caused by an extended, approximately continuous distribution of sources and the one caused by a single, massive object. The former is which is generally dubbed \textit{weak lensing}, and causes distortions on the shape of the background objects, see Ref.~\cite{Bartelmann:2016dvf} for a review of weak lensing and its cosmological applications. When the lens is instead composed by a single massive object along the line of sight between the observer and the source we are instead in a regime of \textit{Strong Lensing}, which will be our main subject during the rest of this chapter.

\subsection{Strong  Lensing by a point mass}
\begin{figure}[tb]
    \centering
    \includegraphics[width=10cm]{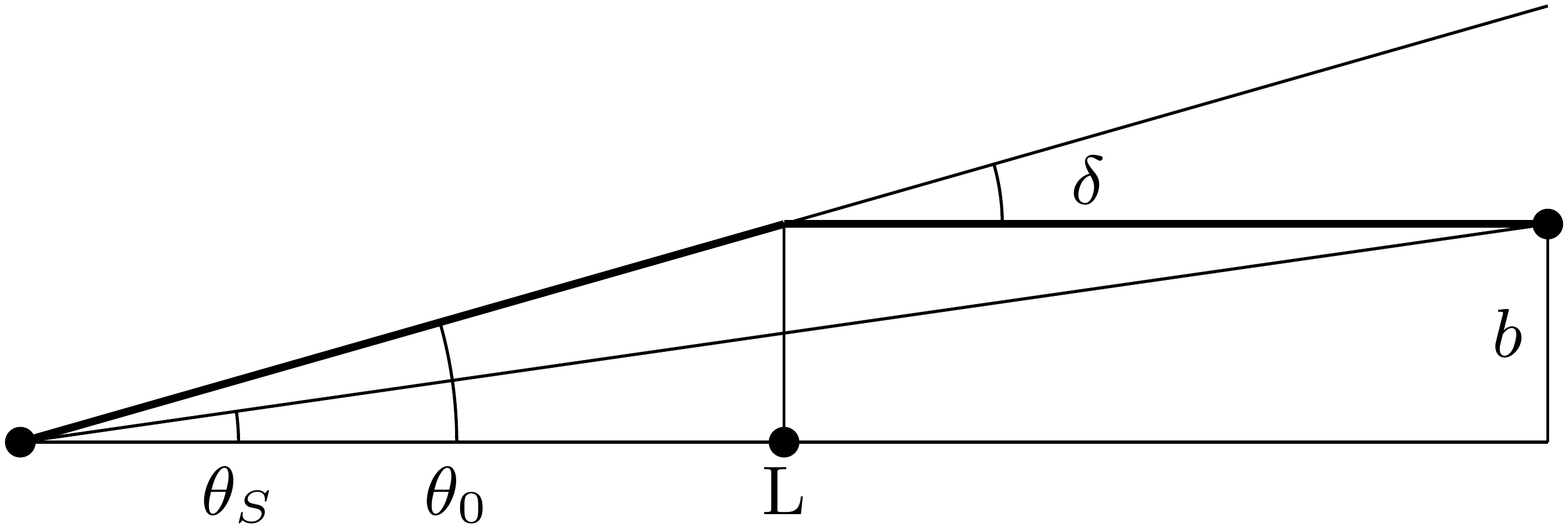}
    \caption{Scheme for strong gravitational lensing induced by a point mass lens \textbf{L}. $\theta_0$ is the apparent angle at which the source is located, $\theta_S$ the true one. The deflection angle 
is $\delta = 4GM/b$, where $b$ is the impact parameter, i.e. the distance on the lens plane between the incoming light ray and the lens itself. Image taken from Ref.~\cite{Piattella:2017uat}}
    \label{lensingscheme}
\end{figure}

To begin with, let us consider gravitational lensing by point masses. A fairly standard configuration is given in Fig.~\ref{lensingscheme}, where $\theta_0$ is the apparent angular position of the source as seen by the observer, and $\theta_S$ indicate the actual position of the source that would be observed in case of no lensing. The above quantities are related by the lens equation in the thin lens approximation~\cite{Weinberg:2008zzc}:

\begin{equation}\label{LEPM}
   \theta_E^2\equiv \theta_0\left(\theta_0 - \theta_S\right) = \frac{4G_N M}{\mathcal{D}_L }\frac{\mathcal{D}_{LS}}{\mathcal{D}_S} =4G_N M\left(1+z_L\right)\left(\frac{1}{\chi_L} - \frac{1}{\chi_S}\right) \; ,
\end{equation}
where the  $\mathcal{D}$'s are the angular diameter distances:
\begin{equation}\label{dS}
	\mathcal{D}_i \equiv a_i\chi_i = \frac{1}{1 + z_i}\int_0^{z_i}\frac{dz'}{H(z')}\;,
\end{equation}
in which the subscript $i = L,S$ refers to the lens or to the source, while $\mathcal{D}_{LS}$ is given by:
\begin{equation}\label{dLS}
	\mathcal{D}_{LS} \equiv a_S(\chi_S - \chi_L) = \frac{1}{1 + z_S}\int_{z_L}^{z_S}\frac{dz'}{H(z')}\;.
\end{equation}
In the above equations $z_{L,S}$ indicates the redshifts and $\chi_{L,S}$ the comoving distances of the lens and  the source respectively.
Using the above definitions it is possible to rewrite the lensing equation as:
\begin{equation}
    \theta{E}^2 = 4G_N M\left(1+z_L\right)\left(\frac{1}{\chi_L} - \frac{1}{\chi_S}\right) \; .
\end{equation}
Being quadratic in $\theta_0$, the lens equation \eqref{LEPM} has the following two roots:
\begin{equation}
\theta_0 = \frac{\theta_S}{2} \pm \sqrt{\frac{\theta_S^2}{4} + \theta_E^2}\;,
\end{equation}
which implies that, because of the lens, the original image of the source is split into two.
Notice that $\theta_S$ is time-independent because of the cosmological principle. In other words, observer, lens and source form a triangle whose sides increase due to the Hubble flow, but whose angles remain unchanged, and therefore $d\theta_S/dt_0 = 0$.

\subsection{Strong Lensing for an extended Lens }
In the thin lens approximation the lens equation \eqref{LEPM} for a general mass distribution is :
\begin{equation}\label{LE}
    \left(\bm{\beta}- \bm{\alpha} \right) = \nabla_{\bm{\theta}}\psi\left(\bm{\beta}\right) \; ,
\end{equation}
where $\bm{\beta} = (\beta_1,\beta_2)$ and $\bm{\alpha} = (\alpha_1,\alpha_2)$ are the position in the sky  of the image and the source respectively, and $\nabla_{\theta}$ is the two-dimensional angular gradient.\footnote{From now on we will restrict the use of the $\theta_i$ notation for point mass lenses, while the $\alpha$ and $\beta$ notation for a general lens profile}
The quantity $\psi(\bm{\beta})$ appearing in Eq.~\eqref{LE} is the \textit{lensing potential} and is defined as:
\begin{equation}\label{LP}
    \psi(\bm{\beta}) \equiv \frac{2}{c^2 }\frac{\mathcal{D}_{LS}}{\mathcal{D}_L\mathcal{D}_S}\int_{\bm{\beta}}  d\lambda \; \Phi \; ,
\end{equation}
where $\Phi$ is the standard Newtonian gravitational potential and the integral is taken along the path of the light ray, which depends on $\bm{\beta}$ and is  parametrized by $\lambda$.
Taking the divergence of Eq.~\eqref{LE}, as long as the extent of the lens is small compared to cosmological distances, we can use the Poisson equation to relate the Laplacian of the lensing potential to the mass distribution of the lens:
\begin{equation}\label{LapPsi}
\nabla^2_{\bm{\theta}}\psi\left(\bm{\beta}\right) = \frac{8\pi G_N}{c^2}\frac{\mathcal{D}_L\mathcal{D}_{LS}}{\mathcal{D}_S} \Sigma(\bm{\beta}) \; ,  
\end{equation}
where we have defined the surface mass density:
\begin{equation}\label{SMD}
    \Sigma (\bm{\beta}) \equiv \int_{\bm{\beta}} d\lambda  \;\rho \;,
\end{equation}
in which appears the mass distribution of the lens $\rho$. For a detailed derivation and an explanation on the assumptions behind Eqs.~\eqref{LE}, \eqref{LP}, \eqref{LapPsi} see for example Ref.~\cite{Bartelmann:2016dvf}.

\section{Strong Lensing observables and Cosmology}

As we saw in the previous section, in a strongly lensed system are present several images of the same source. Thus, of course, a first important observable is the angular separation between the various images. The entity of this variation is typically of the order of few arcseconds. For example, using data from Ref.~\cite{2004mmu..symp..117K}, the quasar QSO0957 + 561  at  redshift $z_S=  1.41$ lensed by a cluster at $z_L = 0.31$  displays two images separated by 6.1''.

On the other hand, even if coming from the same source, these several images at a given moment of time are not necessarily identical. Indeed, the optical path of the photons of each image is in general different, and thus some of the photons will require more time to reach the observer. This effect is called \textit{Time Delay}, which we indicate with $\Delta$, and could be divided into two different contributions~\cite{Weinberg:2008zzc}:
\begin{equation}
    \Delta = \Delta_{geo} + \Delta_{pot} \; ,
\end{equation}
where $\Delta_{geo}$ is the geometrical Time Delay and $\Delta_{pot}$ is the potential Time Delay. 
The former is due to the bending of the trajectory of the photon, whereas the latter is due to the motion into the lens gravitational
field.
For a single image the geometric $\Delta t_{geo}$  induced by a point mass lens is given by:
\begin{equation}
	\Delta t_{ geo} = \frac{(1 + z_L)^2(4GM)^2}{2\theta_0^2}\left(\frac{1}{\chi_L} - \frac{1}{\chi_S}\right)\;,
\end{equation}
from which we obtain $\Delta_{geo} = \Delta t_{geo}(\theta_+) - \Delta t_{geo}(\theta_-)$, where $\theta_{\pm}$ are the two roots of Eq. \eqref{LEPM}. 
The potential Time Delay between the two images due to a point mass in the thin lens approximation is given by:
\begin{equation}
    \Delta_{pot} = \Delta t_{pot}(\theta_+) - \Delta t_{pot}(\theta_-) =  2GM(1 + z_L)\ln\frac{\theta_-}{\theta_+} \; .
\end{equation}

For an extended lens profile the Time Delay between two images $\Delta_{ij}$ is given by, see  Ref.~\cite{Suyu:2012aa} :
\begin{equation}\label{TD}
    \Delta_{ij} = \frac{D_{\Delta_t}}{c}\left( \frac{\left(\bm{\beta}_i - \bm{\alpha}\right)^2}{2} - \frac{\left(\bm{\beta_j} - \bm{\alpha}\right)^2}{2} + \psi\left(\bm{\beta_j}\right) - \psi\left(\bm{\beta_i}\right)\right)\; ,
\end{equation}
where it was defined the Time Delay distance:
\begin{equation}\label{timedelaydistance}
    D_{\Delta_t} \equiv \left(1+ z_L\right)\frac{\mathcal{D}_L \mathcal{D}_S}{\mathcal{D}_{LS}} \; .
\end{equation}
It is straighforward to separate in the right hand side of Eq.~\eqref{TD} the contributions from the geometrical and the potential Time Delay. Indeed, the first two terms inside the brackets are given by the differences between the apparent and the true position of the source, and are thus of geometrical nature. The potential Time Delay is instead given by the difference between the lensing potential at the two apparent positions.

From Eq.~\eqref{TD} we can already understand the importance  of precise Time Delay measurements for cosmological implications. Indeed, if one is able to know the position of the source $\bm{\alpha}$, measures with enough precision the position of the images $\bm{\beta}_i$, and is able to infer the position of the lens and its lensing potential, all the cosmological information is contained in the Time Delay distance~\eqref{timedelaydistance}  through the angular diameter distances~\eqref{dS}.
Thus, from Time Delay measurements (if a reliable description of the lensing profile is given), one is able to reconstruct the Hubble factor without assuming any particular cosmological model. This is precisely the goal of the  H0LiCOW collaboration \footnote{\href{https://shsuyu.github.io/H0LiCOW/site/index.html}{https://shsuyu.github.io/H0LiCOW/site/index.html}}, which within the COSMOGRAIL \footnote{\href{http://www.cosmograil.org}{http://www.cosmograil.org}} program employed Time Delays measurements collected over the last decade to constrain the value of the cosmological parameter $H_0$ to a few percents level~\cite{Suyu:2016qxx,Wong:2019kwg,Bonvin:2016crt}, with competitive precision with respect to other cosmological probes. 
Moreover, Time Delay measurements can also be used to put constraints on the Post-Newtonian parameter $\gamma_{PPN}$, as discussed in Refs.~\cite{Collett:2018gpf,Yu:2018slt,Jyoti:2019pez,Yang:2020eoh}.  Furthermore, with optimistic assumptions on the surveys, in the next years the precision of observations will be enough to provide a smoking gun for Dark Energy~\cite{Shiralilou:2019div}.

In the next sections we will discuss two new observables proposed by us in Ref.~\cite{Piattella:2017uat}: the \textit{Time Delay drift} and the \textit{angular drift}, which could be used to reconstruct $H(z)$  and also to constrain violations of the Equivalence Principle \cite{Giani:2020fpz}. 

\section{Redshift drift of Gravitational Lensing}

The redshift drift is the time variation of the redshift of a source due to the Hubble flow, see Ref.~\cite{Weinberg:2008zzc}. 
In an expanding universe described by the FLRW metric, one straightforwardly obtains the result that a photon is redshifted, and the redshift is given in terms of the scale factor as follows:
\begin{equation}
	1 + z = \frac{a_0}{a_e}\;,
\end{equation}
where $a_0$ is the scale factor evaluated at present time (which is the time of observation) and $a_e$ is the scale factor evaluated at the emission time.

The derivative of the redshift with respect to the observation time $t_0$ is the following:
\begin{equation}\label{reddriftformula}
	\frac{dz}{dt_0} = \frac{1}{a_e}\frac{da_0}{dt_0} - \frac{a_0}{a_e^2}\frac{da_e}{dt_0} = \frac{1}{a_e}\frac{da_0}{dt_0} - \frac{a_0}{a_e^2}\frac{da_e}{dt_e}\frac{dt_e}{dt_0}\;.
\end{equation}
It is not difficult to show that:
\begin{equation}
	\frac{dt_e}{dt_0} = \frac{a_e}{a_0}\;,
\end{equation}
and therefore, the redshift drift formula \eqref{reddriftformula} becomes:
\begin{equation}\label{reddriftformula2}
	\frac{dz}{dt_0} = \frac{1}{a_e}\frac{da_0}{dt_0} - \frac{1}{a_e}\frac{da_e}{dt_e} = (1 + z)H_0 - H(z)\;,
\end{equation}
where the Hubble constant $H_0$ and the Hubble parameter $H(z)$ have appeared, so that measuring $dz/dt_0$ would allow to study the evolution of the Hubble factor.

Since the lens equation contains the angular diameter distances, we realize in Ref.~\cite{Piattella:2017uat} that the apparent positions of the source will acquire a time dependence due to the Hubble flow. Thus, applying Eq.~\eqref{reddriftformula2} to the lens equation we predict the following \textit{angular drift} and the \textit{Time Delay drift} for a point mass:
\begin{equation}\label{ADTDD}
    2\frac{\dot{\theta}_E}{\theta} = H_0 - \frac{H(z_L)}{1+z_L} \; , \qquad \frac{\dot{\Delta}}{\Delta} = K\left(H_0 - \frac{H(z_L)}{1+z_L}\right) \; ,
\end{equation}
where a dot indicates derivation with respect to the observer time $\dot{} = d/dt_0$, and $K$ is a factor of order unity given by:
\begin{equation}
    K = \frac{       \ln\frac{-\theta_-}{\theta_+} + \frac{\theta_- + \theta_+}{\theta_- - \theta_+}}{ \ln\frac{-\theta_-}{\theta_+} + \frac{\theta_+^2 - \theta_-^2 }{\theta_+\theta_-}} \; .
\end{equation}

The above equations tell us that the entity of these drifts is of order $H_0$. Since $H_0 \sim 10^{-18}$ s$^{-1}$, we conclude that the angular drift is of order of $10^{-10}$ arc seconds per year, which with the current precision of observation, $\sim 0.1$ arcseconds, would require $\sim 10^{9}$ yr to be detected.
The Time Delay drift is instad of order $\sim 10^{-3}$ arc seconds per year, which would require $10^{8}$ years to accumulate a drift detectable by current experiments, with sensitivity of order of days.

\section{Test of the Equivalence Principle using Strong Lensing Time Delay}

As we saw in the previous section, drift effects from Strong Lensing in GR are too small to be detected with current experiments, being of order $\sim H_0$. On the other hand, the situation could change if we  move to different theories of gravity, where the presence of a modified Poisson equation will change the time dependence of the lensing potential.
From the phenomenological point of view, many alternative theories of gravity modify the strength of the gravitational interaction by inducing an effective gravitational coupling $G_{eff}$, replacing the Newton constant. For example, we already saw that in VAAS gravity the time dependence of $G_{eff}$ is given by Eq.~\eqref{geffvaas}. In $f(R)$ theories of gravity instead we have $G_{eff} \sim 4G_N/3f_R $, so that $G_{eff}$ acquires a time dependence from the term $(df/dt) \dot{R}^{-1}$. 
However, in the expression for the deflection angle, we have a degeneracy between the mass of the lens and the gravitational coupling $G$.\footnote{From now on we will drop the subscript on $G_{eff}$ to unburden the notation.} Thus, it would not be possible in principle to distinguish between a stronger coupling or a heavier mass.

On the other hand, the situation is different if we consider drift effects. In Ref.~\cite{Giani:2020fpz} we studied the changes induced on the Time Delay drift by a time dependent $G$, and analyzed for illustrative purposes the entity of the constraints on $\dot{G}/G$ which can be extrapolated with the precision of current experiments.

For the case of a point mass, Eqs.~\eqref{ADTDD} become:
\begin{equation} \label{ADPMG}
    2\frac{\dot{\theta}_E}{\theta_E} = H_0 - \frac{H(z_L)}{1+z_L}  +\frac{\dot{G}}{G} \; ,
\end{equation}
\begin{equation} \label{TDDPMG}
\frac{\dot{\Delta}}{\Delta} = K\left(H_0 - \frac{H(z_L)}{1+z_L} +\frac{\dot{G}}{G}\right)\; ,
\end{equation}
where it is understood that $G$ is the gravitational coupling at the redshift of the lens.
The above equations show explicitly that drift effects in the context of Strong Lensing, contrary to what happens with spectroscopic measurements, are sensitive to variations of the gravitational coupling.
Since a signal of order $H_0 - H_L/(1+z_L)$, assuming a realistic cosmological evolution, is beyond the sensitivity of current observations, we can convert the bounds on the drifts in upper bounds on the variation of $G$.

Since Time Delay measurements are generally more precise than angular ones we need to obtain an expression for the time derivative of Eq.~\eqref{TD} for an extended lens profile.

\subsection*{Time Delay drift for extended lenses}
To begin with let us compute the time derivative of the lensing potential:
\begin{equation}
    \dot{\psi}(\bm{\beta})= \frac{2}{c^2 }\frac{\mathcal{D}_{LS}}{\mathcal{D}_L\mathcal{D}_S}\left( \frac{1}{\left(1+z_L\right)}\frac{d z_L}{dt}\int_{\bm{\beta}} d\lambda \; \Phi + \frac{d}{dt}\left(\int_{\bm{\beta}} d\lambda \; \Phi \right)\right) \; .
\end{equation}
The time dependence of the second term in the right hand side of the latter equation comes from the change in time of the light ray path due to a variation of $\bm{\beta}$ and from the Newtonian potential. The former is difficult to compute exactly because it is in general difficult to describe precisely the curve identified by the light ray path. On the other hand, it is reasonable to assume that the induced variation of the curve is small and does not contribute to the support of the integral in Eq.~\eqref{LP}. In particular, this is true if we evaluate the above integral within the  Born approximation, i.e. along the unperturbed light path, as it is customary in Strong Lensing applications where $\Phi/c^2 \ll 1$~\cite{Bartelmann:2016dvf}. Within this assumption we can interchange the operation of integration and time differentiation obtaining:
\begin{equation}
     \dot{\psi}(\bm{\beta}) = \psi(\bm{\beta})\left(H_0 - \frac{H(z_L)}{1+z_L}\right) + \psi(\bm{\beta})\frac{\int_{\bm{\beta}} d\lambda \; \dot{\Phi}}{\int_{\bm{\beta}} d\lambda\; \Phi} \; .
\end{equation}
The above equation can be further simplified if we consider a static distribution of matter. Indeed, in this case the only time dependence of the Newtonian potential is through the effective gravitational coupling $G$,  so that we have:
\begin{equation}\label{TDTDSD}
\dot{\psi}(\bm{\beta}) = \psi(\bm{\beta})\left(\frac{\dot{G}}{G} + H_0 - \frac{H(z_L)}{1+z_L}\right) \; .
\end{equation}

In concrete Time Delay measurements we have to take into account corrections due to the presence of mass distributed along the line of sight. This is done by introducing a parameter called \textit{external convergence}, $\kappa_{ext}$, and defining the \textit{real} Time Delay distance as:
\begin{equation}
    D_{\Delta t}^{real} \equiv\frac{D_{\Delta t}}{1 - \kappa_{ext}} \; ,
\end{equation}
see for example Ref.~\cite{Suyu:2012aa}.  If we assume that the external convergence has a time dependence this would be inherited by the Time Delay distance~\eqref{timedelaydistance}.

Combining the latter and  Eq.~\eqref{TDTDSD} we can finally write down the Time Delay drift for an extended lens with a static distribution of mass:
\begin{equation}\label{TDDk}
    \frac{\dot{\Delta}_{ij}}{\Delta_{ij}} = \left(\frac{\dot{G}}{G}  + H_0 - \frac{H(z_L)}{1+z_L}  \right)\left[1  + \frac{D_{\Delta t}\left(\bm{\beta}_i - \bm{\alpha}\right)^2}{2c \Delta_{ij}} - \frac{D_{\Delta t}\left(\bm{\beta_j} - \bm{\alpha}\right)^2}{2c \Delta_{ij}}\right] -\frac{\dot{\kappa}_{ext}}{1-\kappa_{ext}} \; .
\end{equation}

\subsection*{Estimating the variation of the Time Delay from current data}
We now illustrate how the data can put constraints on the variation of the Time Delay. To this end, we use the package PyCS3 from the COSMOGRAIL program \footnote{Available  \href{http://cosmograil.org/}{here}.}, see Refs.~\cite{Tewes:2012gs,Bonvin:2015jia}. The data used are the simulated light curves used in \cite{Bonvin:2015jia}, produced in the context of the blind Time Delay measurement
competition named Time Delay Challenge 1 (TDC1)~\cite{Dobler:2013rda}, and from the quasar DES J0408-5354~\cite{Courbin:2017yvz}. We split the total time of observations in two equal time periods. Each period consists of 658 days for the trial curves, and of 93 days for DES J0408-5354. The Time Delay between each image is then calculated for each period. The Time Delay estimates are shown in App.~\ref{AppendixD} and summarized in Table \ref{tabletimedelay}. From it we can readily estimate the relative variation \eqref{TDDk} as:
\begin{equation}
     \frac{\dot{\Delta}_{ij}}{\Delta_{ij}}= \frac{\Delta_{ij}\left(t + \delta t\right) - \Delta_{ij}(t)}{\delta t \Delta_{ij}(t)} = \frac{\Delta_{ij}^{I+II} - \Delta_{ij}^{I}}{\Delta_{ij}^{I} \delta t} \; ,
\end{equation}
We display the results in Table \ref{tableredshiftdrift}. 

\begin{table}
    \centering
    \begin{tabular}{c||c|c|c|c|c|c}
          & $\Delta_{AB}$ & $\Delta_{AC}$ & $\Delta_{AD}$ & $\Delta_{BC}$ & $\Delta_{BD}$ & $\Delta_{CD}$ \\
         \hline
         Trial I & $-4.9^{+4.3}_{-5.5}$ & $-18.0^{+6.0}_{-8.0}$ & $-0.7^{+9.6}_{-10.3}$ & $-13.5^{+7.6}_{-7.6}$  & $+4.3^{+9.8}_{-11.5}$ & $+17.5^{+10.7}_{-12.1}$  \\
         %Trial II & -6.83 & -26.45 & -70.39 & -19.61 & -63.56 & -43.94 \\
         Trial I+II & $-4.5^{+2.4}_{-2.0}$ & $-21.3^{+1.4}_{-1.7}$ & $-29.9^{+9.2}_{-7.0}$ & $-16.6^{+1.8}_{-3.4}$  & $-25.2^{+7.5}_{-6.4}$ & $-8.2^{+8.3}_{-6.7}$ \\
         \hline
         %HE 0435 I & -8.56 & -0.24 & -13.83 &  +8.32 & -5.27 & -13.59 \\
         %HE 0435 II & -8.85 & -8.00 & -14.17 & +0.84 &  -5.33 & -6.17 \\
         %HE 0435 I+II & -12.47 & -9.37 & -17.28 & +3.10 & -4.81 & -7.91 \\
         DES 0408 WFI I & $-106.5^{+15.9}_{-14.1}$ & $-110.6^{+50.4}_{-21.7}$ & $-142.2^{+34.3}_{-18.3}$ & $-2.4^{+38.8}_{-19.9}$  & $-37.1^{+25.7}_{-16.2}$ & $-31.7^{+19.8}_{-30.3}$ \\
         DES 0408 WFI I+II & $-112.6^{+6.6}_{-2.2}$ & $-117.2^{+5.9}_{-7.6}$ & $-153.2^{+11.9}_{-9.5}$ & $-7.1^{+8.6}_{-8.6}$  & $-40.5^{+11.3}_{-9.3}$ & $-35.6^{+14.1}_{-10.7}$
         %WFI2033 I & +31.47 & -36.82 & & -68.29  &\\
         %WFI2033 I+II & +31.83 & -34.99 & & -66.82 & 
    \end{tabular} 
    \caption{Time Delay $\Delta$ between four images from a simulated quasar and the DES J0408-5354 quasar \cite{Courbin:2017yvz}. Each image is labeled from $A$ to $D$.  The numbers I and I+II indicate that $\Delta$ was measured over the first half of the period of observations, or over the whole period, respectively. All values are given in days.}
    \label{tabletimedelay}
\end{table}

\begin{table}
    \centering
    \begin{tabular}{c||c|c|c|c|c|c}
          &$|\dot{\Delta}/\Delta|_{AB}$ & $|\dot{\Delta}/\Delta|_{AC}$  & $|\dot{\Delta}/\Delta|_{AD}$& $|\dot{\Delta}/\Delta|_{BC}$& $|\dot{\Delta}/\Delta|_{BD}$ & $|\dot{\Delta}/\Delta|_{CD}$\\
         \hline
         Trial ($\times 10^{-5}$) & X & $27.9 \pm 12.6$ & X & $34.9 \pm 20.9 $ & X & X \\
         \hline
         DES 0408 WFI ($\times 10^{-5}$) & $61.6 \pm 9.9$ & $64.2 \pm 29.6$ & $83.2 \pm 21.1$ & X & $98.5 \pm 73.6$ & X
    \end{tabular}  
    \caption{Estimated absolute Time Delay variation for the simulated quasar and DES J0408-5354. All values are given in $day^{-1}$. The X's represent values with uncertainty bigger than the central value, and are thus omitted.}
    \label{tableredshiftdrift}
\end{table}

\subsection*{Estimated constraint on $\dot{G}/G$}

Through Eq.~\eqref{TDDk} it is possible to relate constraints on the relative time variation of $\Delta_{ij}$ to upper bounds on the variation of $\dot{G}/G$. 
The external convergence time dependence is difficult to evaluate. We expect it to depend explicitly on $\dot{G}/G$, similarly to the potential generated by the lens, with the two contributions having the same sign, but we will assume that it is negligible with respect to the precision of current observations. We want to stress however,  as we discussed in the previous sections, that it is in principle possible to disentangle such contribution by considering differences of Time Delays drifts of multiple images.
As we already saw the drift due to the Hubble flow is of order of $H_0 \sim 10^{-18}$ s$^{-1}$, so we will neglect it as well.
Another effect that could be of relevance is the time variation of the Time Delay due to the peculiar velocity of the lens galaxy and its transverse motion. On the other hand, according to Refs.~\cite{Zitrin:2018let,Wucknitz:2020spz}, this contribution is estimated to be of the order of a few seconds per year, so that the effect, even though it is bigger than the cosmological one due to the Hubble flow, is still not appreciable with current precision. Under these assumptions, and roughly estimating  the term inside the square bracket of Eq.~\eqref{TDDk} to be of order 1,  the values of Table \ref{tableredshiftdrift} can be directly converted into upper bounds on the time variation of the effective gravitational coupling. The results are reported in Table \ref{tableGconst}.
\begin{table}
    \centering
    \begin{tabular}{c||c|c|c|c|c|c}
          &$|\dot{G}/G|_{AB}$&
          $|\dot{G}/G|_{AC} $&
          $|\dot{G}/G|_{AD} $& 
          $|\dot{G}/G|_{BC} $&
          $|\dot{G}/G|_{BD} $&
          $|\dot{G}/G|_{CD} $\\
         \hline
         Trial ($\times 10^{-1}$) & X & $1.0  \pm 0.5$ & X & $1.3 \pm 0.8 $& X & X \\
         DES 0408 WFI ($\times 10^{-1}$) & $2.2 \pm 0.4$& $2.3 \pm 1$ &  $3.0 \pm 0.8$ & X & $3.6\pm 3$ & X
    \end{tabular}  
    \caption{Upper bounds on the absolute value of $\dot{G}/G$ in $yr^{-1}$ from the simulated quasar and DES J0408-5354. The X’s represent values with uncertainty bigger than the central value, and are thus omitted. } 
    \label{tableGconst}
\end{table}
\section{Summary and discussion}
We have introduced two novel observables in the contest of Strong Lensing, i.e. the Time Delay drift and the angular drift. We have shown that measuring these effects one is able to estimate $H(z)$, and thus reconstruct the cosmological evolution and the effective equation of state parameter of the Universe $w_{eff}$. We have shown that in a modified gravity framework where the effective gravitational coupling become time dependent, those drifts earn a contribution $\dot{G}/G$, and thus their measurements could be used to constraint both violations the Equivalence Principle and DE models. Unfortunately, the current precision of the observation is not enough to detect these drifts so that the constraints we obtain are currently not very competitive. On the other hand, they could improve already in the near future with upcoming data, see for example Ref.~\cite{Millon:2020xab}, or simply increasing the observational time. Moreover, as discussed in Refs.~\cite{Zitrin:2018let,Wucknitz:2020spz,Liu:2019jka}, if in the future strongly lensed repeating FRB will be detected, they will provide Time Delay measurements of such extremely high precision, nominally of the order of seconds, that even redshift drift effects due to the Hubble expansion will be appreciable. In this scenario, the impact for cosmological implications of the drifts we proposed in this chapter is very promising, and we keep high expectations for the future.

\chapter{Final Considerations}

\epigraph{\textit{I managed to get a quick PhD — though when I got it I knew almost nothing about physics. But I did learn one big thing: that no one knows everything, and you don't have to.}}{Steven Weinberg}
\section*{Summary}
Research on DE is a tricky business. In principle, one should just try to answer to the question of what is causing the accelerated expansion of the Universe, but in practice this is just the first domino tile of a long chain. The Cosmological Constant $\Lambda$ seems to be the most reasonable and simplest candidate of Dark Energy, in particular because together with the CDM paradigm is able to safely match a number of different observations. However, with the outstanding precision reached in the last years in observational cosmology, the concordance model seems to be not so concordant, with the most uncomfortable question being what is the value of the Hubble parameter now. Curiously, this seems to be an evergreen question in cosmology since from its birth, and it is inspiring the tenacity shown by the scientific community through the last hundred years in looking for a satisfactory answer. We tried to give an overview of the $\Lambda$CDM model in chapter~\ref{chapter:LCDM} of this work; our main goal was to highlight the pillars on which it is built on and the kind of questions that it is able to answer.
Of course our presentation is by no means exhaustive or complete by itself, but we hope it is able at least to address a satisfactory number of references for the curious reader.

DE is so interesting because it could be a window towards new physics or, depending on one's personal perspective, it is already a manifestation of it. The community working on this topic has undoubtedly shown a fair amount of creativity in the last two decades, both in boosting or trying to contain the proliferation of possible candidates of DE. In chapter~\ref{chapter:DE} we tried to give a reasonable classification of the kind of models that were proposed in the last years based on the Lovelock theorem. Keeping into account that a proper treatment of the topic would have required more than a textbook, for each class of models we presented some examples in an attempt of transmitting the flavor and the potential of these proposals.

The ultimate goal of this thesis is to present the results of our research from the last four years. Most of it was oriented on models of DE which does not introduce in the playground new degrees of freedom, but instead attempt to ascribe the accelerated expansion of the Universe to geometrical modifications of the gravitational interaction. 

An interesting phenomenology arises if in such geometrical description we relax the assumption of locality for the field equations. Ultimately, the main motivation for considering these kinds of models is that quantum mechanics seems to be nonlocal at fundamental level, and such could also be a quantum description of gravity.  
In chapter~\ref{chapter:NL} we review the theoretical grounds and the mathematical formalism of nonlocal models of gravity based on the inverse d'Alembertian of the Ricci scalar $\Box^{-1} R$.

In chapter~\ref{chapter:PCNL} we present the results of our research on nonlocal gravity models, in particular~\cite{Giani:2019vjf,Giani:2019xjf}:
\begin{itemize}
    \item We found that the Late-times asymptotic equation of state has a common behavior in those models which contain explicitly a term $\Box^{-1}R$ in the Lagrangian. In particular, even if the Hubble factor does not approach a constant value, we found that $\dot{H}/H^2 \rightarrow 0$ because $H\rightarrow \infty$ due to the divergence of $\Box^{-1}R$. We also show that this does not happen for the DW model if the distortion function $f(\Box^{-1})R$ is chosen in such a way that $f(\infty) \neq \infty$. 
    \item We studied VAAS gravity within a dynamical system approach and found no stable critical points. On the other hand, a qualitative study of its field equations show that it is indeed possible to produce at background level an evolution history compatible with $\Lambda$CDM from the radiation dominated epoch until today.
    \item Still in VAAS gravity, we have shown that on local scales the effects of the nonlocal modifications is encoded in a slip parameter and in a modification of the effective gravitational coupling $\eta, Y \neq 1$. 
    \item We checked consistency of the model with LLR constraints, which are not passed. However, we found that the field equations admit spherically symmetric static solutions which, if realized and stable, would satisfy trivially LLR constraints and for which one of the gravitational potential closely resemble the solution one would obtain in standard GR for the Schwarzschild-deSitter solution. 
\end{itemize}

In chapter~\ref{chapter:IR} we discuss a geometrical theory of gravity containing higher order derivative terms which we proposed in Ref.~\cite{Amendola:2020qho}. The theory is based on the introduction of the anticurvature scalar $A$, which is the trace of the inverse of the Ricci tensor $A_{\mu\nu} = R^{-1}_{\mu\nu}$. We derive the general field equations for an $f(R,A)$ theory and assess their potential for cosmological implications. We found that an interesting phenomenology arise already for the simplest choices of $f$, but also a no-go theorem which claims that polynomial Lagrangians are not able to reproduce a viable cosmological history. Finally, we present some choices of $f$  that evades the no-go theorem and thus are worth further investigation.

Finally, in chapter~\ref{chapter:Lensing} we present the results of our works~\cite{Piattella:2017uat,Giani:2020fpz} about the potential of Strong Lensing observables to study cosmology and DE. In particular, we introduce two new quantities, the \textit{angular drift} and the \textit{Time Delay drift}, which are a manifestation of the Hubble flow in a lensed system.  These drifts, if measured, would allow for a reconstruction of $H(z)$ at the redshift of the lens, and thus potentially very helpful for estimating the effective equation of state parameter of the Universe $w_{eff}$.
We also show that in a modified theory of gravity with a time dependent gravitational coupling, a similar effect arise, and we can use Strong Lensing Time Delay observations to constrain violations of the Equivalence Principle.
The entity of these drifts is unfortunately beyond the sensitivity of current experiment, and results in very weak constraints on $\dot{G}/G$. However, we also discuss some possible future perspectives that would make them appealing. 

\section*{Outlooks}
Most of the nonlocal models of gravity presented in the literature seems to be ruled out by LLR constraints. On the other hand they are motivated by fundamental physics, and as we saw in Ref.\cite{Giani:2019xjf} in most cases they provide a useful mechanism to trigger an accelerated expansion of the 
Universe only at Late-times. For these reasons we are currently trying to understand if they can be used, together with a Cosmological Constant $\Lambda$, to alleviate or solve the $H_0$ tension by changing the rate of expansion of the universe today without affecting CMB measurements or violating LLR constraints. 
Regarding VAAS, a perturbative analysis beyond the small scales regime is still lacking, and is crucial for properly assess the appealing of the model with respect to $\Lambda$CDM. We address this analysis for future works.
Inverse Ricci gravity offers a completely new class of modified gravity theories, and as we saw it can introduce a very interesting phenomenology. On the other hand, containing higher order derivatives, one expect that it is in general unstable unless the Lagrangian is degenerate. We address to future works a research of which kind of $f(A,R)$ will result in a degenerate Lagrangian capable of escaping the no-go theorem.
Finally, the increasing precision of Strong Lensing measurements motivate a deeper investigation of the cosmological information we can extract from them. Moreover, being equivalent to an interferometer with arms of astrophysical size, a strongly lensed system could be sensitive to the passage of a gravitational wave. This could be very interesting for multi-messenger astronomy, and we address to future works a quantitative estimation of the GW impact on Time Delay measurements to assess the potential of this idea.

\appendix

\chapter{Critical points in VAAS gravity}
\label{AppendixA}

We want to study the critical points of the system \eqref{dsfinitedistance2} satisfying the constraint \eqref{modFriedeq2}.

\subsection*{Critical points at finite distance}
Consider first the case where $2\alpha V = 1$, which is special because it eliminates $\dot H$ from the field equations. From the Klein Gordon equations we see that $V = 1/(2\alpha) =$ constant only if $m^2 = 0$. The latter is a parameter of the theory so it is not necessarily vanishing and thus, unless $m^2 = 0$, $V$ cannot be constant. With $V = 1/(2\alpha) =$ constant, then $\dot V = Y = 0$ and thus from Eq.~\eqref{modFriedeq2} we get $\rho = 0$.  We are just left with the Klein Gordon equation for $U$ while $H$ is completely arbitrary, since we have lost the equations ruling its dynamics. Since providing a suitable $H$ is one of the objectives of the model, we do not consider this possibility anymore.

Now we consider $2\alpha V \neq 1$. The right hand sides of the last three equations of the dynamical system~\eqref{dsfinitedistance2} vanish when $X = Y = 0$ and $m^2 = 0$. Again, the latter is a parameter of the theory so it is not necessarily vanishing. Thus if $m^2 \neq 0$ we can already conclude that there are no critical points at finite distance. 

Let us consider now the subclass of theories for which $m^2 = 0$, i.e. with $R_f = 0$. Demanding the vanishing of the right hand sides of the first three equations of system~\eqref{dsfinitedistance2} and from Eq.~\eqref{modFriedeq2} we have, taking into account $X = Y = m^2 = 0$:
\begin{eqnarray}
	3H^2(1 - 2 \alpha V) = \frac{\rho - P}{2 M_{  Pl}^{2}}\;,\\
	-3 H(\rho + P) = 0\;,\\
	H^2(1 - 2\alpha V) = \frac{P - \rho}{2 M_{  Pl}^{2}}\;,\\
	3H^2(1 - 2\alpha V) = \frac{\rho}{M_{  Pl}^{2}}\;.
\end{eqnarray}
Now, from the second equation above we either have that $H = 0$ or $P = -\rho$, i.e. a vacuum energy equation of state is required. If our fluid model has not such equation of state, then the only possibility is $H = 0$ and thus $\rho = P = 0$. This critical point represents Minkowski space. Note that $U$ and $V$ may assume whatever constant value, except $V = 1/(2\alpha)$.

On the other hand, let us assume that indeed the fluid content satisfies a vacuum energy equation of state, i.e. $P = -\rho$. In this case, the above system becomes:
\begin{eqnarray}
	3H^2(1 - 2 \alpha V) = \frac{\rho}{M_{  Pl}^{2}}\;,\\
	H^2(1 - 2\alpha V) = -\frac{\rho}{M_{  Pl}^{2}}\;.
\end{eqnarray}
Summing the two equations we arrive at:
\begin{equation}
	4H^2(1 - 2 \alpha V) = 0\;.
\end{equation}
Since $2\alpha V \neq 1$, we have again $H = 0$ and thus the same Minkowski critical point as before. There is a caveat here. When $m^2 = 0$ the evolution of $U$ is disentangled from the one of $H$. Therefore, we can reduce our dynamical system~\eqref{dsfinitedistance2} to:
\begin{subequations}\label{dsm2zero}
\begin{eqnarray}
    \dot{H} &=&\frac{1}{1 - 2 \alpha V} \left[\frac{\rho - P}{2 M_{  Pl}^{2}} + 2\alpha H Y\right] - 3 H^{2}\;,\\
    \dot{\rho} &=& -3 H(\rho + P)\;,\\
    \dot{Y} &=& - 3 H Y \;,\\
    \dot{V} &=& Y\;.
\end{eqnarray}
\end{subequations}
From this system is not difficult to see that there is a critical point 
\begin{equation}
	P = -\rho = -M^2_{  Pl}\left(1-2\alpha V\right)3H^2 =\mbox{ constant}\;,
\end{equation}
which represents a de Sitter phase. With $H > 0$ constant, the equation for $U$ becomes:
\begin{equation}
	\ddot U + 3H\dot U = -12H^2\;,
\end{equation}
which has the special, non-constant solution $U = -4Ht$.
\subsection*{Critical points at infinite distance}
In order to investigate the critical points at infinity, we build the Poincaré hyperspherein the variable space plus one dimension. The equation of the sphere is the following: 
\begin{eqnarray}\label{sphere}
    h^{2} + r^{2} + p^{2} + x^{2} + y^{2} + u^{2} + v^{2} + z^{2} = 1\;,
\end{eqnarray}
where:
\begin{eqnarray}
    H \equiv \frac{h}{z}\;, \quad \rho \equiv \frac{r}{z}\;, \quad P \equiv \frac{p}{z}\;, \quad X \equiv \frac{x}{z}\;, \quad Y \equiv \frac{y}{z}\;, \quad U \equiv \frac{u}{z}\; \quad \textrm{and} \quad V \equiv \frac{v}{z}\;.
\end{eqnarray}
System \eqref{dsfinitedistance2} thus becomes:
\begin{subequations}\label{dsinfinitedistance1}
\begin{eqnarray}
    z\dot{h} &=& A (1 - h^{2}) -  h \left(r B + x C + y D + u E + v F \right)\;,\\
    z\dot{r} &=& B (1 - r^{2})-  r \left(h A + x C + y D + u E + v F \right)\;,\\
    z\dot{x} &=& C (1 - x^{2}) - x \left(h A + r B + y D + u E + v F \right)\;,\\
    z\dot{y} &=& D (1 - y^{2}) - y \left(h A + r B + x C + u E + v F \right)\;,\\
    z\dot{u} &=& E (1 - u^{2}) - u \left(h A + r B + x C + y D + v F \right)\;,\\
    z\dot{v} &=& F (1 - v^{2}) - v \left(h A + r B + x C + y D + u E \right)\;,\\
    z\dot{z} &=& -\left(h A + r B + x C + y D + u E + v F \right)\;,
\end{eqnarray}
\end{subequations}
where the equation for $\dot{z}$ is obtained from Eq.~\eqref{sphere} and the terms $A$, $B$, $C$, $D$, $E$ and $F$ are defined as follows:
\begin{subequations}
\begin{eqnarray}
    A &\equiv & \frac{z}{z - 2 \alpha v} \left[\frac{z(r - p)}{2M_{  Pl}^{2}} + \frac{m^{2}z}{2}\left(z - u\right) + 2\alpha h y\right] - 3h^2\;,\\
    B &\equiv & - 3 h (r + p)\;,\\
    C &\equiv & - 3 h x - 6 h^{2} -\frac{6 z}{z - 2 \alpha v} \left[\frac{z(r - p)}{2M_{  Pl}^{2}} + \frac{m^{2}z}{2}\left(z - u\right) + 2\alpha h y\right]\;,\\
    D &\equiv & \frac{m^{2}z^2}{2 \alpha} - 3 h y\;,\\
    E &\equiv & zx\;,\\
    F &\equiv & zy\;.
\end{eqnarray}
\end{subequations}
The critical points of the above system corresponding to $z = 0$ are critical points at infinity. In order to find them, let us first define a new time parameter such that $\frac{z}{3}\frac{d}{dt} \equiv \frac{d}{d \tau} \equiv\;'$ and the functions:
\begin{eqnarray}
    G \equiv h \left(h^{2} + \gamma r^{2} + x^{2} + y^{2} + 2 h x -1\right)\;,\\
    \tilde G \equiv h \left(h^{2} + \gamma r^{2} + x^{2} + y^{2} + 2 h x - \gamma\right)\;,
\end{eqnarray}
where we have assumed a barotropic equation of state linking pressure to density:
\begin{equation}
	P = (\gamma - 1)\rho\;.
\end{equation}
The dynamical system \eqref{dsinfinitedistance1} for $z =0$ can thus be written as:
\begin{subequations}\label{dsinfinitedistance2}
\begin{eqnarray}
    h' &=& h G\;, \label{dsinfinitedistance2h}\\
    r' &=& r \tilde G\;,\\
    x' &=& x G - 2 h^{2}\;,\label{dsinfinitedistance2x}\\
    y' &=& y G\;, \label{dsinfinitedistance2y}\\
    u' &=& u\left(G + h\right)\;,\\
    v' &=& v\left(G + h\right)\;.
\end{eqnarray}
\end{subequations}
It is important to emphasise that the solutions of the above system must be compatible with the Friedmann equation, which in terms of the variables on the Poincaré sphere provides:
\begin{eqnarray}\label{Friedmanninfinitedistance}
    \left(1 - 2\alpha \frac{v}{z}\right)\frac{h^{2}}{z^{2}} + \frac{m^{2} u}{6 z} - \frac{\alpha y}{3 z^{2}} (6h + x) - \frac{r}{3 z M_{  Pl}^{2}} = 0\;.
\end{eqnarray}
Multiplying the above equation for $z^{3}$ and then considering $z = 0$, we obtain:
\begin{eqnarray}
    \alpha v h^{2} = 0.
\end{eqnarray}
So that at infinity Friedmann equation imposes that at least one among $\alpha$, $v$, $h$ is vanishing. It is easy to see that when $h = 0$ the function $G$ vanishes identically, so we have a critical hyperplane in the variables space corresponding to Minkowski spacetime. 

We will discuss later the stability of this critical hypersurface, and focus now on the only other interesting case,\footnote{We do not consider $\alpha = 0$ since it simply turns off the nonlocal interacting term.} $v = 0$. In this case, Friedmann equation becomes:
\begin{equation}\label{Friedeqvzero}
    h^2  -\alpha y \left(2h + \frac{x}{3} \right) = 0 \;. 
\end{equation}
Since $h \neq 0$, then from Eqs.~\eqref{dsinfinitedistance2h} and \eqref{dsinfinitedistance2y} we have that:
\begin{equation}
	y = h + K\;,
\end{equation}
where $K$ is an integration constant. From Eq.~\eqref{Friedeqvzero} we then have:
\begin{equation}
	x = \frac{3h^2}{\alpha y} - 6h = \frac{3h^2}{\alpha(h + K)} - 6h\;.
\end{equation}
This result allows us to rewrite Eq.~\eqref{dsinfinitedistance2x} as follows:
\begin{equation}\label{xhequation}
    x' = G\left[\frac{3h^2}{\alpha \left(h + K\right)} - 6h\right] - 2h^2 \; ,
\end{equation}
and we see that it is impossible to have both \eqref{dsinfinitedistance2h} and \eqref{xhequation} vanishing  without $G=h=0$, and so there are no critical points, but $h = 0$, at infinite distance.

\subsection*{Linearisation and stability of the critical point}

In order to calculate the stability of the critical points at $h = 0$, let us linearise system \eqref{dsinfinitedistance2} around the critical point $h_{0} =0$, i.e, $h = h_{0} + \epsilon$, $r = r_{0} + \eta$, $x = x_{0} + \chi$, $y = y_{0} + \varphi$, $u = u_{0} + \lambda$ and $v = v_{0} + \sigma$. Consequently, the linearised dynamical system is given by:
\begin{subequations}\label{linearizedds}
\begin{eqnarray}
    \epsilon ' &=& 0 \; ,\label{linearizeddsepsilon}\\
    \eta ' &=& r_{0} \left(\gamma r_{0}^{2} + x_{0}^{2} + y_{0}^{2} - \gamma \right) \epsilon \; ,\\
    \chi ' &=& x_{0} \left(\gamma r_{0}^{2} + x_{0}^{2} + y_{0}^{2} - 1 \right) \epsilon \; ,\\
    \varphi ' &=& y_{0} \left(\gamma r_{0}^{2} + x_{0}^{2} + y_{0}^{2} - 1 \right) \epsilon \; ,\\
    \sigma ' &=& u_{0} \left(\gamma r_{0}^{2} + x_{0}^{2} + y_{0}^{2}\right) \epsilon \; ,\\
    \lambda ' &=& v_{0} \left(\gamma r_{0}^{2} + x_{0}^{2} + y_{0}^{2}\right) \epsilon \; .
\end{eqnarray}
\end{subequations}
Usually the stability of the critical point is studied by means of the Jacobian matrix, but unfortunately it is degenerate in our case. However, we easily recognise that in the above system of equations the perturbation $\epsilon$ is constrained to be a constant by \eqref{linearizeddsepsilon}. Note also that the combinations $\left(\gamma r_{0}^{2} + x_{0}^{2} + y_{0}^{2} - 1 \right)$ and $\left(\gamma r_{0}^{2} + x_{0}^{2} + y_{0}^{2} - \gamma \right)$ are constant as long as we assume a time-independent equation of state.  Thus all the perturbations with the exception of $\epsilon$, which is constant, go linearly with time showing an unstable behavior.
\chapter{Qualitative dynamic of nonlocal models }
\label{appendixB}

\subsection*{Qualitative behavior of $U$}
The Klein Gordon equation for $U$ in FLRW backround, defining $X \equiv U'$ and using $N= \log a$ as time coordinate is:
\begin{equation}
 X' + \left(3 + \xi\right)X + 6\left(2 + \xi\right) = 0\;.
\end{equation}
The formal solution for $X'$ is:
\begin{equation}
    	X(N) = C_1e^{-F(N)} - 6e^{-F(N)}\int^N_{N_i} d\bar Ne^{F(\bar N)}[2 + \xi(\bar N)]\;,
\end{equation}
which choosing vanishing initial conditions give:
\begin{equation}
    	X(N) = -6e^{-F(N)}\int^N_{N_i} d\bar Ne^{F(\bar N)}[2 + \xi(\bar N)]\;.
\end{equation}
It is straightforward to show that $0 \leq X \leq 6$, indeed rewrite the solution for $X$ as follows:
\begin{equation}
	X(N) = -6e^{-F(N)}\int^N_{N_i} d\bar Ne^{F(\bar N)}[3 + \xi(\bar N)] + 6e^{-F(N)}\int^N_{N_i}d\bar Ne^{F(\bar N)}\;.
\end{equation}
The first integral can be cast as:
\begin{equation}
	X(N) = -6e^{-F(N)}\int^N_{N_i} d\bar N\frac{d(e^{F(\bar N)})}{d\bar N} + 6e^{-F(N)}\int^N_{N_i}d\bar Ne^{F(\bar N)}\;,
\end{equation}
and thus:
\begin{equation}
	X(N) = -6 + 6e^{-F(N)} + 6e^{-F(N)}\int^N_{N_i}d\bar Ne^{F(\bar N)}\;.
\end{equation}
Being the second and third terms on the right hand side strictly positive, we have then that $X(N) > -6$.

\subsection*{Qualitative behavior of VAAS}
 During the radiation-dominated epoch one has $\xi = -2$ and thus:
\begin{equation}
	X = 0\;,
\end{equation}
This implies that $U$ is a constant, and this constant must be zero, because of our initial condition. On the other hand,
\begin{equation}
	Y(N) = -e^{-(N-N_i)}\int^N_{N_i} d\bar Ne^{(\bar N-N_i)}\frac{m^2}{H_0^2\Omega_{  r0}e^{-4\bar N}} = -\frac{m^2}{5H_0^2\Omega_{  r0}}e^{-N}\left(e^{5N} - e^{5N_i}\right)\;,
\end{equation}
from which:
\begin{equation}
	\tilde V = -\frac{m^2}{20H_0^2\Omega_{  r0}}e^{4N} - \frac{m^2}{5H_0^2\Omega_{  r0}}e^{5N_i - N} + C_3\;.
\end{equation}
Since $\tilde V(N_i) = 1$, we then have:
\begin{equation}
	\tilde V = -\frac{m^2}{20H_0^2\Omega_{  r0}}e^{4N} - \frac{m^2}{5H_0^2\Omega_{  r0}}e^{5N_i - N} + \frac{m^2}{4H_0^2\Omega_{  r0}}e^{4N_i} + 1\;.
\end{equation}
This is very small for $N$ large and negative, so one can basically take $\tilde V = 1$.

During the matter-dominated epoch we have $\xi = -3/2$ and thus:
\begin{equation}
		X = -3e^{-3N/2}\int^N_{\tilde N_i} d\bar Ne^{3\bar N/2} = -2 + 2e^{3\tilde N_i/2}\;,
\end{equation}
where $\tilde N_i$ is some new initial value, chosen in the matter-dominated epoch. Being this large and negative, we neglect the exponential with respect to $-2$ and so a linear solution for $U$ follows:
\begin{equation}
	U = C_1 - 2N\;,
\end{equation}
with $C_1$ integration constant. For $Y$ we have a solution similar to the one we found for the radiation-dominated case. We simplify it a little bit, writing:
\begin{equation}
	\tilde V = 1 - \frac{m^2}{12H_0^2\Omega_{  m0}}e^{3N}\;.
\end{equation}
We have neglected all the exponentials contributions containing the initial e-folds number because, even in the matter-dominated epoch, it is very small. It is only at late times that $\tilde V$ starts to grow different from one.

In vacuum, i.e. when matter and radiation dilute, the first Friedmann equation is:
\begin{equation}
	3\tilde V = -\frac{m^2U}{2H^2} - \frac{Y}{2}(6 + X) + \frac{\rho}{M_{Pl}^2H^2}\;.
\end{equation}
Since $U < 0$, $Y < 0$, $X > -6$ and of course $\rho > 0$, we can conclude that $\tilde V > 0$. On the other hand, $Y < 0$ tells us that $\tilde V$ always decreases. So, in order for $\tilde V$ to decrease from one to zero, without becoming negative, we need that $m^2/H^2 \ge 1$ only for a limited interval of e-folds. This, in particular, means that $H$ cannot tend to zero in the far future, for large $N$, but it must increase in order to guarantee that $m^2/H^2 \ll 1$. On the basis of this argument, we can conclude that at late-times, when the matter is completely diluted:
\begin{equation}
	3\tilde V \sim -\frac{m^2U}{2H^2}\;,
\end{equation}
this however provided that $X$ does not diverge, otherwise we cannot neglect the product $XY$ in general.

Combining the two Friedmann equations, we have that:
\begin{equation}
	\xi = -3 + \frac{1}{\tilde V}\left[-Y + \frac{m^2(1 - U)}{2H^2} + \frac{\rho - P}{2M_{ Pl}^2H^2}\right]\;.
\end{equation}
For non-exotic fluids one has $\rho - P > 0$ and therefore we see that definitely $\xi > -3$. At late times, according to our previous discussion, we have that:
\begin{equation}
	\xi \sim -3 - \frac{1}{\tilde V}\frac{m^2U}{2H^2} \sim 0\;.
\end{equation}
Hence, the effective equation of state always tends to $-1$. We can check now directly that $X$ does not diverge from its solution, computed with $\xi = 0$. One obtains:
\begin{equation}
	X = -4\;,
\end{equation}
i.e. a $U$ growing indefinitely negative, say:
\begin{equation}
	U = C_1 - 4N\;.
\end{equation} 
 With this solution, we are left with two equations:
\begin{eqnarray}
Y + 3\tilde V = -\frac{m^2}{2H^2}U\;,\\ 
Y' + 3Y = -\frac{m^2}{H^2}\;.
\end{eqnarray}
The derivative of the left hand side of the first equation is equal to the left hand side of the second equation. Hence, one finds that:
\begin{equation}
	\xi = -\frac{3}{C_1 - 4N} = -\frac{3}{U}\;,
\end{equation}
and we have the solution for $H$:
\begin{equation}
	H = C_2|C_1 - 4N|^{3/4} = 3|U|^{3/4}\;.
\end{equation} The solution for $H'$ is instead:
\begin{equation}
	H' = -\frac{3C_2}{C_1 - 4N}|C_1 - 4N|^{3/4}\;.
\end{equation}

\subsection*{$\Box^{-2}R$ model}

This model was proposed in Ref. \cite{Amendola:2017qge} and is motivated from studies of nonperturbative lattice quantum gravity.
The Lagrangian is:
\begin{equation}
      \mathcal{L} = \mathcal{L}_{EH} -\frac{M^4}{6}\frac{1}{\Box^2}R \; ,
\end{equation}
and is localized introducing the auxiliary fields:
\begin{eqnarray}
\Box U = -R\; , \\
\Box S = -U \; , \\
\Box Q = -1\; ,  \\
\Box L = -Q \; .
\end{eqnarray}
The field equations are:
\begin{eqnarray}
        h^2=\frac{\gamma}{4}\left[V + WU + h^2\left( 6Z + 6Z' -U'Z' -V'W' \right)\right] + \Omega_R^0e^{-4N} + \Omega_M^0 e^{-3N} \; , \label{CosmoB1} \\
        \xi = \frac{1}{2\left(1 - \frac{3}{2}\gamma Z\right)}\left[\frac{-4\Omega_R^0e^{-4N}  -3\Omega_M^0 e^{-3N}}{h^2} + \frac{3}{2}\gamma \left(\frac{W}{h^2} -4Z' + U'Z' + V'W'\right)\right] \; , \label{Cosmob2}
        \end{eqnarray}
        \begin{eqnarray}
        U'' + \left(3 + \xi\right)U = 6\left(2 + \xi\right) \;, \label{KGBU} \\
        V'' + \left(3 + \xi\right)V'= \frac{U}{h^2}\;, \label{KGBV}\\
        W'' + \left(3 + \xi\right)W'= \frac{1}{h^2}\;, \label{KGBW}\\
        Z'' + \left(3 + \xi\right)Z'= \frac{W}{h^2}\;   \label{KGBZ}.
\end{eqnarray}
defining $\tilde{Z}= 1-\frac{3}{2}\gamma Z$ we can rewrite the late-times Friedmann equations \eqref{CosmoB1} \eqref{Cosmob2}, when matter is completely diluted, as:
\begin{eqnarray}
        \tilde{Z} &=& \frac{\gamma}{4h^2}\left(UW + V \right) -\tilde{Z}'\left(1 - \frac{U'}{6}\right) - \frac{\gamma W' V'}{4} \; , \label{CosmoB1A}\\
        \xi &=& \frac{1}{2\tilde{Z}}\left[\frac{3\gamma W}{2h^2} + 4\tilde{Z}' -\tilde{Z}'U' + \frac{3\gamma V'W'}{2} \right] \; \label{CosmoB2A} ,
\end{eqnarray}
while the KG equation \eqref{KGBZ} for $\tilde{Z}$ is:
\begin{equation}
    \tilde{Z}'' + \left(3+\xi \right)\tilde{Z}' = -\frac{3\gamma W}{2h^2} \; ,
\end{equation}
whose formal solution for $\tilde{Z}'$ is given by:
\begin{equation} \label{FormalZB}
 \tilde Z' = -3\gamma e^{-F(N)}\int_{N_i}^{N}d\Bar{N}e^{F(\Bar{N})}\frac{W}{2h^2}  \; .
\end{equation}
We write for convenience also the formal solutions for $V'$ and $W'$:
\begin{eqnarray}
          V' = e^{-F(N)}\int_{N_i}^{N}d\Bar{N}e^{F(\Bar{N})}\frac{U}{h^2}  \; , \label{FormalVB}\\
          W' = e^{-F(N)}\int_{N_i}^{N}d\Bar{N}e^{F(\Bar{N})}\frac{1}{h^2}  \; \label{FormalWB} .
\end{eqnarray}
Since we chose initial conditions $W_i = 0$ and since $W' > 0$ we can conclude that $W > 0$. This implies that $\tilde{Z}'< 0$; on the other hand Eqs. \eqref{Usol} imply $V'>0$, $0< U' <6$ and $U > 0$. 
In order to understand the behavior of $\tilde{Z}$ let us define the function $T$:
\begin{equation}
   T \equiv \frac{UW + V}{h^2} - W'V' \; ,
\end{equation}
 taking its time derivative and using Eqs. \eqref{KGBW} and \eqref{KGBV} we are able to set up a differential equation for $T$:
 \begin{equation}\label{diffeqx}
   T' + 2\xi X = \frac{U'W}{h^2} + 6V'W' \; .
 \end{equation}
 The formal solution of Eq. \eqref{diffeqx} is given by:
\begin{equation} \label{xsol}
    T\left(N\right) = \frac{1}{h^2(N)}\int_{N_i}^{N} d\Bar{N} \left(U'W + 6h^2V'W'\right) - C_X h^2(N) \; ,
\end{equation}
where $C_X$ is an integration constant.
 Since $X(N_i)= 0$ we can conclude from Eq. \eqref{xsol} that $T(N) > 0$.
 This in turns implies that the right hand side of Eq. \eqref{CosmoB1A} it's always positive, and we can conclude that, asymptotically, $\tilde Z > 0 $. 
 
Since $\tilde{Z}$ is positive definite, we must have asymptotically $\tilde{Z}' \rightarrow 0$. It is straightforward to realize from Eq. \eqref{FormalZB} that this is possible only if $h^2$ is a monotonic growing function that grows faster than $W/2$. On the other hand, $W$ is also a monotonic growing function since $W' > 0$. In particular, since $h^2$ grows faster than $W$, it also grows faster then a constant, and so we conclude from Eq. \eqref{FormalWB} that $W' \rightarrow 0$.
Using the latter in Eq. \eqref{CosmoB1A} we are left with: 
\begin{equation}
    \tilde{Z} \sim \frac{\gamma}{4h^2}\left(V + WU \right) \; .
\end{equation}
Using the above result in Eq. \eqref{CosmoB2A} we finally obtain:
\begin{equation}
    \xi \sim \frac{3}{\frac{V}{W} + U} \sim 0 \; ,
\end{equation}
then once again we have $w_{ eff} \rightarrow -1$.

\chapter{Field equations in Inverse Ricci gravity}
\label{Appendix C}
To begin with let us consider the following action:

Consider first the basic  Action 
\begin{equation}
S=\int\sqrt{-g}d^{4}x (R+\alpha A)\label{eq:first} \; ,
\end{equation}
where the anticurvature scalar $A$ is the trace of $A^{\mu\nu}$
\begin{equation}
A^{\mu\nu}=R_{\mu\nu}^{-1} \; .
\end{equation}
 By differentiating Eq. \eqref{Adefinition}, we see that 
\begin{equation}
\delta A^{\mu\tau}  =-A^{\mu\nu}(\delta R_{\nu\sigma})A{}^{\sigma\tau} \; .
\end{equation}
  We have then 
\begin{align}
\delta S & =\int d^{4}x(A\delta\sqrt{-g}+\sqrt{-g}A^{\mu\nu}\delta g_{\mu\nu}+\sqrt{-g}g_{\mu\nu}\delta A^{\mu\nu})\nonumber\\
 & =\int d^{4}x\sqrt{-g}(\frac{1}{2}Ag^{\mu\nu}\delta g_{\mu\nu}+A^{\mu\nu}\delta g_{\mu\nu}+g_{\mu\nu}\delta A^{\mu\nu}) \; ,
\end{align}
and since
\begin{equation}
\delta R_{\alpha\beta}=\nabla_{\rho}\delta\Gamma_{\beta\alpha}^{\rho}-\nabla_{\beta}\delta\Gamma_{\rho\alpha}^{\rho} \; ,
\end{equation}
we obtain 
\begin{align}
\delta A^{\mu\nu} & =-A^{\mu\alpha}(\nabla_{\rho}\delta\Gamma_{\beta\alpha}^{\rho}-\nabla_{\beta}\delta\Gamma_{\rho\alpha}^{\rho})A^{\beta\nu}\nonumber\\
 & =-\frac{1}{2}A^{\mu\alpha}(g^{\rho\lambda}\nabla_{\rho}(\nabla_{\alpha}\delta g_{\beta\lambda}+\nabla_{\beta}\delta g_{\lambda\alpha}-\nabla_{\lambda}\delta g_{\alpha\beta})-g^{\rho\lambda}\nabla_{\beta}(\nabla_{\alpha}\delta g_{\rho\lambda}+\nabla_{\rho}\delta g_{\lambda\alpha}-\nabla_{\lambda}\delta g_{\alpha\rho}))A^{\beta\nu}\nonumber\\
 & =-\frac{1}{2}A^{\mu\alpha}g^{\rho\lambda}(\nabla_{\rho}\nabla_{\alpha}\delta g_{\beta\lambda}-\nabla_{\rho}\nabla_{\lambda}\delta g_{\alpha\beta}-\nabla_{\beta}\nabla_{\alpha}\delta g_{\rho\lambda}+\nabla_{\beta}\nabla_{\lambda}\delta g_{\alpha\rho} + \left[\nabla_{\beta},\nabla_{\rho} \right]\delta g_{\lambda\alpha})A^{\beta\nu} \; .
\end{align}
Using integration by parts, this becomes
\begin{align}
g_{\mu\nu}\delta A^{\mu\nu} & =-\frac{1}{2}g_{\mu\nu}g^{\rho\lambda}(\delta g_{\beta\lambda}\nabla_{\alpha}\nabla_{\rho}(A^{\mu\alpha}A^{\beta\nu})-\delta g_{\alpha\beta}\nabla_{\lambda}\nabla_{\rho}(A^{\mu\alpha}A^{\beta\nu})-\delta g_{\rho\lambda}\nabla_{\alpha}\nabla_{\beta}(A^{\mu\alpha}A^{\beta\nu})+ \nonumber\\ 
&+\delta g_{\alpha\rho}\nabla_{\lambda}\nabla_{\beta}(A^{\mu\alpha}A^{\beta\nu})) +\delta g_{\alpha\rho}\nabla_{\lambda}\nabla_{\beta}(A^{\mu\alpha}A^{\beta\nu})) \nonumber \\
 & =\frac{1}{2}\delta g_{\iota\kappa}(-2g^{\rho\iota}\nabla_{\alpha}\nabla_{\rho}A^{\mu\alpha}A_{\mu}^{\kappa}+\nabla^{2}(A^{\mu\iota}A_{\mu}^{\kappa})+g^{\iota\kappa}\nabla_{\alpha}\nabla_{\beta}(A^{\mu\alpha}A_{\mu}^{\beta})) \; .
\end{align}
So finally the variation is 
\begin{align}
\delta g_{\mu\nu}(\frac{1}{2}Ag^{\mu\nu}+A^{\mu\nu}+\frac{1}{2}(-2g^{\rho\mu}\nabla_{\alpha}\nabla_{\rho}A^{\sigma\alpha}A_{\sigma}^{\nu}+\nabla^{2}(A^{\sigma\mu}A_{\sigma}^{\nu})+g^{\mu\nu}\nabla_{\alpha}\nabla_{\beta}(A^{\sigma\alpha}A_{\sigma}^{\beta}))) \; .
\end{align}
Together with the variation of the standard Hilbert-Einstein Lagrangian  
\begin{align}
\delta g^{\mu\nu}(-\frac{1}{2}Rg_{\mu\nu}+R_{\mu\nu}) & =-\delta g_{\mu\nu}(-\frac{1}{2}Rg^{\mu\nu}+R^{\mu\nu}) \; ,
\end{align}
we obtain finally the equations for the Action (\ref{eq:first})
\begin{align}
R^{\mu\nu}-\frac{1}{2}Rg^{\mu\nu}-\alpha A^{\mu\nu}-\frac{1}{2}\alpha Ag^{\mu\nu}+\frac{\alpha}{2}\left(2g^{\rho\mu}\nabla_{\alpha}\nabla_{\rho}A_{\sigma}^{\alpha}A^{\nu\sigma}-\nabla^{2}A_{\sigma}^{\mu}A^{\nu\sigma}-g^{\mu\nu}\nabla_{\alpha}\nabla_{\rho}A_{\sigma}^{\alpha}A^{\rho\sigma}\right) & =T^{\mu\nu}\label{eq:eom1} \; ,
\end{align}
where we used the fact that $A^{\alpha}_{\sigma }A^{\nu\sigma}=A^{\alpha\tau}g_{\tau\sigma}A^{\sigma\nu}=A^{\alpha\tau}A_{\tau}^{\nu}=A^{\alpha\sigma}A_{\sigma}^{\nu}=A_{\sigma}^{\nu}A^{\alpha\sigma}$ and we employed units in which $8\pi G=1$. It can be show that the left-hand side of Eq.~\eqref{eq:eom1} is divergenceless, as it should be in order to satisfy the Bianchi identities.

The extension to any Lagrangian $f(R,A)$ is quite straightforward: 
\begin{equation}
\delta S=\int d^{4}x\sqrt{-g}(-\frac{1}{2}f(R,A)g_{\mu\nu}\delta g^{\mu\nu}+f_{A}A^{\mu\nu}\delta g_{\mu\nu}+f_{A}g_{\mu\nu}\delta A^{\mu\nu}+f_{R}R_{\mu\nu}\delta g^{\mu\nu}+f_{R}g^{\mu\nu}\delta R_{\mu\nu}) \; ,
\end{equation}
where $f_R=\partial f/\partial R$ and $f_A=\partial f/\partial A$. Then we have
\begin{align}
f_{R}R^{\mu\nu}-f_{A}A^{\mu\nu}&-\frac{1}{2}fg^{\mu\nu}+g^{\rho\mu}\nabla_{\alpha}\nabla_{\rho}f_{A}A^{\alpha}_{\sigma}A^{\nu\sigma}-\frac{1}{2}\nabla^{2}(f_{A}A_{\sigma}^{\mu}A^{\nu\sigma}) &\nonumber\\&-\frac{1}{2}g^{\mu\nu}\nabla_{\alpha}\nabla_{\beta}(f_{A}A_{\sigma}^{\alpha}A^{\beta\sigma})-\nabla^{\mu}\nabla^{\nu}f_{R}+g^{\mu\nu}\nabla^{2}f_{R}  =T^{\mu\nu} \; .
\end{align}

\chapter{Estimating Time delay Uncertainties with PyCS3}
\label{AppendixD}

\begin{figure}[h!]
    \centering
    \includegraphics[scale=0.4]{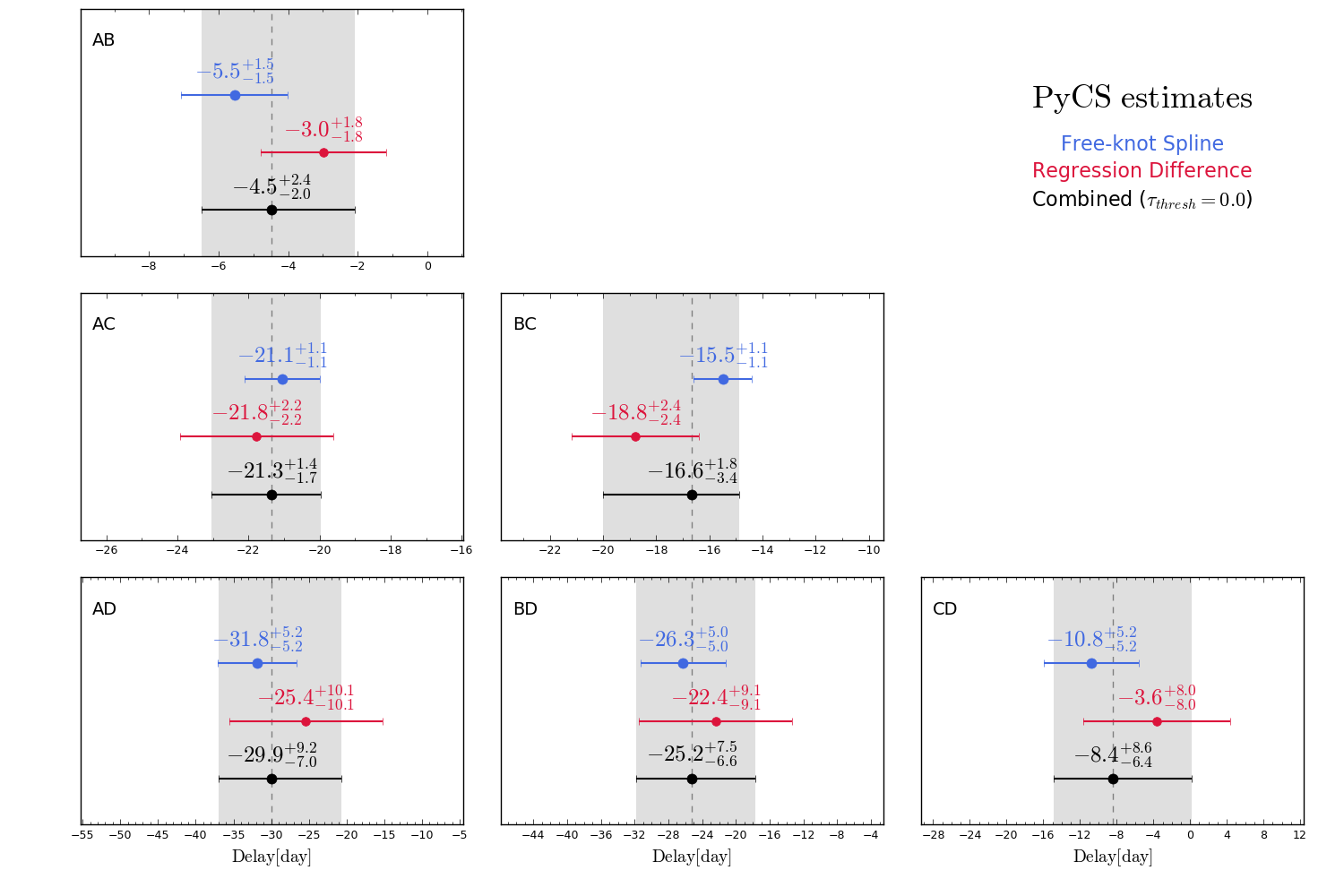}
    \includegraphics[scale=0.45]{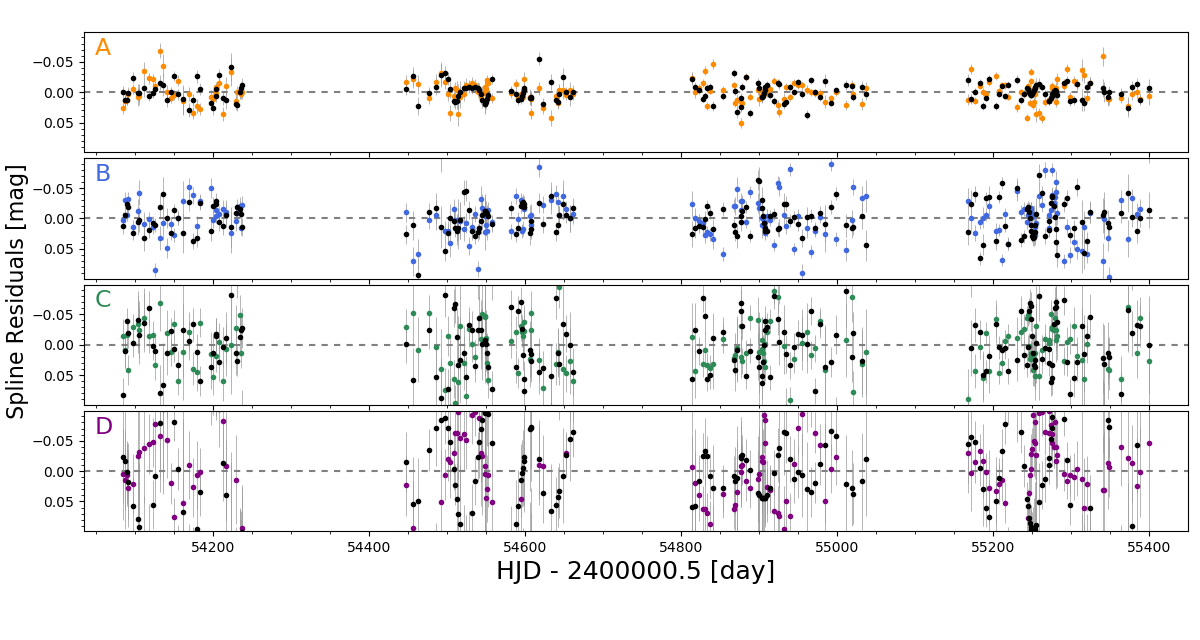}
    \includegraphics[scale=0.45]{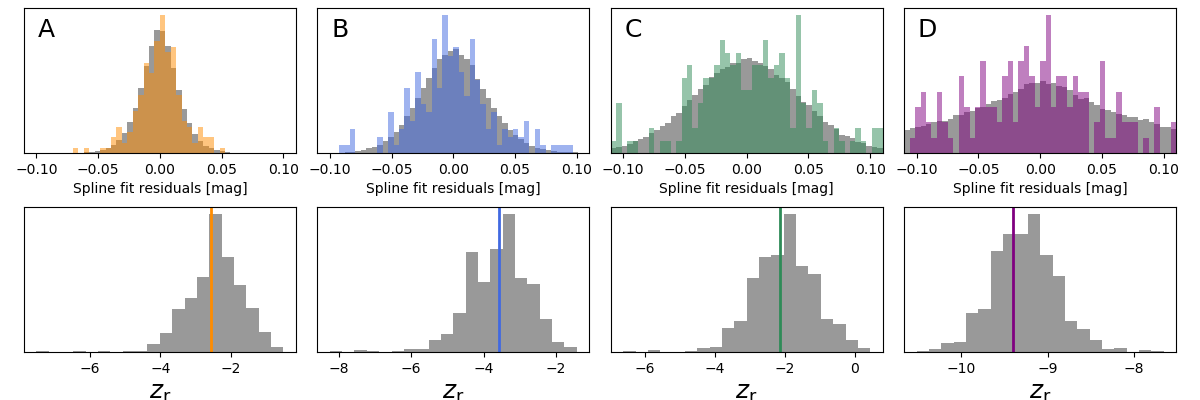}
    \caption{Time delay estimates for the simulated quasar over the full observation period.}
    \label{fig:simul}
\end{figure}

\begin{figure}[htp!]
    \centering
    \includegraphics[scale=0.4]{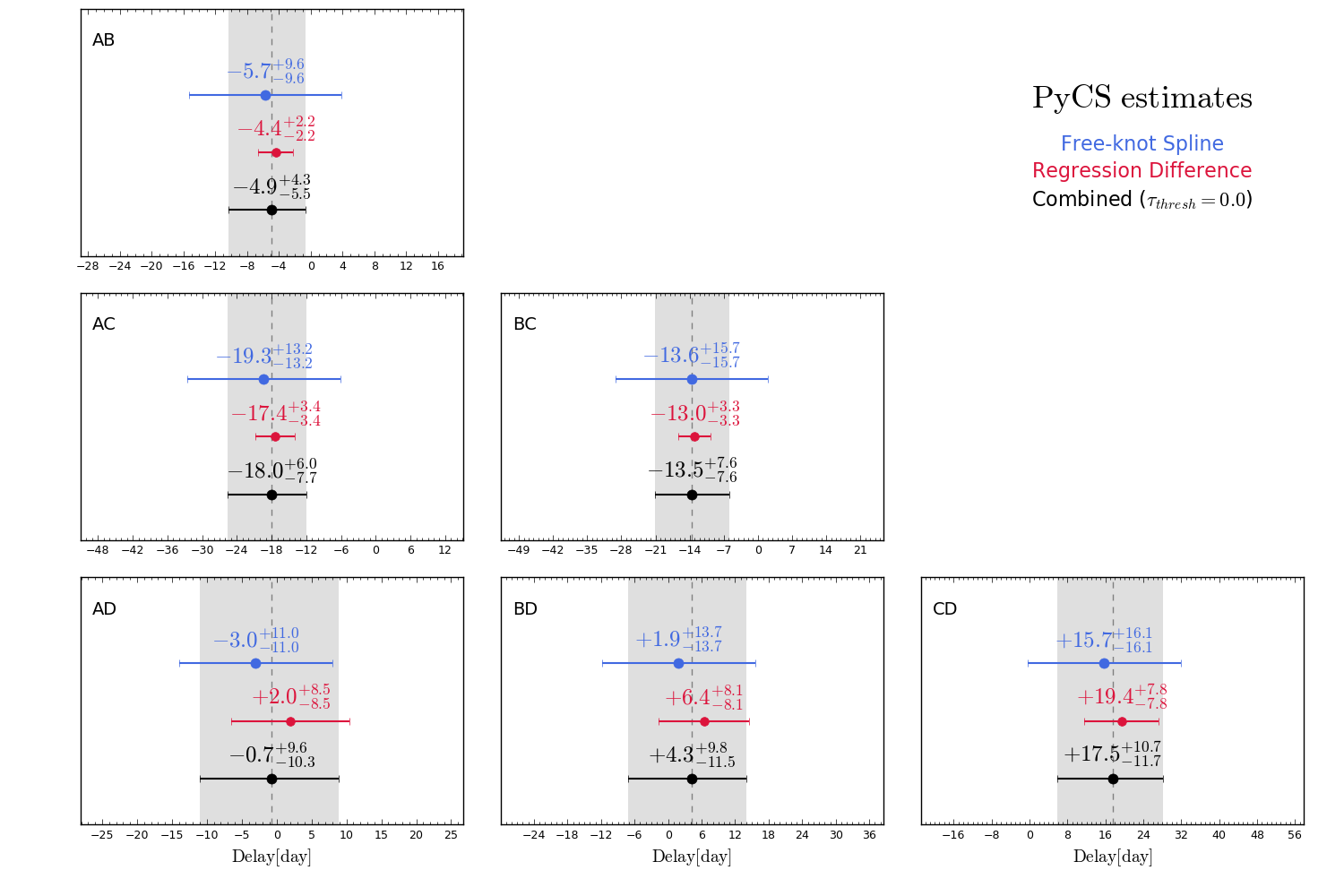}
    \includegraphics[scale=0.4]{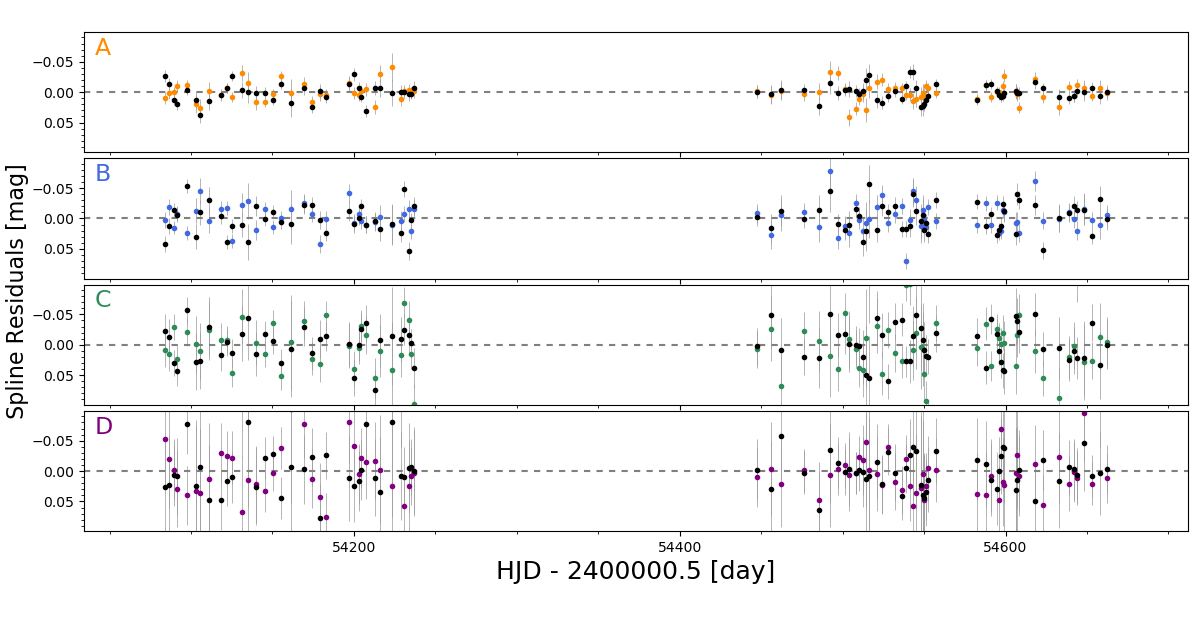}
    \includegraphics[scale=0.4]{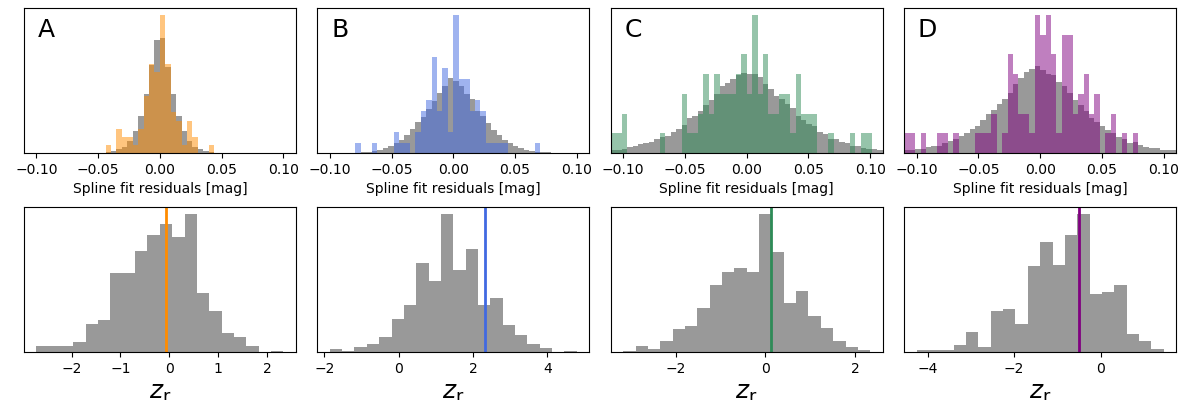}
    \caption{Time delay estimates for the simulated quasar over half of the observation period.}
    \label{fig:simulhalf}
\end{figure}

\begin{figure}[htp!]
    \centering
    \includegraphics[scale=0.4]{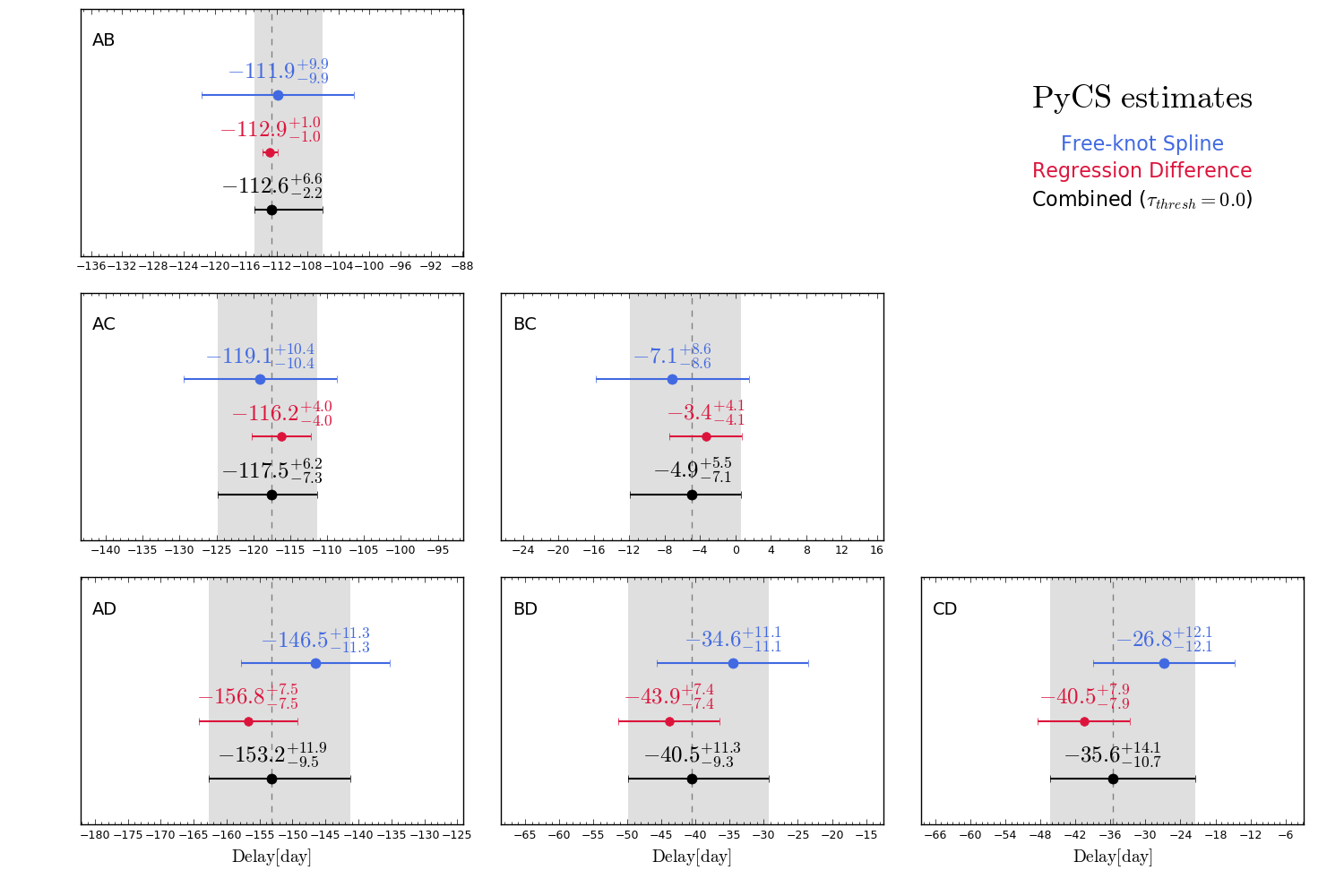}
    \includegraphics[scale=0.4]{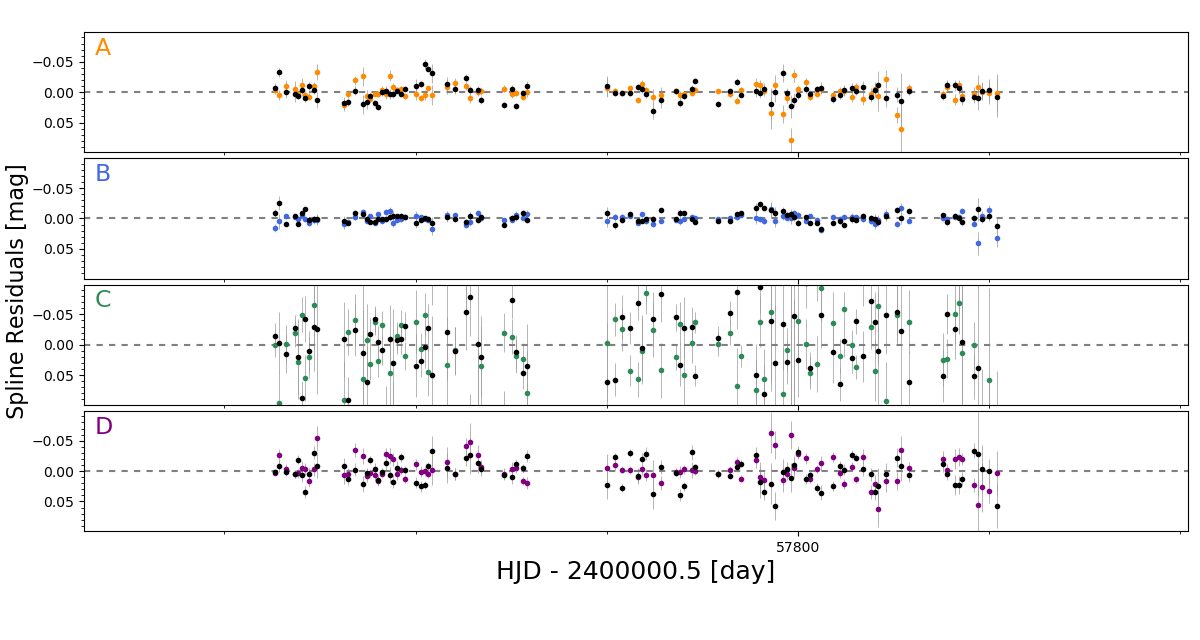}
    \includegraphics[scale=0.4]{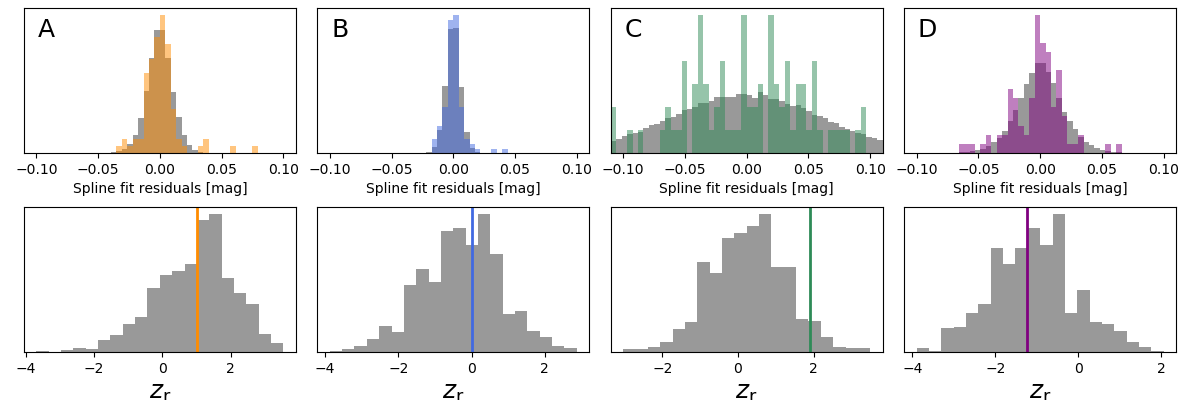}
    \caption{Time delay estimates for the quasar DES J04078-5354 over the full observation period.}
    \label{fig:des}
\end{figure}

\begin{figure}[htp!]
    \centering
    \includegraphics[scale=0.4]{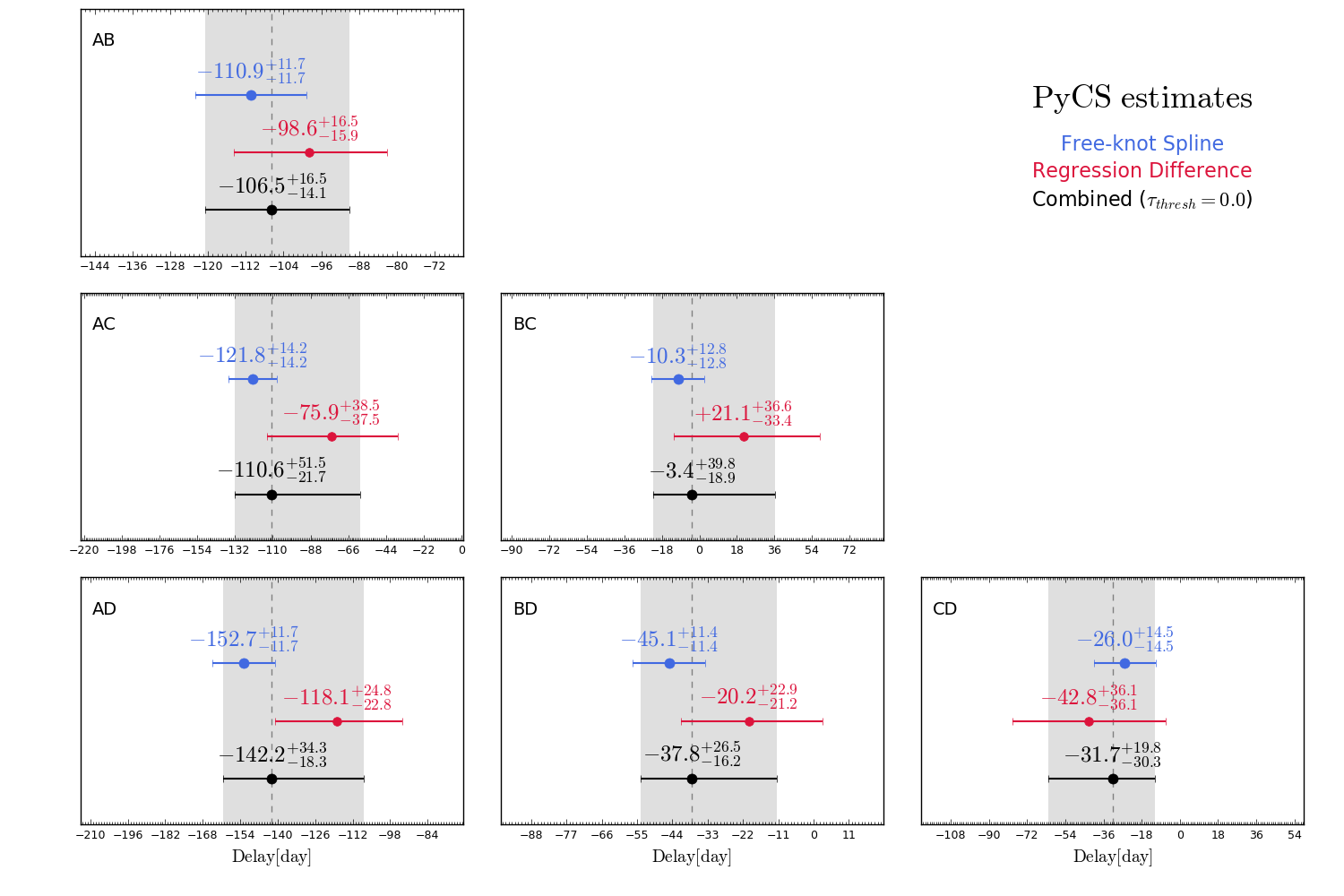}
    \includegraphics[scale=0.4]{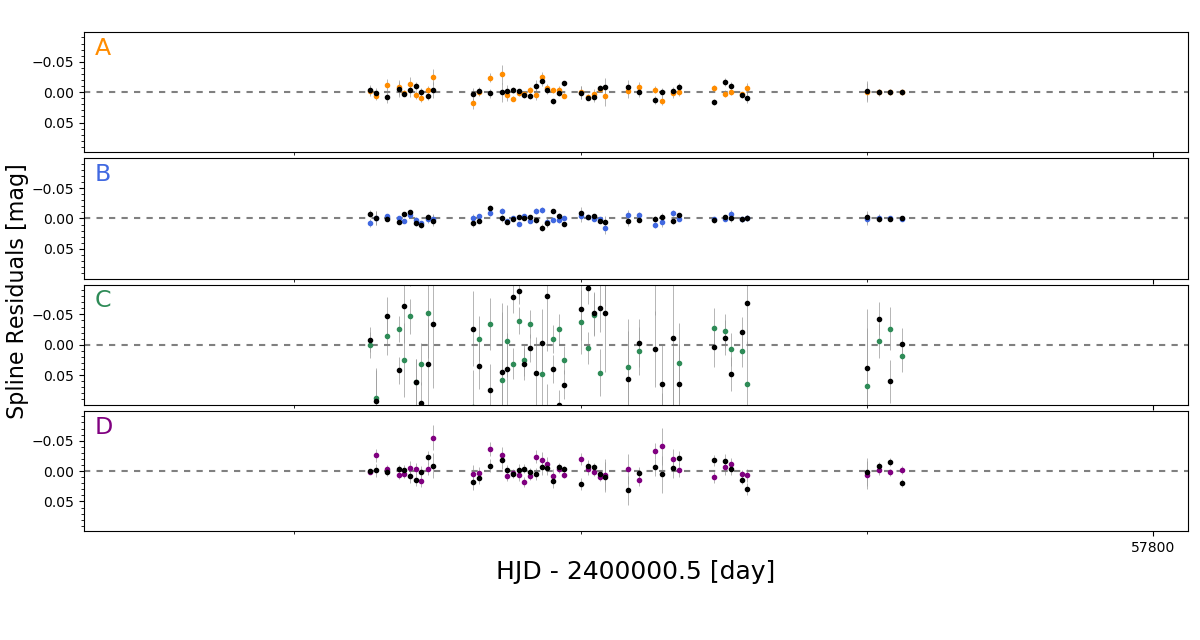}
    \includegraphics[scale=0.4]{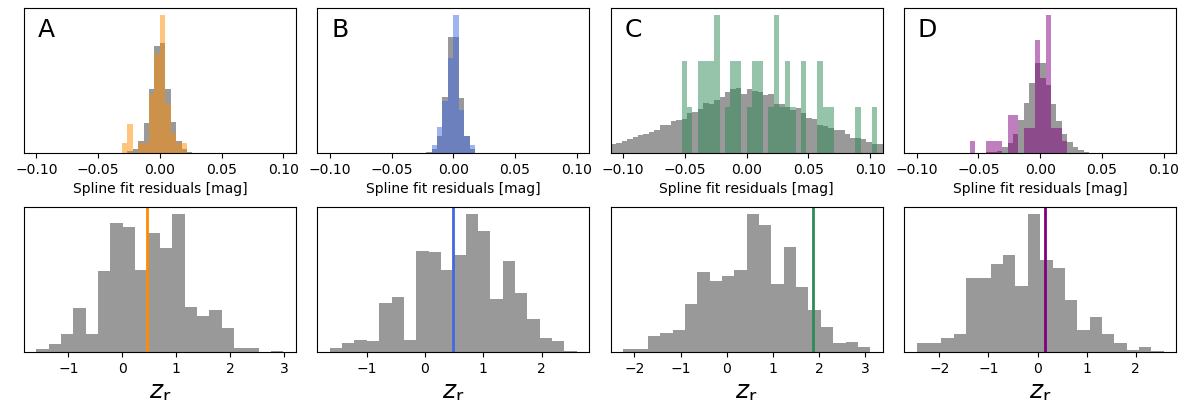}
    \caption{Time delay estimates for the quasar DES J04078-5354 over half of the observation period.}
    \label{fig:deshalf}
\end{figure}

In this appendix, we display the time delays and their uncertainties obtained with PyCS3. Uncertainties are obtained by simulating light curves close to the data, and randomizing the true time delay applied to each curve, see Sec.~(3.2) of Ref.~\cite{Millon:2020xab} for more details. The final marginalization is done performing a hybrid approach between the ``free-knot spline" and the ``regression difference" estimators as explained in Sec.~(3.3) of Ref.~\cite{Millon:2020xab}. The parameter $\tau_{thresh}=0$ indicates the marginalization is done over the two estimators. In each figure, the top panels show the final time delay estimates marginalizing over the two estimators. The middle figure shows the residuals for the spline fit to the data. The top row of the bottom panels show the distribution of data residuals for mock curves (in gray) and data (in colors), whereas the bottom panels show their normalization over the number of runs $z_r$. The time delay estimates for the simulated curves over the whole period of observations ($\approx 1.316$ days) is shown in Fig.~\ref{fig:simul} while Fig.~\ref{fig:simulhalf} is over half the total period. Similarly, Fig.~\ref{fig:des} ($\approx 189$ days) and Fig.~\ref{fig:deshalf} show time delays for the object DES J0408-5354.

\pagebreak
% ATTENTION: check if the file names match those in the \includeonly command.
% All the files must be saved with ".tex", but inside \include the extensions are omitted.

%%%%%%%%%%%%%%%%%%%%%%%%% BIBLIOGRAPHY %%%%%%%%%%%%%%%%%%%%%%%%%%

\pagebreak
\pagestyle{biblioHeader}
{
	\hypersetup{urlcolor=darkblue} 
	\addcontentsline{toc}{chapter}{References}
	\bibliography{thesis.bib}  % put the name of your bib file here.

\begin{thebibliography}{100}

\bibitem{Riess:1998cb}
A.~G. Riess et~al.
\newblock {\em {Observational evidence from supernovae for an accelerating
  universe and a cosmological constant}}.
\newblock
  {\hypersetup{urlcolor=journalcolor}\href{https://doi.org/10.1086/300499}
  {Astron. J.}} {\bfseries 116} 1009 (1998).
\newblock \href {http://arxiv.org/abs/astro-ph/9805201}
  {\path{arXiv:astro-ph/9805201}}.

\bibitem{Perlmutter:1998np}
S. Perlmutter et~al.
\newblock {\em {Measurements of Omega and Lambda from 42 high redshift
  supernovae}}.
\newblock
  {\hypersetup{urlcolor=journalcolor}\href{https://doi.org/10.1086/307221}
  {Astrophys. J.}} {\bfseries 517} 565 (1999).
\newblock \href {http://arxiv.org/abs/astro-ph/9812133}
  {\path{arXiv:astro-ph/9812133}}.

\bibitem{Vardanyan:2017kal}
V. Vardanyan, Y. Akrami, L. Amendola, and A. Silvestri.
\newblock {\em {On nonlocally interacting metrics, and a simple proposal for
  cosmic acceleration}}.
\newblock
  {\hypersetup{urlcolor=journalcolor}\href{https://doi.org/10.1088/1475-7516/2018/03/048}
  {JCAP}} {\bfseries 03} 048 (2018).
\newblock \href {http://arxiv.org/abs/1702.08908} {\path{arXiv:1702.08908}}.

\bibitem{Giani:2019vjf}
L. Giani, T. Miranda, and O.~F. Piattella.
\newblock {\em {Cosmology and Newtonian limit in a model of gravity with
  nonlocally interacting metrics}}.
\newblock
  {\hypersetup{urlcolor=journalcolor}\href{https://doi.org/10.1016/j.dark.2019.100357}
  {Phys. Dark Univ.}} {\bfseries 26} 100357 (2019).
\newblock \href {http://arxiv.org/abs/1905.02720} {\path{arXiv:1905.02720}}.

\bibitem{Giani:2019xjf}
L. Giani and O.~F. Piattella.
\newblock {\em {Late-times asymptotic equation of state for a class of nonlocal
  theories of gravity}}.
\newblock
  {\hypersetup{urlcolor=journalcolor}\href{https://doi.org/10.1103/PhysRevD.100.123508}
  {Phys. Rev. D}} {\bfseries 100} 123508 (2019).
\newblock \href {http://arxiv.org/abs/1906.10480} {\path{arXiv:1906.10480}}.

\bibitem{Amendola:2020qho}
L. Amendola, L. Giani, and G. Laverda.
\newblock {\em {Ricci-inverse gravity: a novel alternative gravity, its flaws,
  and how to cure them}}.
\newblock {\hypersetup{urlcolor=journalcolor}\href{https://doi.org/} {2020}}).
\newblock \href {http://arxiv.org/abs/2006.04209} {\path{arXiv:2006.04209}}.

\bibitem{Piattella:2017uat}
O.~F. Piattella and L. Giani.
\newblock {\em {Redshift drift of gravitational lensing}}.
\newblock
  {\hypersetup{urlcolor=journalcolor}\href{https://doi.org/10.1103/PhysRevD.95.101301}
  {Phys. Rev. D}} {\bfseries 95} 101301 (2017).
\newblock \href {http://arxiv.org/abs/1703.05142} {\path{arXiv:1703.05142}}.

\bibitem{Giani:2020fpz}
L. Giani and E. Frion.
\newblock {\em {Testing the Equivalence Principle with Strong Lensing Time
  Delay Variations}}.
\newblock
  {\hypersetup{urlcolor=journalcolor}\href{https://doi.org/10.1088/1475-7516/2020/09/008}
  {JCAP}} {\bfseries 09} 008 (2020).
\newblock \href {http://arxiv.org/abs/2005.07533} {\path{arXiv:2005.07533}}.

\bibitem{Piattella:2018hvi}
O.~F. Piattella.
\newblock {\em {Lecture Notes in Cosmology}}.
\newblock UNITEXT for Physics. Springer Cham 2018.
\newblock \href {http://arxiv.org/abs/1803.00070} {\path{arXiv:1803.00070}}.

\bibitem{Amendola:2015ksp}
L. Amendola and S. Tsujikawa.
\newblock {\em {Dark Energy}: {Theory and Observations}}.
\newblock Cambridge University Press 2015.

\bibitem{Weinberg:1972kfs}
S. Weinberg.
\newblock {\em {Gravitation and Cosmology}: {Principles and Applications of the
  General Theory of Relativity}}.
\newblock John Wiley and Sons New York 1972.

\bibitem{BeltranJimenez:2019tjy}
J.~B. Jiménez, L. Heisenberg, and T.~S. Koivisto.
\newblock {\em {The Geometrical Trinity of Gravity}}.
\newblock
  {\hypersetup{urlcolor=journalcolor}\href{https://doi.org/10.3390/universe5070173}
  {Universe}} {\bfseries 5} 173 (2019).
\newblock \href {http://arxiv.org/abs/1903.06830} {\path{arXiv:1903.06830}}.

\bibitem{Landau:1982dva}
L. Landau and E. Lifschits.
\newblock {\em {The Classical Theory of Fields}} volume Volume 2 of {\em Course
  of Theoretical Physics}.
\newblock Pergamon Press Oxford 1975.

\bibitem{Friedmann1922}
A. {Friedmann}.
\newblock {\em {{\"U}ber die Kr{\"u}mmung des Raumes}}.
\newblock
  {\hypersetup{urlcolor=journalcolor}\href{https://doi.org/10.1007/BF01332580}
  {Zeitschrift fur Physik}} {\bfseries 10} 377 (1922).

\bibitem{Lematre1927}
G. {Lema{\^\i}tre}.
\newblock {\em {Un Univers homog{\`e}ne de masse constante et de rayon
  croissant rendant compte de la vitesse radiale des n{\'e}buleuses
  extra-galactiques}}.
\newblock {\hypersetup{urlcolor=journalcolor}\href{https://doi.org/} {Annales
  de la Soci\&eacute;t\&eacute; Scientifique de Bruxelles}} {\bfseries 47} 49
  (1927).

\bibitem{Robertson}
H.~P. {Robertson}.
\newblock {\em {Kinematics and World-Structure}}.
\newblock
  {\hypersetup{urlcolor=journalcolor}\href{https://doi.org/10.1086/143681}
  {apj}} {\bfseries 82} 284 (1935).

\bibitem{Walker}
A.~G. {Walker}.
\newblock {\em {On Milne's Theory of World-Structure}}.
\newblock
  {\hypersetup{urlcolor=journalcolor}\href{https://doi.org/10.1112/plms/s2-42.1.90}
  {Proceedings of the London Mathematical Society}} {\bfseries 42} 90 (1937).

\bibitem{Wald:1984rg}
R.~M. Wald.
\newblock {\em {General Relativity}}.
\newblock Chicago Univ. Pr. Chicago, USA 1984.

\bibitem{Aghanim:2018eyx}
N. Aghanim et~al.
\newblock {\em {Planck 2018 results. VI. Cosmological parameters}}.
\newblock
  {\hypersetup{urlcolor=journalcolor}\href{https://doi.org/10.1051/0004-6361/201833910}
  {2018}}).
\newblock \href {http://arxiv.org/abs/1807.06209} {\path{arXiv:1807.06209}}.

\bibitem{Abbott:2018wog}
T. Abbott et~al.
\newblock {\em {First Cosmology Results using Type Ia Supernovae from the Dark
  Energy Survey: Constraints on Cosmological Parameters}}.
\newblock
  {\hypersetup{urlcolor=journalcolor}\href{https://doi.org/10.3847/2041-8213/ab04fa}
  {Astrophys. J. Lett.}} {\bfseries 872} L30 (2019).
\newblock \href {http://arxiv.org/abs/1811.02374} {\path{arXiv:1811.02374}}.

\bibitem{Scolnic:2017caz}
D. Scolnic et~al.
\newblock {\em {The Complete Light-curve Sample of Spectroscopically Confirmed
  SNe Ia from Pan-STARRS1 and Cosmological Constraints from the Combined
  Pantheon Sample}}.
\newblock
  {\hypersetup{urlcolor=journalcolor}\href{https://doi.org/10.3847/1538-4357/aab9bb}
  {Astrophys. J.}} {\bfseries 859} 101 (2018).
\newblock \href {http://arxiv.org/abs/1710.00845} {\path{arXiv:1710.00845}}.

\bibitem{Trenti:2015zja}
M. Trenti, P. Padoan, and R. Jimenez.
\newblock {\em {The Relative and Absolute Ages of old Globular Clusters in the
  Lcdm Framework}}.
\newblock
  {\hypersetup{urlcolor=journalcolor}\href{https://doi.org/10.1088/2041-8205/808/2/L35}
  {Astrophys. J. Lett.}} {\bfseries 808} L35 (2015).
\newblock \href {http://arxiv.org/abs/1502.02670} {\path{arXiv:1502.02670}}.

\bibitem{Navarro:1996gj}
J.~F. Navarro, C.~S. Frenk, and S.~D. White.
\newblock {\em {A Universal density profile from hierarchical clustering}}.
\newblock
  {\hypersetup{urlcolor=journalcolor}\href{https://doi.org/10.1086/304888}
  {Astrophys. J.}} {\bfseries 490} 493 (1997).
\newblock \href {http://arxiv.org/abs/astro-ph/9611107}
  {\path{arXiv:astro-ph/9611107}}.

\bibitem{1948PhRv...74..505G}
G. {Gamow}.
\newblock {\em {The Origin of Elements and the Separation of Galaxies}}.
\newblock
  {\hypersetup{urlcolor=journalcolor}\href{https://doi.org/10.1103/PhysRev.74.505.2}
  {Physical Review}} {\bfseries 74} 505 (1948).

\bibitem{Alpher:1948ve}
R. Alpher, H. Bethe, and G. Gamow.
\newblock {\em {The origin of chemical elements}}.
\newblock
  {\hypersetup{urlcolor=journalcolor}\href{https://doi.org/10.1103/PhysRev.73.803}
  {Phys. Rev.}} {\bfseries 73} 803 (1948).

\bibitem{Penzias:1965wn}
A.~A. Penzias and R.~W. Wilson.
\newblock {\em {A Measurement of excess antenna temperature at 4080-Mc/s}}.
\newblock
  {\hypersetup{urlcolor=journalcolor}\href{https://doi.org/10.1086/148307}
  {Astrophys. J.}} {\bfseries 142} 419 (1965).

\bibitem{Ade:2015xua}
P. Ade et~al.
\newblock {\em {Planck 2015 results. XIII. Cosmological parameters}}.
\newblock
  {\hypersetup{urlcolor=journalcolor}\href{https://doi.org/10.1051/0004-6361/201525830}
  {Astron. Astrophys.}} {\bfseries 594} A13 (2016).
\newblock \href {http://arxiv.org/abs/1502.01589} {\path{arXiv:1502.01589}}.

\bibitem{Lesgourgues:2018ncw}
J. Lesgourgues, G. Mangano, G. Miele, and S. Pastor.
\newblock {\em {Neutrino Cosmology}}.
\newblock Cambridge University Press 2013.

\bibitem{Lesgourgues:2013qba}
J. Lesgourgues.
\newblock {\em {Cosmological Perturbations}}.
\newblock In {\em {Theoretical Advanced Study Institute in Elementary Particle
  Physics}: {Searching for New Physics at Small and Large Scales}} pages 29--97
  2013.
\newblock \href {http://arxiv.org/abs/1302.4640} {\path{arXiv:1302.4640}}.

\bibitem{Weinberg:1988cp}
S. Weinberg.
\newblock {\em {The Cosmological Constant Problem}}.
\newblock
  {\hypersetup{urlcolor=journalcolor}\href{https://doi.org/10.1103/RevModPhys.61.1}
  {Rev. Mod. Phys.}} {\bfseries 61} 1 (1989).

\bibitem{Martin:2012bt}
J. Martin.
\newblock {\em {Everything You Always Wanted To Know About The Cosmological
  Constant Problem (But Were Afraid To Ask)}}.
\newblock
  {\hypersetup{urlcolor=journalcolor}\href{https://doi.org/10.1016/j.crhy.2012.04.008}
  {Comptes Rendus Physique}} {\bfseries 13} 566 (2012).
\newblock \href {http://arxiv.org/abs/1205.3365} {\path{arXiv:1205.3365}}.

\bibitem{Padilla:2015aaa}
A. Padilla.
\newblock {\em {Lectures on the Cosmological Constant Problem}}.
\newblock {\hypersetup{urlcolor=journalcolor}\href{https://doi.org/} {2015}}).
\newblock \href {http://arxiv.org/abs/1502.05296} {\path{arXiv:1502.05296}}.

\bibitem{Velten:2014nra}
H. Velten, R. vom Marttens, and W. Zimdahl.
\newblock {\em {Aspects of the cosmological coincidence problem}}.
\newblock
  {\hypersetup{urlcolor=journalcolor}\href{https://doi.org/10.1140/epjc/s10052-014-3160-4}
  {Eur. Phys. J. C}} {\bfseries 74} 3160 (2014).
\newblock \href {http://arxiv.org/abs/1410.2509} {\path{arXiv:1410.2509}}.

\bibitem{Bullock:2017xww}
J.~S. Bullock and M. Boylan-Kolchin.
\newblock {\em {Small-Scale Challenges to the $\Lambda$CDM Paradigm}}.
\newblock
  {\hypersetup{urlcolor=journalcolor}\href{https://doi.org/10.1146/annurev-astro-091916-055313}
  {Ann. Rev. Astron. Astrophys.}} {\bfseries 55} 343 (2017).
\newblock \href {http://arxiv.org/abs/1707.04256} {\path{arXiv:1707.04256}}.

\bibitem{Kim:2017iwr}
S.~Y. Kim, A.~H.~G. Peter, and J.~R. Hargis.
\newblock {\em {Missing Satellites Problem: Completeness Corrections to the
  Number of Satellite Galaxies in the Milky Way are Consistent with Cold Dark
  Matter Predictions}}.
\newblock
  {\hypersetup{urlcolor=journalcolor}\href{https://doi.org/10.1103/PhysRevLett.121.211302}
  {Phys. Rev. Lett.}} {\bfseries 121} 211302 (2018).
\newblock \href {http://arxiv.org/abs/1711.06267} {\path{arXiv:1711.06267}}.

\bibitem{Handley:2019tkm}
W. Handley.
\newblock {\em {Curvature tension: evidence for a closed universe}}.
\newblock {\hypersetup{urlcolor=journalcolor}\href{https://doi.org/} {2019}}).
\newblock \href {http://arxiv.org/abs/1908.09139} {\path{arXiv:1908.09139}}.

\bibitem{DiValentino:2019qzk}
E. Di~Valentino, A. Melchiorri, and J. Silk.
\newblock {\em {Planck evidence for a closed Universe and a possible crisis for
  cosmology}}.
\newblock
  {\hypersetup{urlcolor=journalcolor}\href{https://doi.org/10.1038/s41550-019-0906-9}
  {Nature Astron.}} {\bfseries 4} 196 (2019).
\newblock \href {http://arxiv.org/abs/1911.02087} {\path{arXiv:1911.02087}}.

\bibitem{Bernal:2016gxb}
J.~L. Bernal, L. Verde, and A.~G. Riess.
\newblock {\em {The trouble with $H_0$}}.
\newblock
  {\hypersetup{urlcolor=journalcolor}\href{https://doi.org/10.1088/1475-7516/2016/10/019}
  {JCAP}} {\bfseries 10} 019 (2016).
\newblock \href {http://arxiv.org/abs/1607.05617} {\path{arXiv:1607.05617}}.

\bibitem{Verde:2019ivm}
L. Verde, T. Treu, and A. Riess.
\newblock {\em {Tensions between the Early and the Late Universe}}.
\newblock 2019.
\newblock \href {http://arxiv.org/abs/1907.10625} {\path{arXiv:1907.10625}}.

\bibitem{Martinelli:2019krf}
M. Martinelli and I. Tutusaus.
\newblock {\em {CMB tensions with low-redshift $H_0$ and $S_8$ measurements:
  impact of a redshift-dependent type-Ia supernovae intrinsic luminosity}}.
\newblock
  {\hypersetup{urlcolor=journalcolor}\href{https://doi.org/10.3390/sym11080986}
  {Symmetry}} {\bfseries 11} 986 (2019).
\newblock \href {http://arxiv.org/abs/1906.09189} {\path{arXiv:1906.09189}}.

\bibitem{DiValentino:2017oaw}
E. Di~Valentino, C. Bøehm, E. Hivon, and F.~R. Bouchet.
\newblock {\em {Reducing the $H_0$ and $\sigma_8$ tensions with Dark
  Matter-neutrino interactions}}.
\newblock
  {\hypersetup{urlcolor=journalcolor}\href{https://doi.org/10.1103/PhysRevD.97.043513}
  {Phys. Rev. D}} {\bfseries 97} 043513 (2018).
\newblock \href {http://arxiv.org/abs/1710.02559} {\path{arXiv:1710.02559}}.

\bibitem{Sola:2018sjf}
J. Sola~Peracaula, A. Gomez-Valent, and J. de Cruz~Pérez.
\newblock {\em {Signs of Dynamical Dark Energy in Current Observations}}.
\newblock
  {\hypersetup{urlcolor=journalcolor}\href{https://doi.org/10.1016/j.dark.2019.100311}
  {Phys. Dark Univ.}} {\bfseries 25} 100311 (2019).
\newblock \href {http://arxiv.org/abs/1811.03505} {\path{arXiv:1811.03505}}.

\bibitem{Barros:2018efl}
B.~J. Barros, L. Amendola, T. Barreiro, and N.~J. Nunes.
\newblock {\em {Coupled quintessence with a $\Lambda$CDM background: removing
  the $\sigma_8$ tension}}.
\newblock
  {\hypersetup{urlcolor=journalcolor}\href{https://doi.org/10.1088/1475-7516/2019/01/007}
  {JCAP}} {\bfseries 01} 007 (2019).
\newblock \href {http://arxiv.org/abs/1802.09216} {\path{arXiv:1802.09216}}.

\bibitem{Sola:2019jek}
J. Solà~Peracaula, A. Gomez-Valent, J. de Cruz~Pérez, and C. Moreno-Pulido.
\newblock {\em {Brans--Dicke Gravity with a Cosmological Constant Smoothes Out
  $\Lambda$CDM Tensions}}.
\newblock
  {\hypersetup{urlcolor=journalcolor}\href{https://doi.org/10.3847/2041-8213/ab53e9}
  {Astrophys. J. Lett.}} {\bfseries 886} L6 (2019).
\newblock \href {http://arxiv.org/abs/1909.02554} {\path{arXiv:1909.02554}}.

\bibitem{Sola:2020lba}
J. Sola, A. Gomez-Valent, J.~d.~C. Perez, and C. Moreno-Pulido.
\newblock {\em {Brans-Dicke cosmology with a $\Lambda$- term: a possible
  solution to $\Lambda$CDM tensions}}.
\newblock {\hypersetup{urlcolor=journalcolor}\href{https://doi.org/} {2020}}).
\newblock \href {http://arxiv.org/abs/2006.04273} {\path{arXiv:2006.04273}}.

\bibitem{Profumo:2017hqp}
S. Profumo.
\newblock {\em {An Introduction to Particle Dark Matter}}.
\newblock World Scientific 2017.

\bibitem{Lovelock:1971yv}
D. Lovelock.
\newblock {\em {The Einstein tensor and its generalizations}}.
\newblock
  {\hypersetup{urlcolor=journalcolor}\href{https://doi.org/10.1063/1.1665613}
  {J. Math. Phys.}} {\bfseries 12} 498 (1971).

\bibitem{Lovelock:1972vz}
D. Lovelock.
\newblock {\em {The four-dimensionality of space and the einstein tensor}}.
\newblock
  {\hypersetup{urlcolor=journalcolor}\href{https://doi.org/10.1063/1.1666069}
  {J. Math. Phys.}} {\bfseries 13} 874 (1972).

\bibitem{Clifton:2011jh}
T. Clifton, P.~G. Ferreira, A. Padilla, and C. Skordis.
\newblock {\em {Modified Gravity and Cosmology}}.
\newblock
  {\hypersetup{urlcolor=journalcolor}\href{https://doi.org/10.1016/j.physrep.2012.01.001}
  {Phys. Rept.}} {\bfseries 513} 1 (2012).
\newblock \href {http://arxiv.org/abs/1106.2476} {\path{arXiv:1106.2476}}.

\bibitem{Horndeski:1974wa}
G.~W. Horndeski.
\newblock {\em {Second-order scalar-tensor field equations in a
  four-dimensional space}}.
\newblock
  {\hypersetup{urlcolor=journalcolor}\href{https://doi.org/10.1007/BF01807638}
  {Int. J. Theor. Phys.}} {\bfseries 10} 363 (1974).

\bibitem{Deffayet:2009mn}
C. Deffayet, S. Deser, and G. Esposito-Farese.
\newblock {\em {Generalized Galileons: All scalar models whose curved
  background extensions maintain second-order field equations and
  stress-tensors}}.
\newblock
  {\hypersetup{urlcolor=journalcolor}\href{https://doi.org/10.1103/PhysRevD.80.064015}
  {Phys. Rev. D}} {\bfseries 80} 064015 (2009).
\newblock \href {http://arxiv.org/abs/0906.1967} {\path{arXiv:0906.1967}}.

\bibitem{Kobayashi:2011nu}
T. Kobayashi, M. Yamaguchi, and J. Yokoyama.
\newblock {\em {Generalized G-inflation: Inflation with the most general
  second-order field equations}}.
\newblock
  {\hypersetup{urlcolor=journalcolor}\href{https://doi.org/10.1143/PTP.126.511}
  {Prog. Theor. Phys.}} {\bfseries 126} 511 (2011).
\newblock \href {http://arxiv.org/abs/1105.5723} {\path{arXiv:1105.5723}}.

\bibitem{Heisenberg:2014rta}
L. Heisenberg.
\newblock {\em {Generalization of the Proca Action}}.
\newblock
  {\hypersetup{urlcolor=journalcolor}\href{https://doi.org/10.1088/1475-7516/2014/05/015}
  {JCAP}} {\bfseries 05} 015 (2014).
\newblock \href {http://arxiv.org/abs/1402.7026} {\path{arXiv:1402.7026}}.

\bibitem{Koivisto:2007bp}
T. Koivisto and D.~F. Mota.
\newblock {\em {Accelerating Cosmologies with an Anisotropic Equation of
  State}}.
\newblock
  {\hypersetup{urlcolor=journalcolor}\href{https://doi.org/10.1086/587451}
  {Astrophys. J.}} {\bfseries 679} 1 (2008).
\newblock \href {http://arxiv.org/abs/0707.0279} {\path{arXiv:0707.0279}}.

\bibitem{Himmetoglu:2008zp}
B. Himmetoglu, C.~R. Contaldi, and M. Peloso.
\newblock {\em {Instability of anisotropic cosmological solutions supported by
  vector fields}}.
\newblock
  {\hypersetup{urlcolor=journalcolor}\href{https://doi.org/10.1103/PhysRevLett.102.111301}
  {Phys. Rev. Lett.}} {\bfseries 102} 111301 (2009).
\newblock \href {http://arxiv.org/abs/0809.2779} {\path{arXiv:0809.2779}}.

\bibitem{DeFelice:2016yws}
A. De~Felice, L. Heisenberg, R. Kase, S. Mukohyama, S. Tsujikawa, and Y.-l.
  Zhang.
\newblock {\em {Cosmology in generalized Proca theories}}.
\newblock
  {\hypersetup{urlcolor=journalcolor}\href{https://doi.org/10.1088/1475-7516/2016/06/048}
  {JCAP}} {\bfseries 06} 048 (2016).
\newblock \href {http://arxiv.org/abs/1603.05806} {\path{arXiv:1603.05806}}.

\bibitem{Heisenberg:2018acv}
L. Heisenberg.
\newblock {\em {Scalar-Vector-Tensor Gravity Theories}}.
\newblock
  {\hypersetup{urlcolor=journalcolor}\href{https://doi.org/10.1088/1475-7516/2018/10/054}
  {JCAP}} {\bfseries 10} 054 (2018).
\newblock \href {http://arxiv.org/abs/1801.01523} {\path{arXiv:1801.01523}}.

\bibitem{Heisenberg:2018mxx}
L. Heisenberg, R. Kase, and S. Tsujikawa.
\newblock {\em {Cosmology in scalar-vector-tensor theories}}.
\newblock
  {\hypersetup{urlcolor=journalcolor}\href{https://doi.org/10.1103/PhysRevD.98.024038}
  {Phys. Rev. D}} {\bfseries 98} 024038 (2018).
\newblock \href {http://arxiv.org/abs/1805.01066} {\path{arXiv:1805.01066}}.

\bibitem{deRham:2014zqa}
C. de Rham.
\newblock {\em {Massive Gravity}}.
\newblock
  {\hypersetup{urlcolor=journalcolor}\href{https://doi.org/10.12942/lrr-2014-7}
  {Living Rev. Rel.}} {\bfseries 17} 7 (2014).
\newblock \href {http://arxiv.org/abs/1401.4173} {\path{arXiv:1401.4173}}.

\bibitem{Hassan:2011zd}
S. Hassan and R.~A. Rosen.
\newblock {\em {Bimetric Gravity from Ghost-free Massive Gravity}}.
\newblock
  {\hypersetup{urlcolor=journalcolor}\href{https://doi.org/10.1007/JHEP02(2012)126}
  {JHEP}} {\bfseries 02} 126 (2012).
\newblock \href {http://arxiv.org/abs/1109.3515} {\path{arXiv:1109.3515}}.

\bibitem{Volkov:2011an}
M.~S. Volkov.
\newblock {\em {Cosmological solutions with massive gravitons in the bigravity
  theory}}.
\newblock
  {\hypersetup{urlcolor=journalcolor}\href{https://doi.org/10.1007/JHEP01(2012)035}
  {JHEP}} {\bfseries 01} 035 (2012).
\newblock \href {http://arxiv.org/abs/1110.6153} {\path{arXiv:1110.6153}}.

\bibitem{Akrami:2015qga}
Y. Akrami, S. Hassan, F. K\"onnig, A. Schmidt-May, and A.~R. Solomon.
\newblock {\em {Bimetric gravity is cosmologically viable}}.
\newblock
  {\hypersetup{urlcolor=journalcolor}\href{https://doi.org/10.1016/j.physletb.2015.06.062}
  {Phys. Lett. B}} {\bfseries 748} 37 (2015).
\newblock \href {http://arxiv.org/abs/1503.07521} {\path{arXiv:1503.07521}}.

\bibitem{Akrami:2013pna}
Y. Akrami, T.~S. Koivisto, and M. Sandstad.
\newblock {\em {Cosmological constraints on ghost-free bigravity: background
  dynamics and late-time acceleration}}.
\newblock In {\em {13th Marcel Grossmann Meeting on Recent Developments in
  Theoretical and Experimental General Relativity, Astrophysics, and
  Relativistic Field Theories}} pages 1252--1254 2015.
\newblock \href {http://arxiv.org/abs/1302.5268} {\path{arXiv:1302.5268}}.

\bibitem{Motohashi:2014opa}
H. Motohashi and T. Suyama.
\newblock {\em {Third order equations of motion and the Ostrogradsky
  instability}}.
\newblock
  {\hypersetup{urlcolor=journalcolor}\href{https://doi.org/10.1103/PhysRevD.91.085009}
  {Phys. Rev. D}} {\bfseries 91} 085009 (2015).
\newblock \href {http://arxiv.org/abs/1411.3721} {\path{arXiv:1411.3721}}.

\bibitem{Woodard:2015zca}
R.~P. Woodard.
\newblock {\em {Ostrogradsky's theorem on Hamiltonian instability}}.
\newblock
  {\hypersetup{urlcolor=journalcolor}\href{https://doi.org/10.4249/scholarpedia.32243}
  {Scholarpedia}} {\bfseries 10} 32243 (2015).
\newblock \href {http://arxiv.org/abs/1506.02210} {\path{arXiv:1506.02210}}.

\bibitem{Gleyzes:2014dya}
J. Gleyzes, D. Langlois, F. Piazza, and F. Vernizzi.
\newblock {\em {Healthy theories beyond Horndeski}}.
\newblock
  {\hypersetup{urlcolor=journalcolor}\href{https://doi.org/10.1103/PhysRevLett.114.211101}
  {Phys. Rev. Lett.}} {\bfseries 114} 211101 (2015).
\newblock \href {http://arxiv.org/abs/1404.6495} {\path{arXiv:1404.6495}}.

\bibitem{Langlois:2017mdk}
D. Langlois.
\newblock {\em {Degenerate Higher-Order Scalar-Tensor (DHOST) theories}}.
\newblock In {\em {52nd Rencontres de Moriond on Gravitation}} pages 221--228
  2017.
\newblock \href {http://arxiv.org/abs/1707.03625} {\path{arXiv:1707.03625}}.

\bibitem{Heisenberg:2016eld}
L. Heisenberg, R. Kase, and S. Tsujikawa.
\newblock {\em {Beyond generalized Proca theories}}.
\newblock
  {\hypersetup{urlcolor=journalcolor}\href{https://doi.org/10.1016/j.physletb.2016.07.052}
  {Phys. Lett. B}} {\bfseries 760} 617 (2016).
\newblock \href {http://arxiv.org/abs/1605.05565} {\path{arXiv:1605.05565}}.

\bibitem{Kaluza:1921tu}
T. Kaluza.
\newblock {\em {Zum Unitätsproblem der Physik}}.
\newblock
  {\hypersetup{urlcolor=journalcolor}\href{https://doi.org/10.1142/S0218271818700017}
  {Int. J. Mod. Phys. D}} {\bfseries 27} 1870001 (2018).
\newblock \href {http://arxiv.org/abs/1803.08616} {\path{arXiv:1803.08616}}.

\bibitem{Nordstrom:1988fi}
G. Nordstrom.
\newblock {\em {On the possibility of unifying the electromagnetic and the
  gravitational fields}}.
\newblock {\hypersetup{urlcolor=journalcolor}\href{https://doi.org/} {Phys.
  Z.}} {\bfseries 15} 504 (1914).
\newblock \href {http://arxiv.org/abs/physics/0702221}
  {\path{arXiv:physics/0702221}}.

\bibitem{Klein:1926tv}
O. Klein.
\newblock {\em {Quantum Theory and Five-Dimensional Theory of Relativity. (In
  German and English)}}.
\newblock
  {\hypersetup{urlcolor=journalcolor}\href{https://doi.org/10.1007/BF01397481}
  {Z. Phys.}} {\bfseries 37} 895 (1926).

\bibitem{Padmanabhan:2013xyr}
T. Padmanabhan and D. Kothawala.
\newblock {\em {Lanczos-Lovelock models of gravity}}.
\newblock
  {\hypersetup{urlcolor=journalcolor}\href{https://doi.org/10.1016/j.physrep.2013.05.007}
  {Phys. Rept.}} {\bfseries 531} 115 (2013).
\newblock \href {http://arxiv.org/abs/1302.2151} {\path{arXiv:1302.2151}}.

\bibitem{Rastall:1973nw}
P. Rastall.
\newblock {\em {Generalization of the einstein theory}}.
\newblock
  {\hypersetup{urlcolor=journalcolor}\href{https://doi.org/10.1103/PhysRevD.6.3357}
  {Phys. Rev. D}} {\bfseries 6} 3357 (1972).

\bibitem{Batista:2011nu}
C.~E. Batista, M.~H. Daouda, J.~C. Fabris, O.~F. Piattella, and D.~C.
  Rodrigues.
\newblock {\em {Rastall Cosmology and the Lambda CDM Model}}.
\newblock
  {\hypersetup{urlcolor=journalcolor}\href{https://doi.org/10.1103/PhysRevD.85.084008}
  {Phys. Rev. D}} {\bfseries 85} 084008 (2012).
\newblock \href {http://arxiv.org/abs/1112.4141} {\path{arXiv:1112.4141}}.

\bibitem{1984NCimB..80...42S}
L.~L. {Smalley}.
\newblock {\em {Variational principle for a prototype Rastall theory of
  gravitation}}.
\newblock
  {\hypersetup{urlcolor=journalcolor}\href{https://doi.org/10.1007/BF02899371}
  {Nuovo Cimento B Serie}} {\bfseries 80} 42 (1984).

\bibitem{Santos:2017nxm}
R.~V.~d. Santos and J.~A.~C. Nogales.
\newblock {\em {Cosmology from a Lagrangian formulation for Rastall's theory}}.
\newblock {\hypersetup{urlcolor=journalcolor}\href{https://doi.org/} {2017}}).
\newblock \href {http://arxiv.org/abs/1701.08203} {\path{arXiv:1701.08203}}.

\bibitem{DeMoraes:2019mef}
W. De~Moraes and A. Santos.
\newblock {\em {Lagrangian formalism for Rastall theory of gravity and
  G\"odel-type universe}}.
\newblock
  {\hypersetup{urlcolor=journalcolor}\href{https://doi.org/10.1007/s10714-019-2652-9}
  {Gen. Rel. Grav.}} {\bfseries 51} 167 (2019).
\newblock \href {http://arxiv.org/abs/1912.06471} {\path{arXiv:1912.06471}}.

\bibitem{Fisher:2019ekh}
S.~B. Fisher and E.~D. Carlson.
\newblock {\em {Reexamining $f(R,T)$ gravity}}.
\newblock
  {\hypersetup{urlcolor=journalcolor}\href{https://doi.org/10.1103/PhysRevD.100.064059}
  {Phys. Rev. D}} {\bfseries 100} 064059 (2019).
\newblock \href {http://arxiv.org/abs/1908.05306} {\path{arXiv:1908.05306}}.

\bibitem{Visser:2017gpz}
M. Visser.
\newblock {\em {Rastall gravity is equivalent to Einstein gravity}}.
\newblock
  {\hypersetup{urlcolor=journalcolor}\href{https://doi.org/10.1016/j.physletb.2018.05.028}
  {Phys. Lett. B}} {\bfseries 782} 83 (2018).
\newblock \href {http://arxiv.org/abs/1711.11500} {\path{arXiv:1711.11500}}.

\bibitem{Darabi:2017coc}
F. Darabi, H. Moradpour, I. Licata, Y. Heydarzade, and C. Corda.
\newblock {\em {Einstein and Rastall Theories of Gravitation in Comparison}}.
\newblock
  {\hypersetup{urlcolor=journalcolor}\href{https://doi.org/10.1140/epjc/s10052-017-5502-5}
  {Eur. Phys. J. C}} {\bfseries 78} 25 (2018).
\newblock \href {http://arxiv.org/abs/1712.09307} {\path{arXiv:1712.09307}}.

\bibitem{Maggiore:2013mea}
M. Maggiore.
\newblock {\em {Phantom dark energy from nonlocal infrared modifications of
  general relativity}}.
\newblock
  {\hypersetup{urlcolor=journalcolor}\href{https://doi.org/10.1103/PhysRevD.89.043008}
  {Phys. Rev. D}} {\bfseries 89} 043008 (2014).
\newblock \href {http://arxiv.org/abs/1307.3898} {\path{arXiv:1307.3898}}.

\bibitem{Belgacem:2018wtb}
E. Belgacem, A. Finke, A. Frassino, and M. Maggiore.
\newblock {\em {Testing nonlocal gravity with Lunar Laser Ranging}}.
\newblock
  {\hypersetup{urlcolor=journalcolor}\href{https://doi.org/10.1088/1475-7516/2019/02/035}
  {JCAP}} {\bfseries 02} 035 (2019).
\newblock \href {http://arxiv.org/abs/1812.11181} {\path{arXiv:1812.11181}}.

\bibitem{Einstein:1916vd}
A. Einstein.
\newblock {\em {The Foundation of the General Theory of Relativity}}.
\newblock
  {\hypersetup{urlcolor=journalcolor}\href{https://doi.org/10.1002/andp.200590044}
  {Annalen Phys.}} {\bfseries 49} 769 (1916).

\bibitem{1952prel.book..189E}
A. {Einstein}.
\newblock {\em {Do gravitational fields play an essential part in the structure
  of the elementary particles of matter?}} pages 189--198.
\newblock 1952.

\bibitem{Henneaux:1989zc}
M. Henneaux and C. Teitelboim.
\newblock {\em {The Cosmological Constant and General Covariance}}.
\newblock
  {\hypersetup{urlcolor=journalcolor}\href{https://doi.org/10.1016/0370-2693(89)91251-3}
  {Phys. Lett. B}} {\bfseries 222} 195 (1989).

\bibitem{Barvinsky:2017pmm}
A. Barvinsky and A.~Y. Kamenshchik.
\newblock {\em {Darkness without dark matter and energy -- generalized
  unimodular gravity}}.
\newblock
  {\hypersetup{urlcolor=journalcolor}\href{https://doi.org/10.1016/j.physletb.2017.09.045}
  {Phys. Lett. B}} {\bfseries 774} 59 (2017).
\newblock \href {http://arxiv.org/abs/1705.09470} {\path{arXiv:1705.09470}}.

\bibitem{Alvarez:2015pla}
E. \'Alvarez, S. Gonz\'alez-Mart\'\i{}n, M. Herrero-Valea, and C. Mart\'\i{}n.
\newblock {\em {Unimodular Gravity Redux}}.
\newblock
  {\hypersetup{urlcolor=journalcolor}\href{https://doi.org/10.1103/PhysRevD.92.061502}
  {Phys. Rev. D}} {\bfseries 92} 061502 (2015).
\newblock \href {http://arxiv.org/abs/1505.00022} {\path{arXiv:1505.00022}}.

\bibitem{Smolin:2009ti}
L. Smolin.
\newblock {\em {The Quantization of unimodular gravity and the cosmological
  constant problems}}.
\newblock
  {\hypersetup{urlcolor=journalcolor}\href{https://doi.org/10.1103/PhysRevD.80.084003}
  {Phys. Rev. D}} {\bfseries 80} 084003 (2009).
\newblock \href {http://arxiv.org/abs/0904.4841} {\path{arXiv:0904.4841}}.

\bibitem{Padilla:2014yea}
A. Padilla and I.~D. Saltas.
\newblock {\em {A note on classical and quantum unimodular gravity}}.
\newblock
  {\hypersetup{urlcolor=journalcolor}\href{https://doi.org/10.1140/epjc/s10052-015-3767-0}
  {Eur. Phys. J. C}} {\bfseries 75} 561 (2015).
\newblock \href {http://arxiv.org/abs/1409.3573} {\path{arXiv:1409.3573}}.

\bibitem{Fiol:2008vk}
B. Fiol and J. Garriga.
\newblock {\em {Semiclassical Unimodular Gravity}}.
\newblock
  {\hypersetup{urlcolor=journalcolor}\href{https://doi.org/10.1088/1475-7516/2010/08/015}
  {JCAP}} {\bfseries 08} 015 (2010).
\newblock \href {http://arxiv.org/abs/0809.1371} {\path{arXiv:0809.1371}}.

\bibitem{Shaposhnikov:2008xb}
M. Shaposhnikov and D. Zenhausern.
\newblock {\em {Scale invariance, unimodular gravity and dark energy}}.
\newblock
  {\hypersetup{urlcolor=journalcolor}\href{https://doi.org/10.1016/j.physletb.2008.11.054}
  {Phys. Lett. B}} {\bfseries 671} 187 (2009).
\newblock \href {http://arxiv.org/abs/0809.3395} {\path{arXiv:0809.3395}}.

\bibitem{Horava:2009uw}
P. Horava.
\newblock {\em {Quantum Gravity at a Lifshitz Point}}.
\newblock
  {\hypersetup{urlcolor=journalcolor}\href{https://doi.org/10.1103/PhysRevD.79.084008}
  {Phys. Rev. D}} {\bfseries 79} 084008 (2009).
\newblock \href {http://arxiv.org/abs/0901.3775} {\path{arXiv:0901.3775}}.

\bibitem{Mukohyama:2009mz}
S. Mukohyama.
\newblock {\em {Dark matter as integration constant in Horava-Lifshitz
  gravity}}.
\newblock
  {\hypersetup{urlcolor=journalcolor}\href{https://doi.org/10.1103/PhysRevD.80.064005}
  {Phys. Rev. D}} {\bfseries 80} 064005 (2009).
\newblock \href {http://arxiv.org/abs/0905.3563} {\path{arXiv:0905.3563}}.

\bibitem{Mukohyama:2010xz}
S. Mukohyama.
\newblock {\em {Horava-Lifshitz Cosmology: A Review}}.
\newblock
  {\hypersetup{urlcolor=journalcolor}\href{https://doi.org/10.1088/0264-9381/27/22/223101}
  {Class. Quant. Grav.}} {\bfseries 27} 223101 (2010).
\newblock \href {http://arxiv.org/abs/1007.5199} {\path{arXiv:1007.5199}}.

\bibitem{Brandenberger:2009yt}
R. Brandenberger.
\newblock {\em {Matter Bounce in Horava-Lifshitz Cosmology}}.
\newblock
  {\hypersetup{urlcolor=journalcolor}\href{https://doi.org/10.1103/PhysRevD.80.043516}
  {Phys. Rev. D}} {\bfseries 80} 043516 (2009).
\newblock \href {http://arxiv.org/abs/0904.2835} {\path{arXiv:0904.2835}}.

\bibitem{Saridakis:2009bv}
E.~N. Saridakis.
\newblock {\em {Horava-Lifshitz Dark Energy}}.
\newblock
  {\hypersetup{urlcolor=journalcolor}\href{https://doi.org/10.1140/epjc/s10052-010-1294-6}
  {Eur. Phys. J. C}} {\bfseries 67} 229 (2010).
\newblock \href {http://arxiv.org/abs/0905.3532} {\path{arXiv:0905.3532}}.

\bibitem{Carloni:2010nx}
S. Carloni, M. Chaichian, S. Nojiri, S.~D. Odintsov, M. Oksanen, and A.
  Tureanu.
\newblock {\em {Modified first-order Horava-Lifshitz gravity: Hamiltonian
  analysis of the general theory and accelerating FRW cosmology in power-law
  F(R) model}}.
\newblock
  {\hypersetup{urlcolor=journalcolor}\href{https://doi.org/10.1103/PhysRevD.82.065020}
  {Phys. Rev. D}} {\bfseries 82} 065020 (2010).
\newblock [Erratum: Phys.Rev.D 85, 129904 (2012)].
\newblock \href {http://arxiv.org/abs/1003.3925} {\path{arXiv:1003.3925}}.

\bibitem{Calcagni:2009ar}
G. Calcagni.
\newblock {\em {Cosmology of the Lifshitz universe}}.
\newblock
  {\hypersetup{urlcolor=journalcolor}\href{https://doi.org/10.1088/1126-6708/2009/09/112}
  {JHEP}} {\bfseries 09} 112 (2009).
\newblock \href {http://arxiv.org/abs/0904.0829} {\path{arXiv:0904.0829}}.

\bibitem{Amendola:2019laa}
L. Amendola, D. Bettoni, A.~M. Pinho, and S. Casas.
\newblock {\em {Measuring gravity at cosmological scales}}.
\newblock
  {\hypersetup{urlcolor=journalcolor}\href{https://doi.org/10.3390/universe6020020}
  {Universe}} {\bfseries 6} 20 (2020).
\newblock \href {http://arxiv.org/abs/1902.06978} {\path{arXiv:1902.06978}}.

\bibitem{Will:2014kxa}
C.~M. Will.
\newblock {\em {The Confrontation between General Relativity and Experiment}}.
\newblock
  {\hypersetup{urlcolor=journalcolor}\href{https://doi.org/10.12942/lrr-2014-4}
  {Living Rev. Rel.}} {\bfseries 17} 4 (2014).
\newblock \href {http://arxiv.org/abs/1403.7377} {\path{arXiv:1403.7377}}.

\bibitem{Uzan:2010pm}
J.-P. Uzan.
\newblock {\em {Varying Constants, Gravitation and Cosmology}}.
\newblock
  {\hypersetup{urlcolor=journalcolor}\href{https://doi.org/10.12942/lrr-2011-2}
  {Living Rev. Rel.}} {\bfseries 14} 2 (2011).
\newblock \href {http://arxiv.org/abs/1009.5514} {\path{arXiv:1009.5514}}.

\bibitem{Bertotti:2003rm}
B. Bertotti, L. Iess, and P. Tortora.
\newblock {\em {A test of general relativity using radio links with the Cassini
  spacecraft}}.
\newblock
  {\hypersetup{urlcolor=journalcolor}\href{https://doi.org/10.1038/nature01997}
  {Nature}} {\bfseries 425} 374 (2003).

\bibitem{Hofmann:2018myc}
F. Hofmann and J. M\"uller.
\newblock {\em {Relativistic tests with lunar laser ranging}}.
\newblock
  {\hypersetup{urlcolor=journalcolor}\href{https://doi.org/10.1088/1361-6382/aa8f7a}
  {Class. Quant. Grav.}} {\bfseries 35} 035015 (2018).

\bibitem{DeglInnocenti:1995hbi}
S. Degl'Innocenti, G. Fiorentini, G. Raffelt, B. Ricci, and A. Weiss.
\newblock {\em {Time variation of Newton's constant and the age of globular
  clusters}}.
\newblock {\hypersetup{urlcolor=journalcolor}\href{https://doi.org/} {Astron.
  Astrophys.}} {\bfseries 312} 345 (1996).
\newblock \href {http://arxiv.org/abs/astro-ph/9509090}
  {\path{arXiv:astro-ph/9509090}}.

\bibitem{Verbiest:2008gy}
J. Verbiest, M. Bailes, W. van Straten, G. Hobbs, R. Edwards, R. Manchester, N.
  Bhat, J. Sarkissian, B. Jacoby, and S. Kulkarni.
\newblock {\em {Precision timing of PSR J0437-4715: an accurate pulsar
  distance, a high pulsar mass and a limit on the variation of Newton's
  gravitational constant}}.
\newblock
  {\hypersetup{urlcolor=journalcolor}\href{https://doi.org/10.1086/529576}
  {Astrophys. J.}} {\bfseries 679} 675 (2008).
\newblock \href {http://arxiv.org/abs/0801.2589} {\path{arXiv:0801.2589}}.

\bibitem{Bellini:2014fua}
E. Bellini and I. Sawicki.
\newblock {\em {Maximal freedom at minimum cost: linear large-scale structure
  in general modifications of gravity}}.
\newblock
  {\hypersetup{urlcolor=journalcolor}\href{https://doi.org/10.1088/1475-7516/2014/07/050}
  {JCAP}} {\bfseries 07} 050 (2014).
\newblock \href {http://arxiv.org/abs/1404.3713} {\path{arXiv:1404.3713}}.

\bibitem{Creminelli:2017sry}
P. Creminelli and F. Vernizzi.
\newblock {\em {Dark Energy after GW170817 and GRB170817A}}.
\newblock
  {\hypersetup{urlcolor=journalcolor}\href{https://doi.org/10.1103/PhysRevLett.119.251302}
  {Phys. Rev. Lett.}} {\bfseries 119} 251302 (2017).
\newblock \href {http://arxiv.org/abs/1710.05877} {\path{arXiv:1710.05877}}.

\bibitem{Kase:2018aps}
R. Kase and S. Tsujikawa.
\newblock {\em {Dark energy in Horndeski theories after GW170817: A review}}.
\newblock
  {\hypersetup{urlcolor=journalcolor}\href{https://doi.org/10.1142/S0218271819420057}
  {Int. J. Mod. Phys. D}} {\bfseries 28} 1942005 (2019).
\newblock \href {http://arxiv.org/abs/1809.08735} {\path{arXiv:1809.08735}}.

\bibitem{Bordin:2020fww}
L. Bordin, E.~J. Copeland, and A. Padilla.
\newblock {\em {Dark energy loopholes some time after GW170817}}.
\newblock {\hypersetup{urlcolor=journalcolor}\href{https://doi.org/} {2020}}).
\newblock \href {http://arxiv.org/abs/2006.06652} {\path{arXiv:2006.06652}}.

\bibitem{Ezquiaga:2017ekz}
J.~M. Ezquiaga and M. Zumalacárregui.
\newblock {\em {Dark Energy After GW170817: Dead Ends and the Road Ahead}}.
\newblock
  {\hypersetup{urlcolor=journalcolor}\href{https://doi.org/10.1103/PhysRevLett.119.251304}
  {Phys. Rev. Lett.}} {\bfseries 119} 251304 (2017).
\newblock \href {http://arxiv.org/abs/1710.05901} {\path{arXiv:1710.05901}}.

\bibitem{Dalang:2019rke}
C. Dalang, P. Fleury, and L. Lombriser.
\newblock {\em {Horndeski gravity and standard sirens}}.
\newblock
  {\hypersetup{urlcolor=journalcolor}\href{https://doi.org/10.1103/PhysRevD.102.044036}
  {Phys. Rev. D}} {\bfseries 102} 044036 (2020).
\newblock \href {http://arxiv.org/abs/1912.06117} {\path{arXiv:1912.06117}}.

\bibitem{Amendola:2018ltt}
L. Amendola, D. Bettoni, G. Domènech, and A.~R. Gomes.
\newblock {\em {Doppelgänger dark energy: modified gravity with non-universal
  couplings after GW170817}}.
\newblock
  {\hypersetup{urlcolor=journalcolor}\href{https://doi.org/10.1088/1475-7516/2018/06/029}
  {JCAP}} {\bfseries 06} 029 (2018).
\newblock \href {http://arxiv.org/abs/1803.06368} {\path{arXiv:1803.06368}}.

\bibitem{Amendola:2016saw}
L. Amendola et~al.
\newblock {\em {Cosmology and fundamental physics with the Euclid satellite}}.
\newblock
  {\hypersetup{urlcolor=journalcolor}\href{https://doi.org/10.1007/s41114-017-0010-3}
  {Living Rev. Rel.}} {\bfseries 21} 2 (2018).
\newblock \href {http://arxiv.org/abs/1606.00180} {\path{arXiv:1606.00180}}.

\bibitem{Amendola:2012ky}
L. Amendola, M. Kunz, M. Motta, I.~D. Saltas, and I. Sawicki.
\newblock {\em {Observables and unobservables in dark energy cosmologies}}.
\newblock
  {\hypersetup{urlcolor=journalcolor}\href{https://doi.org/10.1103/PhysRevD.87.023501}
  {Phys. Rev. D}} {\bfseries 87} 023501 (2013).
\newblock \href {http://arxiv.org/abs/1210.0439} {\path{arXiv:1210.0439}}.

\bibitem{Dubovsky:2005xd}
S. Dubovsky, T. Gregoire, A. Nicolis, and R. Rattazzi.
\newblock {\em {Null energy condition and superluminal propagation}}.
\newblock
  {\hypersetup{urlcolor=journalcolor}\href{https://doi.org/10.1088/1126-6708/2006/03/025}
  {JHEP}} {\bfseries 03} 025 (2006).
\newblock \href {http://arxiv.org/abs/hep-th/0512260}
  {\path{arXiv:hep-th/0512260}}.

\bibitem{Bekenstein:1982eu}
J.~D. Bekenstein.
\newblock {\em {Fine Structure Constant: Is It Really a Constant?}}
\newblock
  {\hypersetup{urlcolor=journalcolor}\href{https://doi.org/10.1103/PhysRevD.25.1527}
  {Phys. Rev.}} {\bfseries D25} 1527 (1982).

\bibitem{Leal:2014yqa}
P.~M.~M. Leal, C.~J. A.~P. Martins, and L.~B. Ventura.
\newblock {\em {Fine-structure constant constraints on Bekenstein-type
  models}}.
\newblock
  {\hypersetup{urlcolor=journalcolor}\href{https://doi.org/10.1103/PhysRevD.90.027305}
  {Phys. Rev.}} {\bfseries D90} 027305 (2014).
\newblock \href {http://arxiv.org/abs/1407.4099} {\path{arXiv:1407.4099}}.

\bibitem{Holanda:2015oda}
R.~F.~L. Holanda, S.~J. Landau, J.~S. Alcaniz, I.~E. Sanchez~G., and V.~C.
  Busti.
\newblock {\em {Constraints on a possible variation of the fine structure
  constant from galaxy cluster data}}.
\newblock
  {\hypersetup{urlcolor=journalcolor}\href{https://doi.org/10.1088/1475-7516/2016/05/047}
  {JCAP}} {\bfseries 1605} 047 (2016).
\newblock \href {http://arxiv.org/abs/1510.07240} {\path{arXiv:1510.07240}}.

\bibitem{Pinho:2016mkm}
A. Pinho and C. Martins.
\newblock {\em {Updated constraints on spatial variations of the fine-structure
  constant}}.
\newblock
  {\hypersetup{urlcolor=journalcolor}\href{https://doi.org/10.1016/j.physletb.2016.03.014}
  {Phys. Lett. B}} {\bfseries 756} 121 (2016).
\newblock \href {http://arxiv.org/abs/1603.04498} {\path{arXiv:1603.04498}}.

\bibitem{Hees:2020gda}
A. Hees et~al.
\newblock {\em {Search for a Variation of the Fine Structure Constant around
  the Supermassive Black Hole in Our Galactic Center}}.
\newblock
  {\hypersetup{urlcolor=journalcolor}\href{https://doi.org/10.1103/PhysRevLett.124.081101}
  {Phys. Rev. Lett.}} {\bfseries 124} 081101 (2020).
\newblock \href {http://arxiv.org/abs/2002.11567} {\path{arXiv:2002.11567}}.

\bibitem{Murphy:2003hw}
M.~T. Murphy, J. Webb, and V. Flambaum.
\newblock {\em {Further evidence for a variable fine-structure constant from
  Keck/HIRES QSO absorption spectra}}.
\newblock
  {\hypersetup{urlcolor=journalcolor}\href{https://doi.org/10.1046/j.1365-8711.2003.06970.x}
  {Mon. Not. Roy. Astron. Soc.}} {\bfseries 345} 609 (2003).
\newblock \href {http://arxiv.org/abs/astro-ph/0306483}
  {\path{arXiv:astro-ph/0306483}}.

\bibitem{Webb:2010hc}
J.~K. Webb, J.~A. King, M.~T. Murphy, V.~V. Flambaum, R.~F. Carswell, and M.~B.
  Bainbridge.
\newblock {\em {Indications of a spatial variation of the fine structure
  constant}}.
\newblock
  {\hypersetup{urlcolor=journalcolor}\href{https://doi.org/10.1103/PhysRevLett.107.191101}
  {Phys. Rev. Lett.}} {\bfseries 107} 191101 (2011).
\newblock \href {http://arxiv.org/abs/1008.3907} {\path{arXiv:1008.3907}}.

\bibitem{Tino:2020nla}
G. Tino, L. Cacciapuoti, S. Capozziello, G. Lambiase, and F. Sorrentino.
\newblock {\em {Precision Gravity Tests and the Einstein Equivalence
  Principle}}.
\newblock
  {\hypersetup{urlcolor=journalcolor}\href{https://doi.org/10.1016/j.ppnp.2020.103772}
  {Prog. Part. Nucl. Phys.}} {\bfseries 112} 103772 (2020).
\newblock \href {http://arxiv.org/abs/2002.02907} {\path{arXiv:2002.02907}}.

\bibitem{Bekenstein:2009fq}
J.~D. Bekenstein and M. Schiffer.
\newblock {\em {Varying fine structure 'constant' and charged black holes}}.
\newblock
  {\hypersetup{urlcolor=journalcolor}\href{https://doi.org/10.1103/PhysRevD.80.123508}
  {Phys. Rev.}} {\bfseries D80} 123508 (2009).
\newblock \href {http://arxiv.org/abs/0906.4557} {\path{arXiv:0906.4557}}.

\bibitem{Shaw:2006zs}
D.~J. Shaw and J.~D. Barrow.
\newblock {\em {The Local effects of cosmological variations in physical
  'constants' and scalar fields. II. Quasi-spherical spacetimes}}.
\newblock
  {\hypersetup{urlcolor=journalcolor}\href{https://doi.org/10.1103/PhysRevD.73.123506}
  {Phys. Rev.}} {\bfseries D73} 123506 (2006).
\newblock \href {http://arxiv.org/abs/gr-qc/0601056}
  {\path{arXiv:gr-qc/0601056}}.

\bibitem{Barrow:2014vva}
J.~D. Barrow and J. Magueijo.
\newblock {\em {Local Varying-Alpha Theories}}.
\newblock
  {\hypersetup{urlcolor=journalcolor}\href{https://doi.org/10.1142/S0217732315400295}
  {Mod. Phys. Lett.}} {\bfseries A30} 1540029 (2015).
\newblock \href {http://arxiv.org/abs/1412.3278} {\path{arXiv:1412.3278}}.

\bibitem{Olive:2001vz}
K.~A. Olive and M. Pospelov.
\newblock {\em {Evolution of the fine structure constant driven by dark matter
  and the cosmological constant}}.
\newblock
  {\hypersetup{urlcolor=journalcolor}\href{https://doi.org/10.1103/PhysRevD.65.085044}
  {Phys. Rev.}} {\bfseries D65} 085044 (2002).
\newblock \href {http://arxiv.org/abs/hep-ph/0110377}
  {\path{arXiv:hep-ph/0110377}}.

\bibitem{Marra:2005yt}
V. Marra and F. Rosati.
\newblock {\em {Cosmological evolution of alpha driven by a general coupling
  with quintessence}}.
\newblock
  {\hypersetup{urlcolor=journalcolor}\href{https://doi.org/10.1088/1475-7516/2005/05/011}
  {JCAP}} {\bfseries 0505} 011 (2005).
\newblock \href {http://arxiv.org/abs/astro-ph/0501515}
  {\path{arXiv:astro-ph/0501515}}.

\bibitem{Barrow:2009nt}
J.~D. Barrow.
\newblock {\em {Varying Alpha}}.
\newblock
  {\hypersetup{urlcolor=journalcolor}\href{https://doi.org/10.1002/andp.201010416}
  {Annalen Phys.}} {\bfseries 19} 202 (2010).
\newblock \href {http://arxiv.org/abs/0912.5510} {\path{arXiv:0912.5510}}.

\bibitem{Barrow:2011kr}
J.~D. Barrow and S.~Z.~W. Lip.
\newblock {\em {A Generalized Theory of Varying Alpha}}.
\newblock
  {\hypersetup{urlcolor=journalcolor}\href{https://doi.org/10.1103/PhysRevD.85.023514}
  {Phys. Rev.}} {\bfseries D85} 023514 (2012).
\newblock \href {http://arxiv.org/abs/1110.3120} {\path{arXiv:1110.3120}}.

\bibitem{Barrow:2013uza}
J.~D. Barrow and A.~A.~H. Graham.
\newblock {\em {General Dynamics of Varying-Alpha Universes}}.
\newblock
  {\hypersetup{urlcolor=journalcolor}\href{https://doi.org/10.1103/PhysRevD.88.103513}
  {Phys. Rev.}} {\bfseries D88} 103513 (2013).
\newblock \href {http://arxiv.org/abs/1307.6816} {\path{arXiv:1307.6816}}.

\bibitem{Sloan:2013wya}
D. Sloan.
\newblock {\em {Loop Quantum Cosmology and the Fine Structure Constant}}.
\newblock
  {\hypersetup{urlcolor=journalcolor}\href{https://doi.org/10.1088/0264-9381/31/2/025014}
  {Class. Quant. Grav.}} {\bfseries 31} 025014 (2014).
\newblock \href {http://arxiv.org/abs/1307.5527} {\path{arXiv:1307.5527}}.

\bibitem{Graham:2014hva}
A.~A.~H. Graham.
\newblock {\em {Varying-alpha and K-essence}}.
\newblock
  {\hypersetup{urlcolor=journalcolor}\href{https://doi.org/10.1088/0264-9381/32/1/015019}
  {Class. Quant. Grav.}} {\bfseries 32} 015019 (2015).
\newblock \href {http://arxiv.org/abs/1408.2788} {\path{arXiv:1408.2788}}.

\bibitem{vandeBruck:2015rma}
C. van~de Bruck, J. Mifsud, and N.~J. Nunes.
\newblock {\em {The variation of the fine-structure constant from disformal
  couplings}}.
\newblock
  {\hypersetup{urlcolor=journalcolor}\href{https://doi.org/10.1088/1475-7516/2015/12/018}
  {JCAP}} {\bfseries 1512} 018 (2015).
\newblock \href {http://arxiv.org/abs/1510.00200} {\path{arXiv:1510.00200}}.

\bibitem{Fritzsch:2016ewd}
H. Fritzsch, R.~C. Nunes, and J. Sola.
\newblock {\em {Running vacuum in the Universe and the time variation of the
  fundamental constants of Nature}}.
\newblock {\hypersetup{urlcolor=journalcolor}\href{https://doi.org/} {2016}}).
\newblock \href {http://arxiv.org/abs/1605.06104} {\path{arXiv:1605.06104}}.

\bibitem{Calabrese:2013lga}
E. Calabrese, M. Martinelli, S. Pandolfi, V. Cardone, C. Martins, S. Spiro, and
  P. Vielzeuf.
\newblock {\em {Dark Energy coupling with electromagnetism as seen from future
  low-medium redshift probes}}.
\newblock
  {\hypersetup{urlcolor=journalcolor}\href{https://doi.org/10.1103/PhysRevD.89.083509}
  {Phys. Rev. D}} {\bfseries 89} 083509 (2014).
\newblock \href {http://arxiv.org/abs/1311.5841} {\path{arXiv:1311.5841}}.

\bibitem{Martins:2019ebg}
C. Martins and M. Prat~Colomer.
\newblock {\em {Fine-structure constant constraints on late-time dark energy
  transitions}}.
\newblock
  {\hypersetup{urlcolor=journalcolor}\href{https://doi.org/10.1016/j.physletb.2019.02.039}
  {Phys. Lett. B}} {\bfseries 791} 230 (2019).
\newblock \href {http://arxiv.org/abs/1903.04310} {\path{arXiv:1903.04310}}.

\bibitem{Martins:2015jta}
C. Martins, A. Pinho, R. Alves, M. Pino, C. Rocha, and M. von Wietersheim.
\newblock {\em {Dark energy and Equivalence Principle constraints from
  astrophysical tests of the stability of the fine-structure constant}}.
\newblock
  {\hypersetup{urlcolor=journalcolor}\href{https://doi.org/10.1088/1475-7516/2015/08/047}
  {JCAP}} {\bfseries 08} 047 (2015).
\newblock \href {http://arxiv.org/abs/1508.06157} {\path{arXiv:1508.06157}}.

\bibitem{Einstein:1935rr}
A. Einstein, B. Podolsky, and N. Rosen.
\newblock {\em {Can quantum mechanical description of physical reality be
  considered complete?}}
\newblock
  {\hypersetup{urlcolor=journalcolor}\href{https://doi.org/10.1103/PhysRev.47.777}
  {Phys. Rev.}} {\bfseries 47} 777 (1935).

\bibitem{Wiseman_2006}
H.~M. Wiseman.
\newblock {\em From Einstein’s theorem to Bell’s theorem: a history of
  quantum non-locality}.
\newblock
  {\hypersetup{urlcolor=journalcolor}\href{https://doi.org/10.1080/00107510600581011}
  {Contemporary Physics}} {\bfseries 47} 79–88 (2006).
\newblock URL: \url{http://dx.doi.org/10.1080/00107510600581011}.

\bibitem{Belgacem:2017cqo}
E. Belgacem, Y. Dirian, S. Foffa, and M. Maggiore.
\newblock {\em {Nonlocal gravity. Conceptual aspects and cosmological
  predictions}}.
\newblock
  {\hypersetup{urlcolor=journalcolor}\href{https://doi.org/10.1088/1475-7516/2018/03/002}
  {JCAP}} {\bfseries 03} 002 (2018).
\newblock \href {http://arxiv.org/abs/1712.07066} {\path{arXiv:1712.07066}}.

\bibitem{Dalvit:1994gf}
D.~A. Dalvit and F.~D. Mazzitelli.
\newblock {\em {Running coupling constants, Newtonian potential and
  nonlocalities in the effective action}}.
\newblock
  {\hypersetup{urlcolor=journalcolor}\href{https://doi.org/10.1103/PhysRevD.50.1001}
  {Phys. Rev. D}} {\bfseries 50} 1001 (1994).
\newblock \href {http://arxiv.org/abs/gr-qc/9402003}
  {\path{arXiv:gr-qc/9402003}}.

\bibitem{Nojiri:2007uq}
S. Nojiri and S.~D. Odintsov.
\newblock {\em {Modified non-local-F(R) gravity as the key for the inflation
  and dark energy}}.
\newblock
  {\hypersetup{urlcolor=journalcolor}\href{https://doi.org/10.1016/j.physletb.2007.12.001}
  {Phys. Lett.}} {\bfseries B659} 821 (2008).
\newblock \href {http://arxiv.org/abs/0708.0924} {\path{arXiv:0708.0924}}.

\bibitem{Amendola:2017qge}
L. Amendola, N. Burzilla, and H. Nersisyan.
\newblock {\em {Quantum Gravity inspired nonlocal gravity model}}.
\newblock
  {\hypersetup{urlcolor=journalcolor}\href{https://doi.org/10.1103/PhysRevD.96.084031}
  {Phys. Rev. D}} {\bfseries 96} 084031 (2017).
\newblock \href {http://arxiv.org/abs/1707.04628} {\path{arXiv:1707.04628}}.

\bibitem{Dodelson:2013sma}
S. Dodelson and S. Park.
\newblock {\em {Nonlocal Gravity and Structure in the Universe}}.
\newblock
  {\hypersetup{urlcolor=journalcolor}\href{https://doi.org/10.1103/PhysRevD.90.043535}
  {Phys. Rev. D}} {\bfseries 90} 043535 (2014).
\newblock [Erratum: Phys.Rev.D 98, 029904 (2018)].
\newblock \href {http://arxiv.org/abs/1310.4329} {\path{arXiv:1310.4329}}.

\bibitem{Park:2012cp}
S. Park and S. Dodelson.
\newblock {\em {Structure formation in a nonlocally modified gravity model}}.
\newblock
  {\hypersetup{urlcolor=journalcolor}\href{https://doi.org/10.1103/PhysRevD.87.024003}
  {Phys. Rev. D}} {\bfseries 87} 024003 (2013).
\newblock \href {http://arxiv.org/abs/1209.0836} {\path{arXiv:1209.0836}}.

\bibitem{Nersisyan:2017mgj}
H. Nersisyan, A.~F. Cid, and L. Amendola.
\newblock {\em {Structure formation in the Deser-Woodard nonlocal gravity
  model: a reappraisal}}.
\newblock
  {\hypersetup{urlcolor=journalcolor}\href{https://doi.org/10.1088/1475-7516/2017/04/046}
  {JCAP}} {\bfseries 04} 046 (2017).
\newblock \href {http://arxiv.org/abs/1701.00434} {\path{arXiv:1701.00434}}.

\bibitem{Park:2017zls}
S. Park.
\newblock {\em {Revival of the Deser-Woodard nonlocal gravity model: Comparison
  of the original nonlocal form and a localized formulation}}.
\newblock
  {\hypersetup{urlcolor=journalcolor}\href{https://doi.org/10.1103/PhysRevD.97.044006}
  {Phys. Rev.}} {\bfseries D97} 044006 (2018).
\newblock \href {http://arxiv.org/abs/1711.08759} {\path{arXiv:1711.08759}}.

\bibitem{Park:2019btx}
S. Park and R.~P. Woodard.
\newblock {\em {Exciting the scalar ghost mode through time evolution}}.
\newblock
  {\hypersetup{urlcolor=journalcolor}\href{https://doi.org/10.1103/PhysRevD.99.024014}
  {Phys. Rev.}} {\bfseries D99} 024014 (2019).
\newblock \href {http://arxiv.org/abs/1809.06841} {\path{arXiv:1809.06841}}.

\bibitem{Dvali:2006su}
G. Dvali.
\newblock {\em {Predictive Power of Strong Coupling in Theories with Large
  Distance Modified Gravity}}.
\newblock
  {\hypersetup{urlcolor=journalcolor}\href{https://doi.org/10.1088/1367-2630/8/12/326}
  {New J. Phys.}} {\bfseries 8} 326 (2006).
\newblock \href {http://arxiv.org/abs/hep-th/0610013}
  {\path{arXiv:hep-th/0610013}}.

\bibitem{Deser:2007jk}
S. Deser and R. Woodard.
\newblock {\em {Nonlocal Cosmology}}.
\newblock
  {\hypersetup{urlcolor=journalcolor}\href{https://doi.org/10.1103/PhysRevLett.99.111301}
  {Phys. Rev. Lett.}} {\bfseries 99} 111301 (2007).
\newblock \href {http://arxiv.org/abs/0706.2151} {\path{arXiv:0706.2151}}.

\bibitem{Woodard:2014iga}
R.~P. Woodard.
\newblock {\em {Nonlocal Models of Cosmic Acceleration}}.
\newblock
  {\hypersetup{urlcolor=journalcolor}\href{https://doi.org/10.1007/s10701-014-9780-6}
  {Found. Phys.}} {\bfseries 44} 213 (2014).
\newblock \href {http://arxiv.org/abs/1401.0254} {\path{arXiv:1401.0254}}.

\bibitem{Koivisto:2008xfa}
T. Koivisto.
\newblock {\em {Dynamics of Nonlocal Cosmology}}.
\newblock
  {\hypersetup{urlcolor=journalcolor}\href{https://doi.org/10.1103/PhysRevD.77.123513}
  {Phys. Rev.}} {\bfseries D77} 123513 (2008).
\newblock \href {http://arxiv.org/abs/0803.3399} {\path{arXiv:0803.3399}}.

\bibitem{Koivisto:2008dh}
T.~S. Koivisto.
\newblock {\em {Newtonian limit of nonlocal cosmology}}.
\newblock
  {\hypersetup{urlcolor=journalcolor}\href{https://doi.org/10.1103/PhysRevD.78.123505}
  {Phys. Rev. D}} {\bfseries 78} 123505 (2008).
\newblock \href {http://arxiv.org/abs/0807.3778} {\path{arXiv:0807.3778}}.

\bibitem{Amendola:2019fhc}
L. Amendola, Y. Dirian, H. Nersisyan, and S. Park.
\newblock {\em {Observational Constraints in Nonlocal Gravity: the
  Deser-Woodard Case}}.
\newblock
  {\hypersetup{urlcolor=journalcolor}\href{https://doi.org/10.1088/1475-7516/2019/03/045}
  {JCAP}} {\bfseries 1903} 045 (2019).
\newblock \href {http://arxiv.org/abs/1901.07832} {\path{arXiv:1901.07832}}.

\bibitem{Deser:2019lmm}
S. Deser and R.~P. Woodard.
\newblock {\em {Nonlocal Cosmology II --- Cosmic acceleration without fine
  tuning or dark energy}}.
\newblock {\hypersetup{urlcolor=journalcolor}\href{https://doi.org/} {2019}}).
\newblock \href {http://arxiv.org/abs/1902.08075} {\path{arXiv:1902.08075}}.

\bibitem{Deffayet:2009ca}
C. Deffayet and R.~P. Woodard.
\newblock {\em {Reconstructing the Distortion Function for Nonlocal
  Cosmology}}.
\newblock
  {\hypersetup{urlcolor=journalcolor}\href{https://doi.org/10.1088/1475-7516/2009/08/023}
  {JCAP}} {\bfseries 0908} 023 (2009).
\newblock \href {http://arxiv.org/abs/0904.0961} {\path{arXiv:0904.0961}}.

\bibitem{Maggiore:2014sia}
M. Maggiore and M. Mancarella.
\newblock {\em {Nonlocal gravity and dark energy}}.
\newblock
  {\hypersetup{urlcolor=journalcolor}\href{https://doi.org/10.1103/PhysRevD.90.023005}
  {Phys. Rev.}} {\bfseries D90} 023005 (2014).
\newblock \href {http://arxiv.org/abs/1402.0448} {\path{arXiv:1402.0448}}.

\bibitem{Dirian:2014ara}
Y. Dirian, S. Foffa, N. Khosravi, M. Kunz, and M. Maggiore.
\newblock {\em {Cosmological perturbations and structure formation in nonlocal
  infrared modifications of general relativity}}.
\newblock
  {\hypersetup{urlcolor=journalcolor}\href{https://doi.org/10.1088/1475-7516/2014/06/033}
  {JCAP}} {\bfseries 1406} 033 (2014).
\newblock \href {http://arxiv.org/abs/1403.6068} {\path{arXiv:1403.6068}}.

\bibitem{Nersisyan:2016hjh}
H. Nersisyan, Y. Akrami, L. Amendola, T.~S. Koivisto, and J. Rubio.
\newblock {\em {Dynamical analysis of $R\dfrac{1}{\Box^{2}}R$ cosmology: Impact
  of initial conditions and constraints from supernovae}}.
\newblock
  {\hypersetup{urlcolor=journalcolor}\href{https://doi.org/10.1103/PhysRevD.94.043531}
  {Phys. Rev.}} {\bfseries D94} 043531 (2016).
\newblock \href {http://arxiv.org/abs/1606.04349} {\path{arXiv:1606.04349}}.

\bibitem{Dirian:2016puz}
Y. Dirian, S. Foffa, M. Kunz, M. Maggiore, and V. Pettorino.
\newblock {\em {Non-local gravity and comparison with observational datasets.
  II. Updated results and Bayesian model comparison with $\Lambda$CDM}}.
\newblock
  {\hypersetup{urlcolor=journalcolor}\href{https://doi.org/10.1088/1475-7516/2016/05/068}
  {JCAP}} {\bfseries 1605} 068 (2016).
\newblock \href {http://arxiv.org/abs/1602.03558} {\path{arXiv:1602.03558}}.

\bibitem{Migkas:2020fza}
K. Migkas, G. Schellenberger, T. Reiprich, F. Pacaud, M. Ramos-Ceja, and L.
  Lovisari.
\newblock {\em {Probing cosmic isotropy with a new X-ray galaxy cluster sample
  through the $L_{\text{X}}-T$ scaling relation}}.
\newblock
  {\hypersetup{urlcolor=journalcolor}\href{https://doi.org/10.1051/0004-6361/201936602}
  {Astron. Astrophys.}} {\bfseries 636} A15 (2020).
\newblock \href {http://arxiv.org/abs/2004.03305} {\path{arXiv:2004.03305}}.

\bibitem{Bartelmann:2016dvf}
M. Bartelmann and M. Maturi.
\newblock {\em {Weak gravitational lensing}}.
\newblock 2016.
\newblock \href {http://arxiv.org/abs/1612.06535} {\path{arXiv:1612.06535}}.

\bibitem{Weinberg:2008zzc}
S. Weinberg.
\newblock {\em {Cosmology}}.
\newblock 2008.

\bibitem{2004mmu..symp..117K}
C.~S. {Kochanek} and P.~L. {Schechter}.
\newblock {\em {The Hubble Constant from Gravitational Lens Time Delays}}.
\newblock In W.~L. {Freedman}, editor, {\em Measuring and Modeling the
  Universe} page 117 2004.
\newblock \href {http://arxiv.org/abs/astro-ph/0306040}
  {\path{arXiv:astro-ph/0306040}}.

\bibitem{Suyu:2012aa}
S. Suyu et~al.
\newblock {\em {Two accurate time-delay distances from strong lensing:
  Implications for cosmology}}.
\newblock
  {\hypersetup{urlcolor=journalcolor}\href{https://doi.org/10.1088/0004-637X/766/2/70}
  {Astrophys. J.}} {\bfseries 766} 70 (2013).
\newblock \href {http://arxiv.org/abs/1208.6010} {\path{arXiv:1208.6010}}.

\bibitem{Suyu:2016qxx}
S. Suyu et~al.
\newblock {\em {H0LiCOW -- I. H0 Lenses in COSMOGRAIL's Wellspring: program
  overview}}.
\newblock
  {\hypersetup{urlcolor=journalcolor}\href{https://doi.org/10.1093/mnras/stx483}
  {Mon. Not. Roy. Astron. Soc.}} {\bfseries 468} 2590 (2017).
\newblock \href {http://arxiv.org/abs/1607.00017} {\path{arXiv:1607.00017}}.

\bibitem{Wong:2019kwg}
K.~C. Wong et~al.
\newblock {\em {H0LiCOW XIII. A 2.4\% measurement of $H_{0}$ from lensed
  quasars: $5.3\sigma$ tension between early and late-Universe probes}}.
\newblock {\hypersetup{urlcolor=journalcolor}\href{https://doi.org/} {2019}}).
\newblock \href {http://arxiv.org/abs/1907.04869} {\path{arXiv:1907.04869}}.

\bibitem{Bonvin:2016crt}
V. Bonvin et~al.
\newblock {\em {H0LiCOW -- V. New COSMOGRAIL time delays of HE 0435$-$1223:
  $H_0$ to 3.8 per cent precision from strong lensing in a flat $\Lambda$CDM
  model}}.
\newblock
  {\hypersetup{urlcolor=journalcolor}\href{https://doi.org/10.1093/mnras/stw3006}
  {Mon. Not. Roy. Astron. Soc.}} {\bfseries 465} 4914 (2017).
\newblock \href {http://arxiv.org/abs/1607.01790} {\path{arXiv:1607.01790}}.

\bibitem{Collett:2018gpf}
T.~E. Collett, L.~J. Oldham, R.~J. Smith, M.~W. Auger, K.~B. Westfall, D.
  Bacon, R.~C. Nichol, K.~L. Masters, K. Koyama, and R. van~den Bosch.
\newblock {\em {A precise extragalactic test of General Relativity}}.
\newblock
  {\hypersetup{urlcolor=journalcolor}\href{https://doi.org/10.1126/science.aao2469}
  {Science}} {\bfseries 360} 1342 (2018).
\newblock \href {http://arxiv.org/abs/1806.08300} {\path{arXiv:1806.08300}}.

\bibitem{Yu:2018slt}
H. Yu and F. Wang.
\newblock {\em {Testing Weak Equivalence Principle with Strongly Lensed Cosmic
  Transients}}.
\newblock
  {\hypersetup{urlcolor=journalcolor}\href{https://doi.org/10.1140/epjc/s10052-018-6162-9}
  {Eur. Phys. J. C}} {\bfseries 78} 692 (2018).
\newblock \href {http://arxiv.org/abs/1801.01257} {\path{arXiv:1801.01257}}.

\bibitem{Jyoti:2019pez}
D. Jyoti, J.~B. Munoz, R.~R. Caldwell, and M. Kamionkowski.
\newblock {\em {Cosmic Time Slip: Testing Gravity on Supergalactic Scales with
  Strong-Lensing Time Delays}}.
\newblock
  {\hypersetup{urlcolor=journalcolor}\href{https://doi.org/10.1103/PhysRevD.100.043031}
  {Phys. Rev. D}} {\bfseries 100} 043031 (2019).
\newblock \href {http://arxiv.org/abs/1906.06324} {\path{arXiv:1906.06324}}.

\bibitem{Yang:2020eoh}
T. Yang, S. Birrer, and B. Hu.
\newblock {\em {The first simultaneous measurement of Hubble constant and
  post-Newtonian parameter from Time-Delay Strong Lensing}}.
\newblock {\hypersetup{urlcolor=journalcolor}\href{https://doi.org/} {2020}}).
\newblock \href {http://arxiv.org/abs/2003.03277} {\path{arXiv:2003.03277}}.

\bibitem{Shiralilou:2019div}
B. Shiralilou, M. Martinelli, G. Papadomanolakis, S. Peirone, F. Renzi, and A.
  Silvestri.
\newblock {\em {Strong Lensing Time Delay Constraints on Dark Energy: a
  Forecast}}.
\newblock {\hypersetup{urlcolor=journalcolor}\href{https://doi.org/} {2019}}).
\newblock \href {http://arxiv.org/abs/1910.03566} {\path{arXiv:1910.03566}}.

\bibitem{Tewes:2012gs}
M. Tewes, F. Courbin, and G. Meylan.
\newblock {\em {COSMOGRAIL XI: Techniques for time delay measurement in
  presence of microlensing}}.
\newblock
  {\hypersetup{urlcolor=journalcolor}\href{https://doi.org/10.1051/0004-6361/201220123}
  {Astron. Astrophys.}} {\bfseries 553} A120 (2013).
\newblock \href {http://arxiv.org/abs/1208.5598} {\path{arXiv:1208.5598}}.

\bibitem{Bonvin:2015jia}
V. Bonvin, M. Tewes, F. Courbin, T. Kuntzer, D. Sluse, and G. Meylan.
\newblock {\em {COSMOGRAIL: the COSmological MOnitoring of GRAvItational Lenses
  XV. Assessing the achievability and precision of time-delay measurements}}.
\newblock
  {\hypersetup{urlcolor=journalcolor}\href{https://doi.org/10.1051/0004-6361/201526704}
  {Astron. Astrophys.}} {\bfseries 585} A88 (2016).
\newblock \href {http://arxiv.org/abs/1506.07524} {\path{arXiv:1506.07524}}.

\bibitem{Dobler:2013rda}
G. Dobler, C. Fassnacht, T. Treu, P.~J. Marshall, K. Liao, A. Hojjati, E.
  Linder, and N. Rumbaugh.
\newblock {\em {Strong Lens Time Delay Challenge: I. Experimental Design}}.
\newblock
  {\hypersetup{urlcolor=journalcolor}\href{https://doi.org/10.1088/0004-637X/799/2/168}
  {Astrophys. J.}} {\bfseries 799} 168 (2015).
\newblock \href {http://arxiv.org/abs/1310.4830} {\path{arXiv:1310.4830}}.

\bibitem{Courbin:2017yvz}
F. Courbin et~al.
\newblock {\em {COSMOGRAIL: the COSmological MOnitoring of GRAvItational Lenses
  - XVI. Time delays for the quadruply imaged quasar DES J0408$-$5354 with
  high-cadence photometric monitoring}}.
\newblock
  {\hypersetup{urlcolor=journalcolor}\href{https://doi.org/10.1051/0004-6361/201731461}
  {Astron. Astrophys.}} {\bfseries 609} A71 (2018).
\newblock \href {http://arxiv.org/abs/1706.09424} {\path{arXiv:1706.09424}}.

\bibitem{Zitrin:2018let}
A. Zitrin and D. Eichler.
\newblock {\em {Observing Cosmological Processes in Real Time with Repeating
  Fast Radio Bursts}}.
\newblock
  {\hypersetup{urlcolor=journalcolor}\href{https://doi.org/10.3847/1538-4357/aad6a2}
  {Astrophys. J.}} {\bfseries 866} 101 (2018).
\newblock \href {http://arxiv.org/abs/1807.03287} {\path{arXiv:1807.03287}}.

\bibitem{Wucknitz:2020spz}
O. Wucknitz, L. Spitler, and U.-L. Pen.
\newblock {\em {Cosmology with gravitationally lensed repeating Fast Radio
  Bursts}}.
\newblock {\hypersetup{urlcolor=journalcolor}\href{https://doi.org/} {2020}}).
\newblock \href {http://arxiv.org/abs/2004.11643} {\path{arXiv:2004.11643}}.

\bibitem{Millon:2020xab}
M. Millon et~al.
\newblock {\em {COSMOGRAIL XIX: Time delays in 18 strongly lensed quasars from
  15 years of optical monitoring}}.
\newblock
  {\hypersetup{urlcolor=journalcolor}\href{https://doi.org/10.1051/0004-6361/202037740}
  {Astron. Astrophys.}} {\bfseries 640} A105 (2020).
\newblock \href {http://arxiv.org/abs/2002.05736} {\path{arXiv:2002.05736}}.

\bibitem{Liu:2019jka}
B. Liu, Z. Li, H. Gao, and Z. Zhu.
\newblock {\em {Prospects of strongly lensed repeating fast radio bursts:
  Complementary constraints on dark energy evolution}}.
\newblock
  {\hypersetup{urlcolor=journalcolor}\href{https://doi.org/10.1103/PhysRevD.99.123517}
  {Phys. Rev. D}} {\bfseries 99} 123517 (2019).
\newblock \href {http://arxiv.org/abs/1907.10488} {\path{arXiv:1907.10488}}.

\end{thebibliography}
}

\end{document}